\documentclass[numberedappendix]{emulateapj}
\usepackage{epsfig,lscape}
\begin{document}
\newcommand{\Lya}{Ly$\alpha$}
\newcommand{\Rc}{$R_{\rm C}$}
\newcommand{\galex} {{\it GALEX}}
\newcommand{\nuv}{{\it NUV}}
\newcommand{\EBV}{E(B-V)}
\newcommand{\fcont}{$f_{\rm contam}$}
\newcommand{\color}{[{\it See the electronic edition of the Journal for a color version of this figure}.]}

\title{Lyman Break Galaxies at $z\approx1.8-2.8$: \galex/\nuv~Imaging of the Subaru Deep
  Field\altaffilmark{1}}
\author{Chun Ly,\altaffilmark{2} Matthew A. Malkan,\altaffilmark{2} Tommaso Treu,\altaffilmark{3}
  Jong-Hak Woo,\altaffilmark{2,4} Thayne Currie,\altaffilmark{5} Masao Hayashi,\altaffilmark{6}
  Nobunari Kashikawa,\altaffilmark{7,8} Kentaro Motohara,\altaffilmark{9}
  Kazuhiro Shimasaku,\altaffilmark{6,10} and Makiko Yoshida\altaffilmark{6}}
\shorttitle{Lyman Break Galaxies at $z=1.8-2.8$}
\shortauthors{Ly et al.}
\email{chun@astro.ucla.edu}
\altaffiltext{1}{Based on data obtained at the W.M. Keck Observatory (operated as a scientific partnership
  among the California Institute of Technology, the University of California, and NASA), the Subaru
  Telescope (operated by the National Astronomical Observatory of Japan), and the MMT Observatory (a joint
  facility of the University of Arizona and the Smithsonian Institution).}
\altaffiltext{2}{Department of Physics and Astronomy, UCLA, Los Angeles, CA.}
\altaffiltext{3}{Department of Physics, UCSB, Santa Barbara, CA.}
\altaffiltext{4}{Hubble fellow.}
\altaffiltext{5}{Harvard-Smithsonian Center for Astrophysics, Cambridge, MA.}
\altaffiltext{6}{Department of Astronomy, School of Science, University of Tokyo, Bunkyo, Tokyo, Japan.}
\altaffiltext{7}{Optical and Infrared Astronomy Division, National Astronomical Observatory, Mitaka,
  Tokyo, Japan.}
\altaffiltext{8}{Department of Astronomy, School of Science, Graduate University for Advanced Studies,
  Mitaka, Tokyo, Japan.}
\altaffiltext{9}{Institute of Astronomy, University of Tokyo, Mitaka, Tokyo, Japan.}
\altaffiltext{10}{Research Center for the Early Universe, School of Science, University of Tokyo, Tokyo,
  Japan.}

\begin{abstract}
  A photometric sample of $\sim$7100 $V<25.3$ Lyman break galaxies (LBGs) has been selected
  by combining Subaru/Suprime-Cam $BV$\Rc$i$\arcmin$z$\arcmin\ optical data with deep \galex/\nuv\ imaging
  of the Subaru Deep Field. Follow-up spectroscopy
  confirmed 24 LBGs at $1.5\lesssim z\lesssim2.7$. Among the optical spectra, 12 have \Lya~emission with
  rest-frame equivalent widths of $\approx5-60$\AA. The success rate for identifying LBGs as \nuv-dropouts
  at $1.5<z<2.7$ is 86\%. The rest-frame UV (1700\AA) luminosity function (LF) is constructed from the
  photometric sample with corrections for stellar contamination and $z<1.5$ interlopers (lower limits). The
  LF is $1.7\pm0.1$ ($1.4\pm0.1$ with a hard upper limit on stellar contamination) times higher than those
  of $z\sim2$ BXs and $z\sim3$ LBGs. Three explanations were considered, and it is argued that significantly
  underestimating low-$z$ contamination or effective comoving volume is unlikely: the former would be
  inconsistent with the spectroscopic sample at 93\% confidence, and the second explanation would not
  resolve the discrepancy. The third scenario is that different photometric selection of the samples yields
  non-identical galaxy populations, such that some BX galaxies are LBGs and vice versa. This argument is
  supported by a higher surface density of LBGs at all magnitudes while the redshift distribution of the two
  populations is nearly identical. This study, when combined with other star-formation rate (SFR) density UV
  measurements from LBG surveys, indicates that there is a rise in the SFR density: a factor of $3-6$ ($3-10$)
  increase from $z\sim5$ ($z\sim6$) to $z\sim2$, followed by a decrease to $z\sim0$. This result, along with
  past sub-mm studies that find a peak at $z\sim2$ in their redshift distribution, suggest that $z\sim2$ is
  the epoch of peak star-formation. Additional spectroscopy is required to characterize the complete shape of
  the $z\sim2$ LBG UV LF via measurements of AGN, stellar, and low-$z$ contamination and accurate distances.
\end{abstract}

\keywords{
  galaxies: photometry --- galaxies: high redshift ---
  galaxies: luminosity function --- galaxies: evolution
}
  
\section{INTRODUCTION}\label{1}
Over the past decade, the number of Lyman break galaxies \citep[LBGs; for a review, see][]{giavalisco02}
identified at $z\sim3-6$ has grown rapidly from deep, wide-field optical imaging surveys
\citep[e.g.,][]{steidel99,bouwens06,yoshida06}. Follow-up spectroscopy on large telescopes has shown
that this method (called the Lyman break technique or the ``drop-out'' method) is efficient at
identifying high-$z$ star-forming galaxies. Furthermore, these studies have measured the cosmic
star-formation history (SFH) at $z>3$, which is key for understanding galaxy evolution. It indicates
that the star-formation rate (SFR) density is 10 or more times higher in the past than at $z\sim0$.\\
\indent Extending the Lyman break technique to $z<3$ requires deep, wide-field UV imaging from space,
which is difficult. In addition, [\ion{O}{2}] (the bluest optical nebular emission line) is redshifted
into the near-infrared (NIR) for $z\gtrsim1.5$ where high background and lower sensitivity limit surveys
to small samples \citep[e.g.,][]{malkan96,moorwood00,werf00,erb03}. The combination of these observational
limitations has made it difficult to probe $z\approx1.5-2.5$.\\
\indent One solution to the problem is the `BX' method developed by \cite{adelberger04}. This technique
identifies blue galaxies that are detected in $U$, but show a moderately red $U-G$ color when the Lyman
continuum break begins to enter into the $U$-band at $z\sim2$.\\
\indent Other methods have used NIR imaging to identify galaxies at $z=1-3$ via the Balmer/4000\AA\ break.
For example, selection of objects with $J-K>2.3$ (Vega) has yielded ``distant red galaxies'' at
$z\sim2-3$ \citep{dokkum04}, and the `{\it BzK}' method has found passive and star-forming (dusty and
less dusty) galaxies at $z\approx1.5-2.5$ \citep{daddi04,hayashi07}. The completeness of these methods is
not as well understood as UV-selected techniques, since limited spectra have been obtained.\\
\indent In this paper, the Lyman break technique is extended down to $z\sim1.8$ with wide-field, deep
\nuv~imaging of the Subaru Deep Field (SDF) with the Galaxy Evolution Explorer \citep[\galex;][]{martin05}.
This survey has the advantage of sampling a large contiguous area, which allows for large scale
structure studies (to be discussed in future work), an accurate measurement of a large portion of the
luminosity function, and determining if the SFH peaks at $z\sim2$.\\
\indent In \S~\ref{2}, the photometric and spectroscopic observations are described. Section~\ref{3}
presents the color selection criteria to produce a photometric sample of \nuv-dropouts, which are objects
undetected or very faint in the \nuv, but present in the optical. The removal of foreground stars and
low-$z$ galaxy contaminants, and the sample completeness are discussed in \S~\ref{4}. In \S~\ref{5}, the
observed UV luminosity function (LF) is constructed from $\sim$7100 \nuv-dropouts in the SDF, and the
comoving star-formation rate (SFR) density at $z=1.8-2.8$ is determined. Comparisons of these results
with previous surveys are described in \S~\ref{6}, and a discussion is provided in \S~\ref{7}. The
appendix includes a description of objects with unusual spectral properties. 
A flat cosmology with [$\Omega_{\Lambda}$, $\Omega_M$, $h_{70}$] = [0.7, 0.3, 1.0] is adopted for
consistency with recent LBG studies. All magnitudes are reported on the AB system \citep{oke74}.

\section{OBSERVATIONS}\label{2}
This section describes the deep \nuv~data obtained (\S~\ref{2.1}), followed by the spectroscopic
observations (\S~\ref{2.2} and \ref{2.4}) from Keck, Subaru, and MMT (Multiple Mirror Telescope). An
objective method for obtaining redshifts, cross-correlating spectra with templates, is presented
(\S~\ref{2.3}) and confirms that most \nuv-dropouts are at $z\sim2$. These spectra are later used
in \S~\ref{3.2} to define the final empirical selection criteria for $z\sim2$ LBGs. A summary of the
success rate for finding $z\sim2$ galaxies as \nuv-dropouts is included.

\subsection{\galex/\nuv~Imaging of the SDF}\label{2.1}
The SDF \citep{kashik04}, centered at $\alpha$(J2000) = 13$^{\rm h}$24$^{\rm m}$38\fs9, $\delta$(J2000) =
+27\arcdeg29\arcmin25\farcs9, is a deep wide-field (857.5 arcmin$^2$) extragalactic survey with optical
data obtained from Suprime-Cam \citep{miyazaki02}, the prime-focus camera mounted on the Subaru Telescope
\citep{iye04}. It was imaged with \galex~in the \nuv~($1750-2750$\AA) between 2005 March 10 and 2007 May 29
(GI1-065) with a total integration time of 138176 seconds.
A total of 37802 objects are detected in the full \nuv~image down to a depth of $\approx$27.0 mag
(3$\sigma$, 7.5\arcsec~diameter aperture). The \galex-SDF photometric catalog will be presented in future
work. For now, objects undetected or faint ($NUV>25.5$) in the \nuv~are discussed.\\
\indent The \nuv~image did not require mosaicking to cover the SDF, since the \galex~field-of-view (FOV) is
larger and the center of the SDF is located at (+3.87\arcmin, +3.72\arcsec) from the center of the
\nuv~image. The \nuv~spatial resolution (FWHM) is 5.25\arcsec, and was found to vary by no more than 6\%
across the region of interest \citep{morrissey07}.

\subsection{Follow-up Spectroscopy}\label{2.2}
\subsubsection{Keck/LRIS}\label{2.2.1}
\begin{figure} 
  \epsscale{1.0}
  \plotone{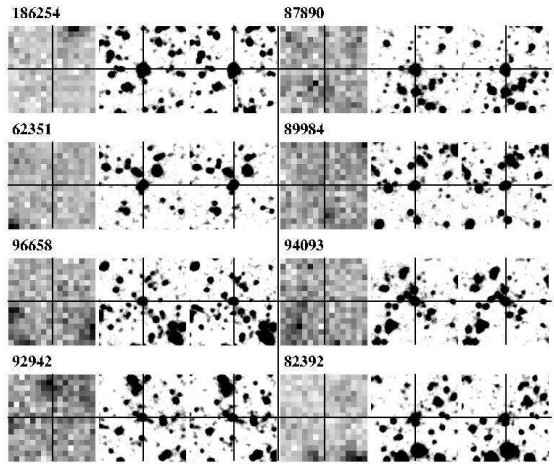}
  \caption{Postage stamps for some \nuv-dropouts targeted with LRIS. From left to right is \nuv, $B$, and
    $V$. Each image is 24\arcsec~on a side and reveals that optical sources do not have a
    \nuv~counterpart. Photometric and spectroscopic information are provided in Table~\ref{table1}.}
  \label{postage}
\end{figure}

When objects for Keck spectroscopy were selected, the \nuv~observations had accumulated 79598 seconds.
Although the selection criteria and photometric catalog are revised later in this paper, a brief
description of the original selection is provided, since it is the basis for the Keck sample. An
{\it initial} \nuv-dropout catalog (hereafter ver. 1) of sources with $NUV-B>1.5$ and $B-V<0.5$ was
obtained. No aperture correction was applied to the 7.5\arcsec~aperture \nuv~flux and the
2\arcsec~aperture was used for optical photometry. These differ from the final selection discussed in
\S~\ref{3.2}. The \nuv~3$\sigma$ limiting magnitude for the ver. 1 catalog is 27.0 within a
3.39\arcsec~radius aperture. Postage stamps (see Figure~\ref{postage}) were examined for follow-up targets
to ensure that they are indeed \nuv-dropouts.
%
%
The Keck Low Resolution Imaging and Spectrograph \citep[LRIS;][]{oke95} was used to target
candidate LBGs in multi-slit mode on 2007 January 23$-$25. The total integration times were either 3400,
3600, or 4833 seconds, and 36 \nuv-dropouts were targeted within 3 slitmasks. A dichroic beam splitter was
used with the 600 lines mm$^{-1}$ grism blazed at 4000\AA\ and the 400 lines mm$^{-1}$ grating blazed at
8500\AA, yielding blue (red) spectral coverage of $3500-5300$\AA\ ($6600-9000$\AA), although the coverages
varied with location along the dispersion axis. The slits were 4\arcsec~to 8\arcsec~in length and
1\arcsec~in width, yielding spectral resolution of $\approx$0.9\AA\ at 4300\AA\ and $\approx$1.2\AA\ at
8000\AA.\\
\indent Standard methods for reducing optical spectra were followed in PyRAF where an IRAF script,
developed by K. Adelberger to reduce LRIS data, was used. When reducing the blue spectra, dome flat-fields
were not used due to the known LRIS ghosting problem. Other LRIS users have avoided flat-fielding their
blue spectra, since the CCD response is mostly flat (D. Stern, priv. comm).\\
\indent HgNe arc-lamps were used for wavelength calibration of the blue side while OH sky-lines were used
for the red side. Typical wavelength RMS was less than 0.1\AA. For flux calibration, long-slit spectra of
BD+26 2606 \citep{oke83} were obtained following the last observation for each night.\\
\indent In the first mask, three of five alignment stars had coordinates that were randomly off by as much
as 1\arcsec~from the true coordinates. These stars were taken from the USNO catalog, where as the better
alignment stars were from the 2MASS catalog with a few tenths of an arcsecond offsets. This hindered
accurate alignment, and resulted in a lower success rate of detection: the first mask had 7 of 12
\nuv-dropouts that were {\it not} identified, while the other two masks had 2/10 and 3/14.

\subsubsection{MMT/Hectospec}\label{2.2.2}
Spectra of \nuv-dropouts from the final photometric catalog were obtained with the multifiber optical
spectrograph Hectospec \citep{fabricant05} on the 6.5m MMT on
2008 March 13 and April 10, 11, and 14. Compared to Keck/LRIS, MMT/Hectospec has a smaller collecting
area and lower throughput in the blue, so fewer detections were anticipated. Therefore, observations
were restricted to bright ($V_{\texttt{auto}} = 22.0-23.0$) sources, which used 21 of 943 fibers from
four configurations. Each source was observed in four, six, or seven 20-minute exposures using the
270 mm$^{-1}$ grating. This yielded a spectral coverage of $4000-9000$\AA\ with 6\AA\ resolution. The
spectra were wavelength calibrated, and sky-subtracted using the standard Hectospec reduction pipeline
\citep{fabricant05}. A more detailed discussion of these observations is deferred to a forthcoming
paper (Ly et al. 2008, in prep.).

\subsection{Spectroscopic Identification of Sources}\label{2.3}
\begin{figure} 
  \plotone{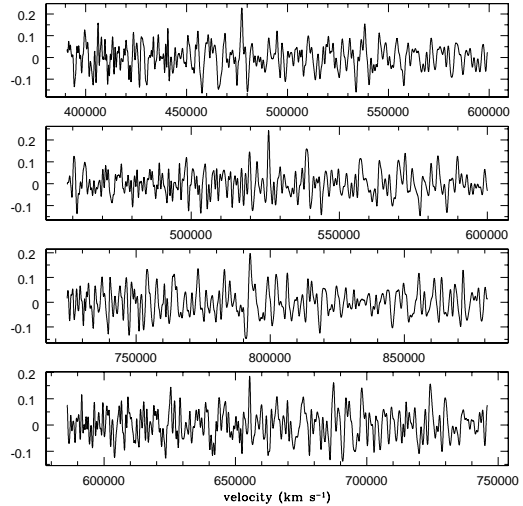}
  \caption{Cross-correlation spectra for targets that yielded $R=2.5-3.2$ without an emission line. From
    top to bottom shows \nuv-dropout ID 182284, 186254, 96927, and 92942. The top two have $R$-values of
    $\gtrsim3.1$ and are identified as LBGs while the latter two have $R=2.6-3.0$ and are classified as
    ambiguous. The peak near the center of the plots represents the strongest peak in the cross-correlation.}
  \label{xcplot}
\end{figure}

The IRAF task, \texttt{xcsao} from the \textsc{rvsao} package \citep[ver. 2.5.0]{kurtz98}, was used to
cross-correlate with six UV spectral templates of LBGs. For cases with \Lya~in emission, the composite
of 811 LBGs from \cite{shapley03} and the two top quartile bins (in \Lya~equivalent width) of
\cite{steidel03} were used. For sources lacking \Lya~emission (i.e., pure absorption-line systems), the
spectra of MS 1512-cB58 (hereafter `cB58') from \cite{pettini00}, and the two lowest quartile bins of
\cite{steidel03} were used.\\
\indent When no blue features were present, the red end of the spectrum was examined. An object could still
be at $z>1.5$, but at a low enough redshift for \Lya~to be shortward of the spectral window. In this case,
rest-frame NUV features, such as \ion{Fe}{2} and \ion{Mg}{2}, are available. \citet{savaglio04} provided a
composite rest-frame NUV spectrum of 13 star-forming galaxies at $1.3<z<2$. For objects below
$z\approx1.5$, optical features are available to determine redshift. The composite SDSS spectra
($3500-7000$\AA\ coverage) from \citet{yip04} and those provided with \textsc{rvsao} ($3000-7000$\AA)
are used for low-$z$ cases. Note that in computing redshifts, several different initial guesses were made
to determine the global peak of the cross-correlation. In most cases, the solutions converged to the same
redshift when the initial guesses are very different. The exceptions are classified as `ambiguous'.\\
\indent Where spectra had poor S/N, although a redshift was obtained for the source, the reliability of
identification (as given by \texttt{xcsao}'s $R$-value) was low ($R=2-3$). An objective test, which was
performed to determine what $R$-values are reliable, was to remove the \Lya~emission from those spectra
and templates, and then re-run \texttt{xcsao} to see what $R$-values are obtained based on absorption
line features in the spectra. Among 10 cases (from LRIS spectroscopy), 6 were reconfirmed at a similar
redshift ($\Delta z = 2.4\times10^{-4}-1.5\times10^{-3}$)\footnote[11]{This is lower, but still
  consistent with differences between emission and absorption redshifts of 650 km s$^{-1}$ for LBGs
  \citep{shapley03}.} with $R$-values of 2.30$-$7.07. This test indicates that a threshold of $R=2.5$ is
reasonable for defining whether the redshift of a source (lacking emission lines) was determined.
This cut is further supported by \cite{kurtz98}, who found that the success of determining redshifts
at $R=2.5-3$ is $\sim$90\%. However, to obtain more reliable redshifts, a more stricter $R=3.0$ threshold is adopted.
If a $R=2.5$ threshold is adopted, then seven sources with $R=2.5-3$ (ID 86765, 92942, 96927, 153628, 169090, 190498,
and 190947) are re-classified as `identified'. These
redshifts are marginally significant: a few to several absorption features coincide with the expected
UV lines for the best-fit redshifts of $\sim$2, but a few additional absorption lines are not evident
in the low S/N data. Statistics presented below are provided for both adopted $R$-value cuts.\\
\indent While some sources are classified ambiguous, it is likely that they could be at high-$z$. For example,
185177 (classified as ambiguous) could be a LBG, since it shows a weak emission line
at $\sim$4500\AA\ ($z\sim2.7$ if \Lya) and a few absorption lines. This source, statistics ($\sim$50\%
successful identification for $R=2.0-2.5$) from \cite{kurtz98}, and NUV-78625 (with $R=2.3$ but identified
`by eye' to be a $z\approx1.6$ AGN) suggest that while a cut is placed at $R=2.5$ or $R=3.0$, it could be
that some solutions with $R=2.0-3.0$ are correctly identified. An $R=3.0$ ($R=2.5$) typically corresponds to a peak of
$\sim$0.25 ($\sim$0.2) in the cross correlation spectra, which is typically 3$\sigma$ ($2-3\sigma$) above the RMS in the
cross-correlation (see Figure~\ref{xcplot}).

\subsubsection{LRIS Results}\label{2.3.1}
\begin{figure*} 
  \epsscale{1.0}
  \plotone{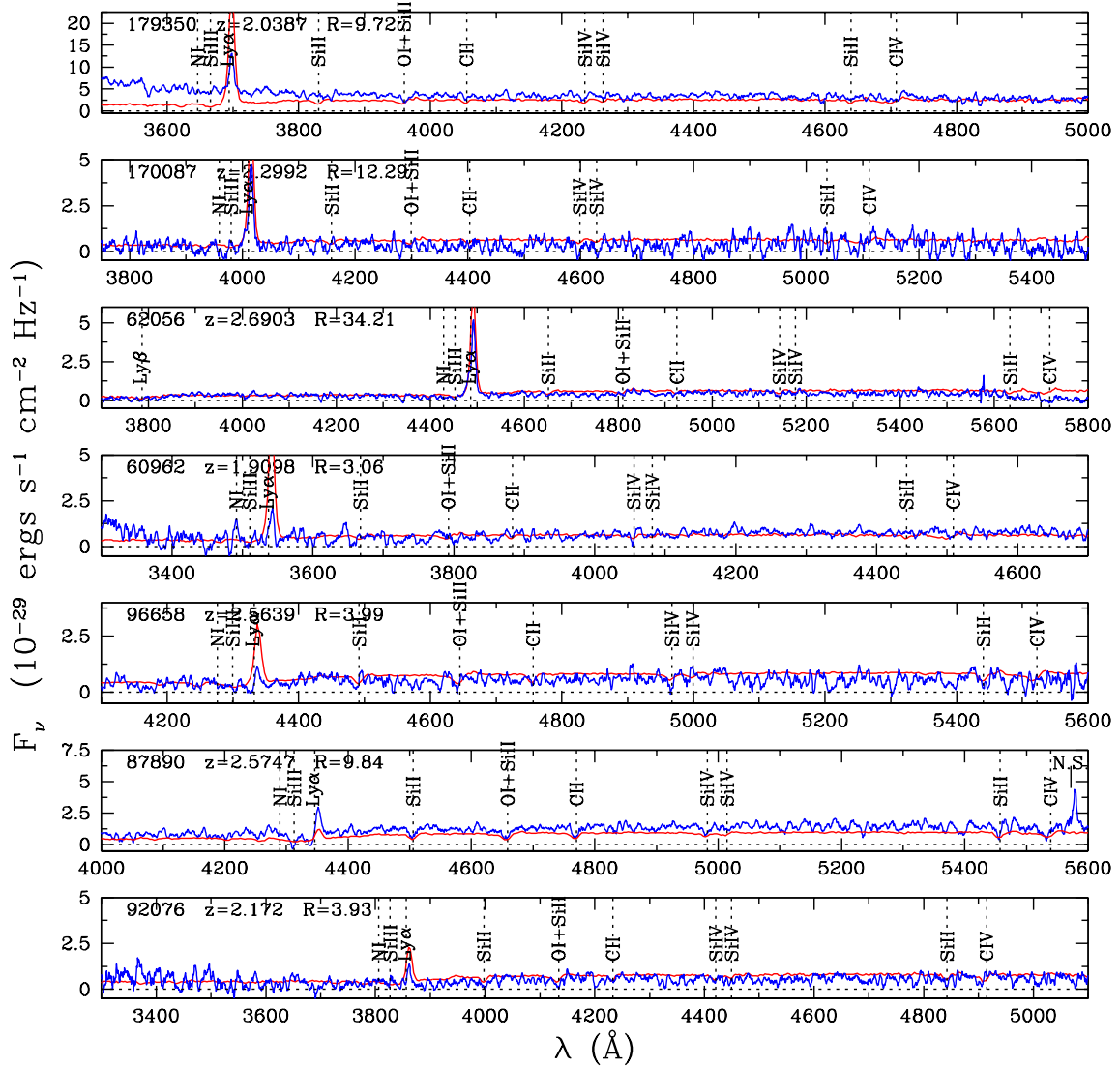}
  \caption{LRIS spectra of confirmed \nuv-dropouts from the ver. 1 catalog with known redshifts. Most of the LBGs
    with \Lya~emission are shown here with the remaining in Figure~\ref{spec2}. Overlayed on these
    spectra is the composite template (shown as grey) with the highest $R$-value (see Table~\ref{table1})
    from
    cross-correlation. Note that these overlayed templates are intended to show the location of spectral
    features, and is not meant to compare the flux and/or the spectral index differences between the
    spectra and the templates. The ID number, redshifts, and $R$-values are shown in the upper
    left-hand corner of each panel. \color}
  \label{spec1}
\end{figure*}

\begin{figure*} 
  \epsscale{1.0}
  \plotone{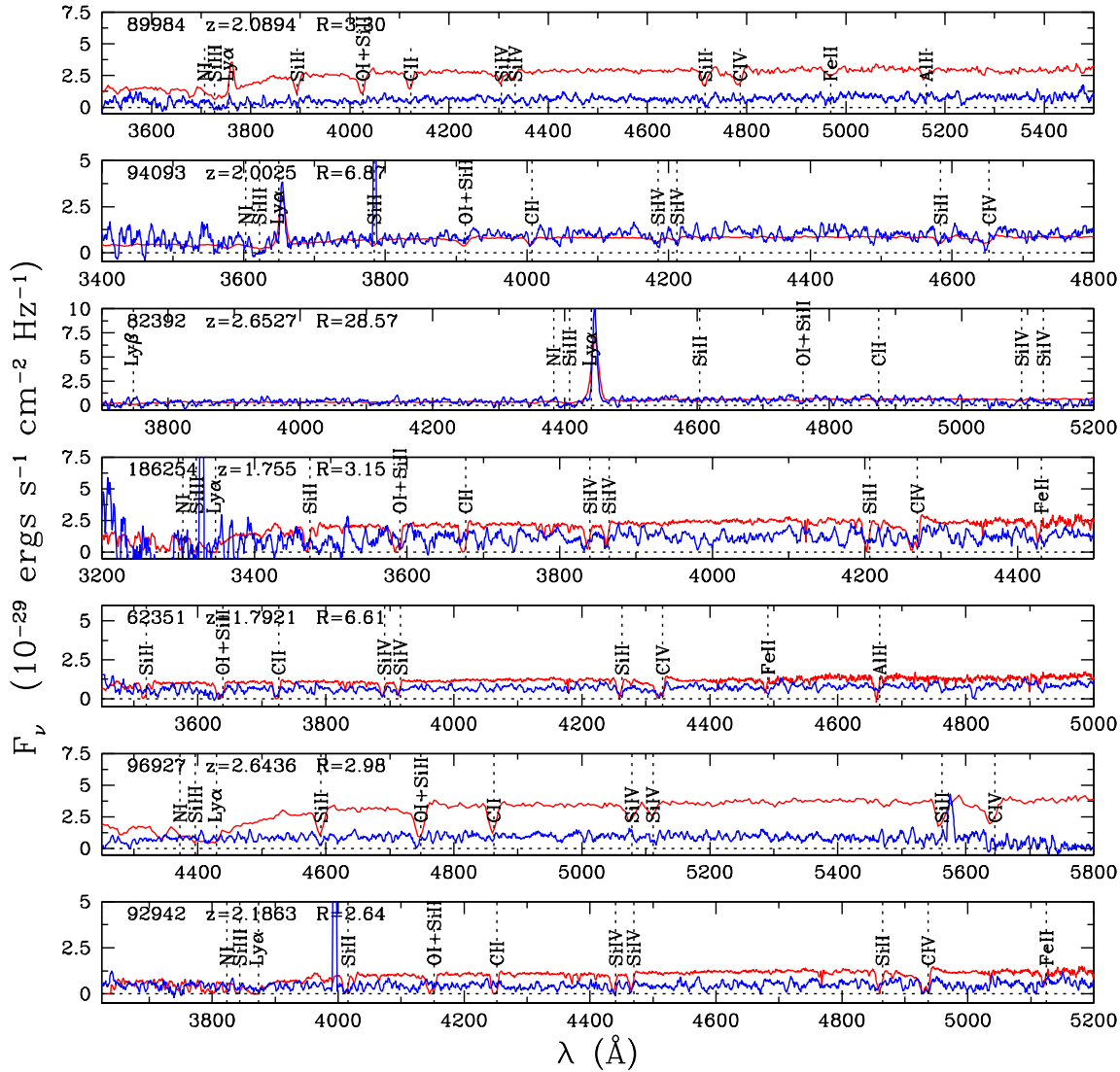}
  \caption{Same as Figure~\ref{spec1}, but some spectra do not have \Lya~emission. The strong line seen
    in the spectrum of 96927 at $\sim$5570\AA\ is a sky subtraction artifact, and cosmic rays are seen in
    the spectra of 94093 (at 3780\AA), 186254 (at 3325\AA), and 92942
    (at 3990\AA). These features are removed in the cross-correlation process. \color}
  \label{spec2}
\end{figure*}

\begin{figure*} 
  \epsscale{0.57}\plotone{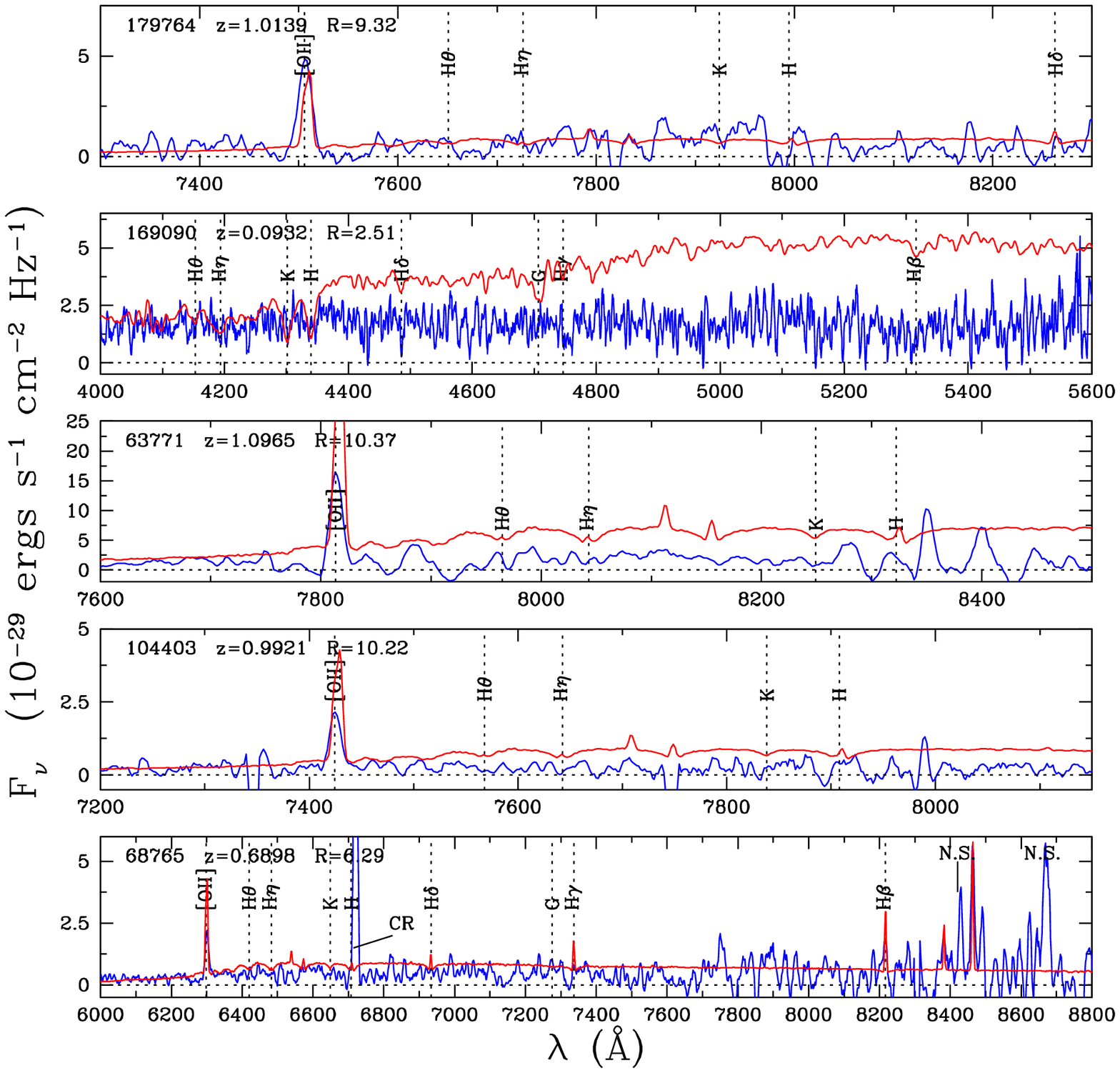}\epsscale{0.5478}\plotone{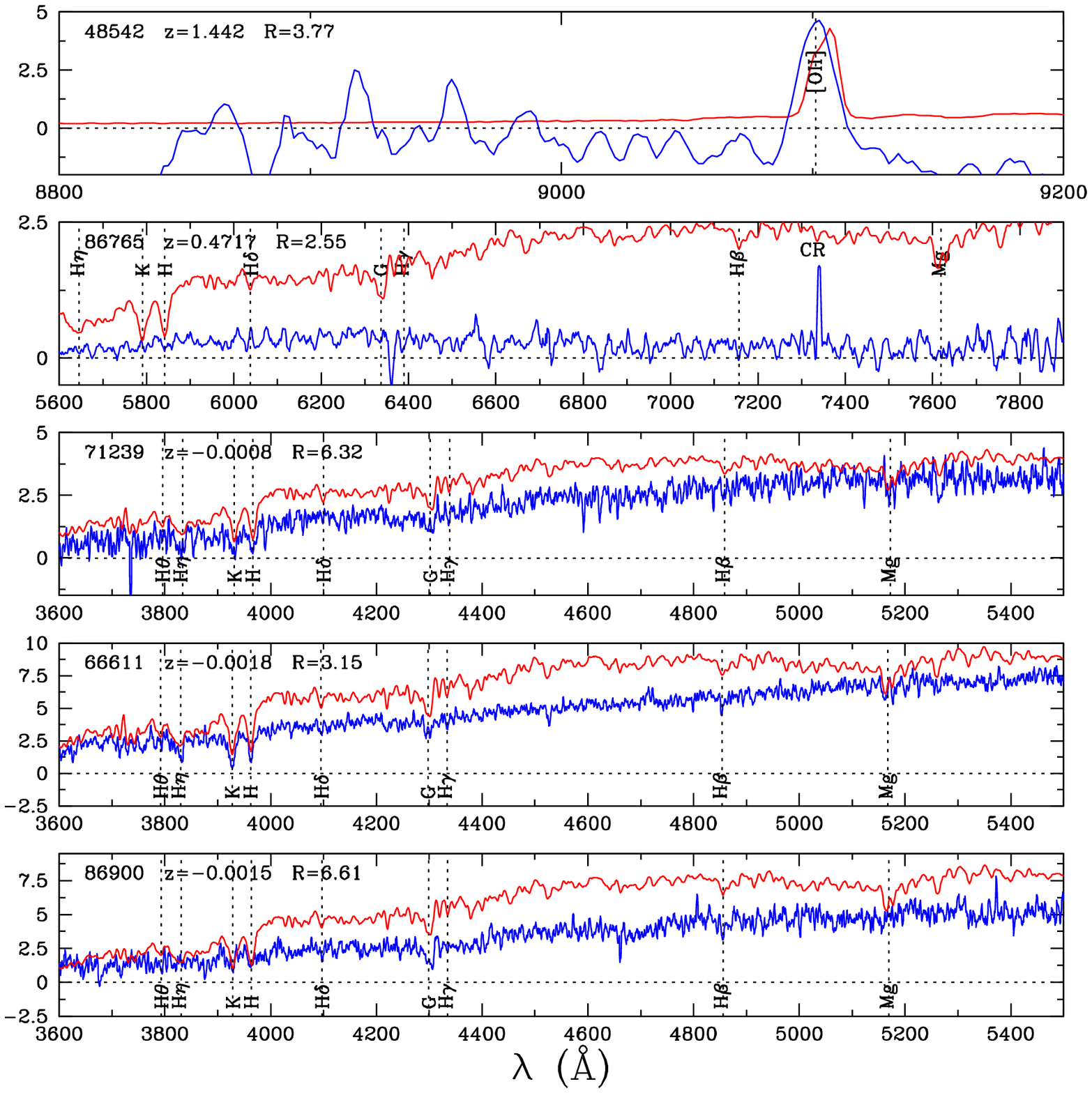}
  \caption{Same as Figures~\ref{spec1} and \ref{spec2}, but this shows the $z<1.5$ interlopers
    and galactic stars. \color }
  \label{spec3}
\end{figure*}

12 (14 with $R\geq2.5$) LBGs are found at $1.7\lesssim z\lesssim2.7$ out of 36 attempts. Among those, 10 show
\Lya~in emission, while 2 (4 with $R\geq2.5$) are identified purely by UV absorption lines. Their spectra are shown
in Figures~\ref{spec1} and \ref{spec2}, and Table~\ref{table1} summarizes their photometric and spectroscopic
properties. Contamination was found from 3 stars and 5 (7 with $R\geq2.5$) low-$z$ galaxies (shown in
Figure~\ref{spec3}), corresponding to a 60\% success rate (58\% if $R>2.5$ is adopted). Four sources showed a
single emission line, which is believed to be [\ion{O}{2}] at $z\sim1-1.5$, one source showed [\ion{O}{2}],
H$\beta$, and [\ion{O}{3}] at $z\sim0.7$, and two sources with absorption lines have $R\sim2.5$ results with
$z\sim0.1$ and $\sim0.5$ (these would be ``ambiguous'' with the $R\geq3.0$ criterion). The success of identifying $z\sim2$
LBGs improves with different color selection criteria that remove most interlopers (see \S~\ref{3.2}).\\
\indent Of the remaining 16 spectra (12 with $R>2.5$ cut), 8 (4 with $R>2.5$ cut) were detected, but the S/N of the
spectra was too low, and the other 8 were undetected. These objects were unsuccessful due to the short integration time
of about one hour and their faintness (average $V$ magnitude of 24.2).\\
\indent It is worthwhile to indicate that the fraction of LRIS spectra with \Lya\ emission is high (83\%). In comparison,
\cite{shapley03} reported that 68\% of their $z\sim3$ spectroscopic sample contained \Lya\ in emission. If the fraction
of LBGs with \Lya\ emission does not increase from $z\sim3$ to $z\sim2$, it would imply that 5 $z\sim2$ galaxies would
not show \Lya\ in emission. Considering the difficulties with detecting \Lya\ in absorption with relatively short
integration times, the above 83\% is not surprising, and suggests that most of the $z>1.5$ ambiguous LRIS redshifts
listed in Table~\ref{table1} are correct.

\subsubsection{Hectospec Results}\label{2.3.2}
\begin{figure*} 
  \epsscale{1.0}
  \plotone{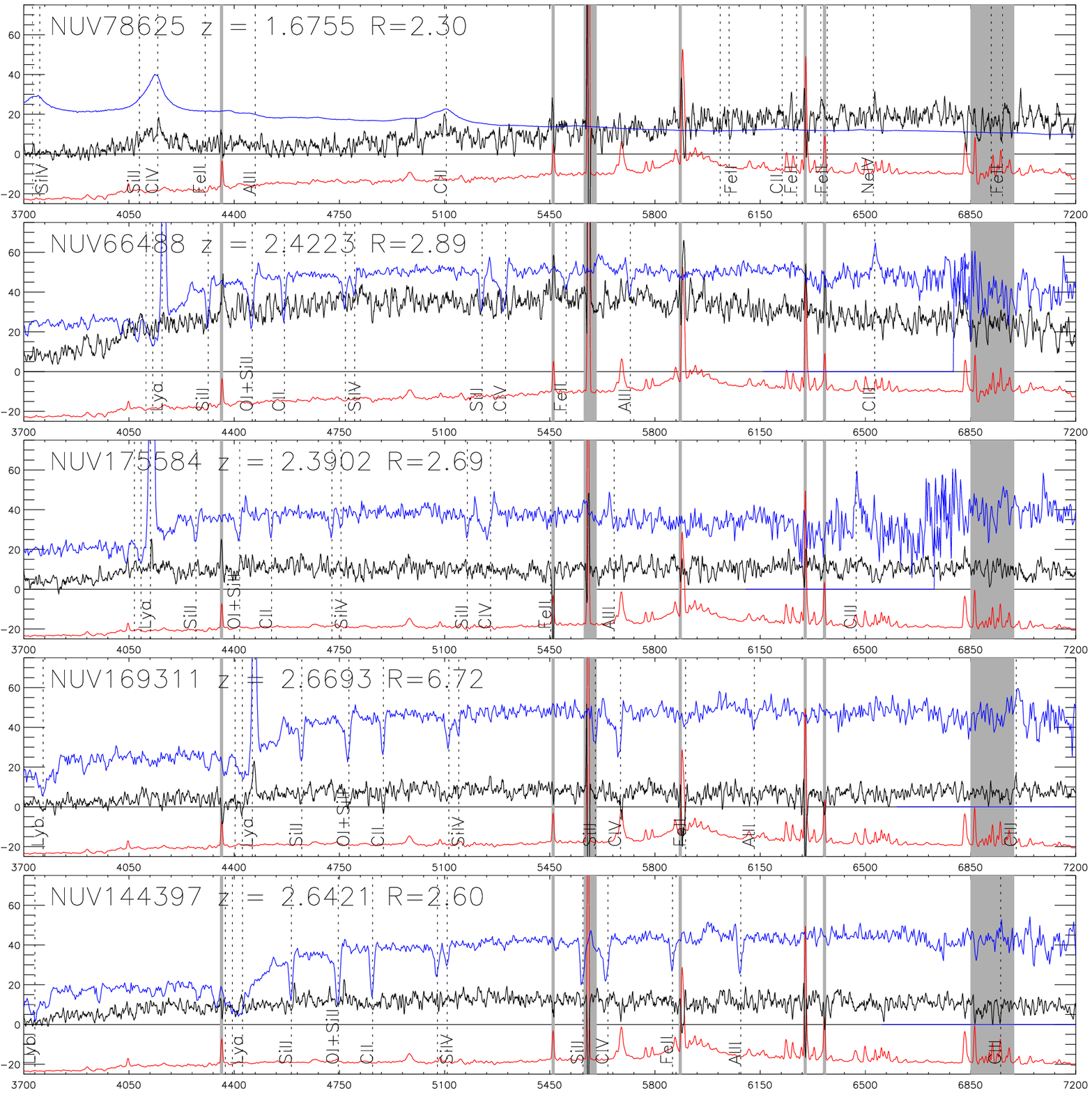}
  \caption{Same as Figures~\ref{spec1} and \ref{spec2}, but these are Hectospec observations of
    LBGs in the final photometric catalog. The cross-correlation template and the typical
    sky spectrum are shown above and below the spectrum of the source, respectively. \color}
  \label{spec4}
\end{figure*}

\begin{figure*} 
  \epsscale{1.0}
  \plotone{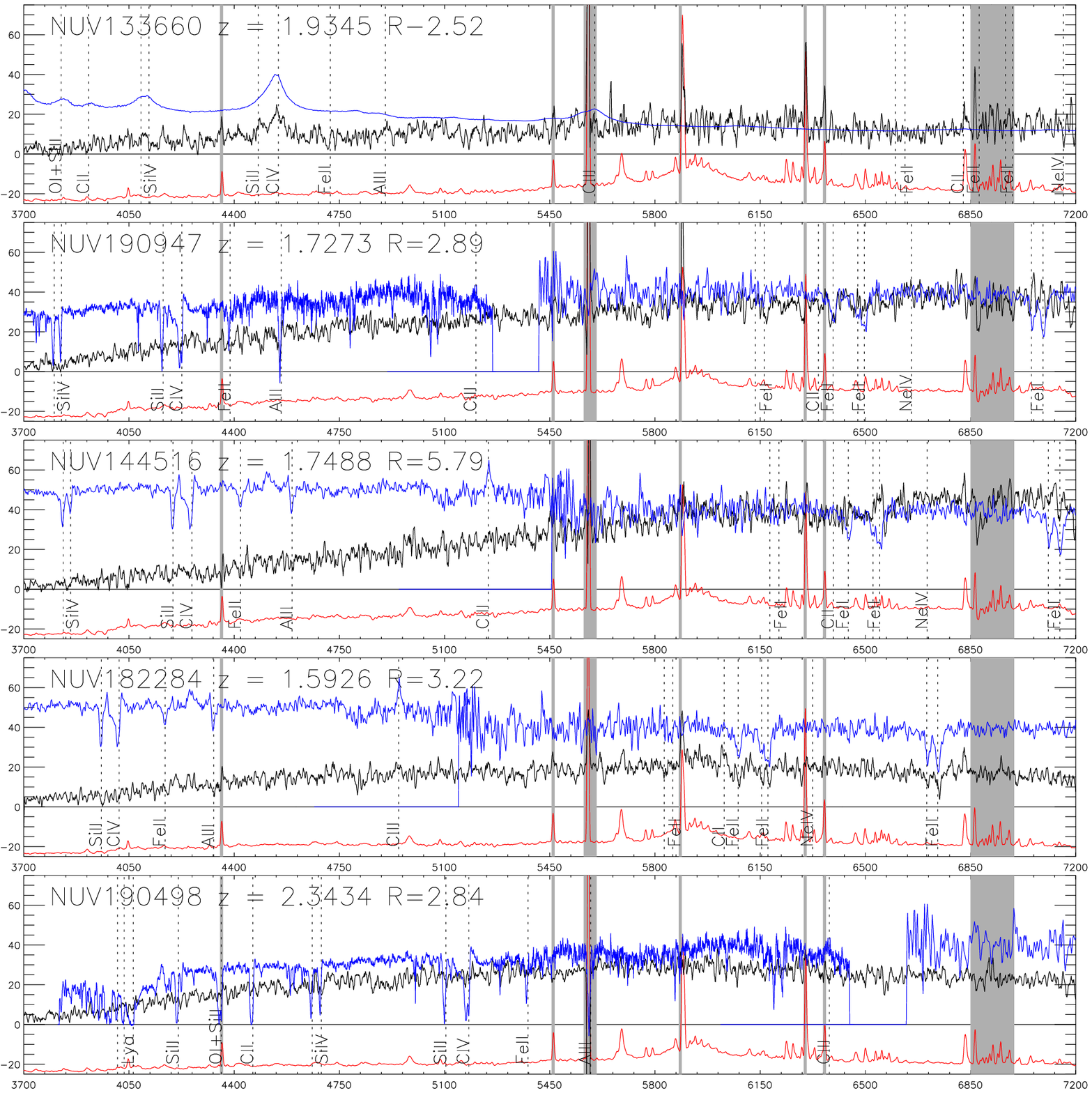}
  \caption{Same as Figure~\ref{spec4}. \color }
  \label{spec5}
\end{figure*}

\begin{figure*} 
  \epsscale{1.0}
  \plotone{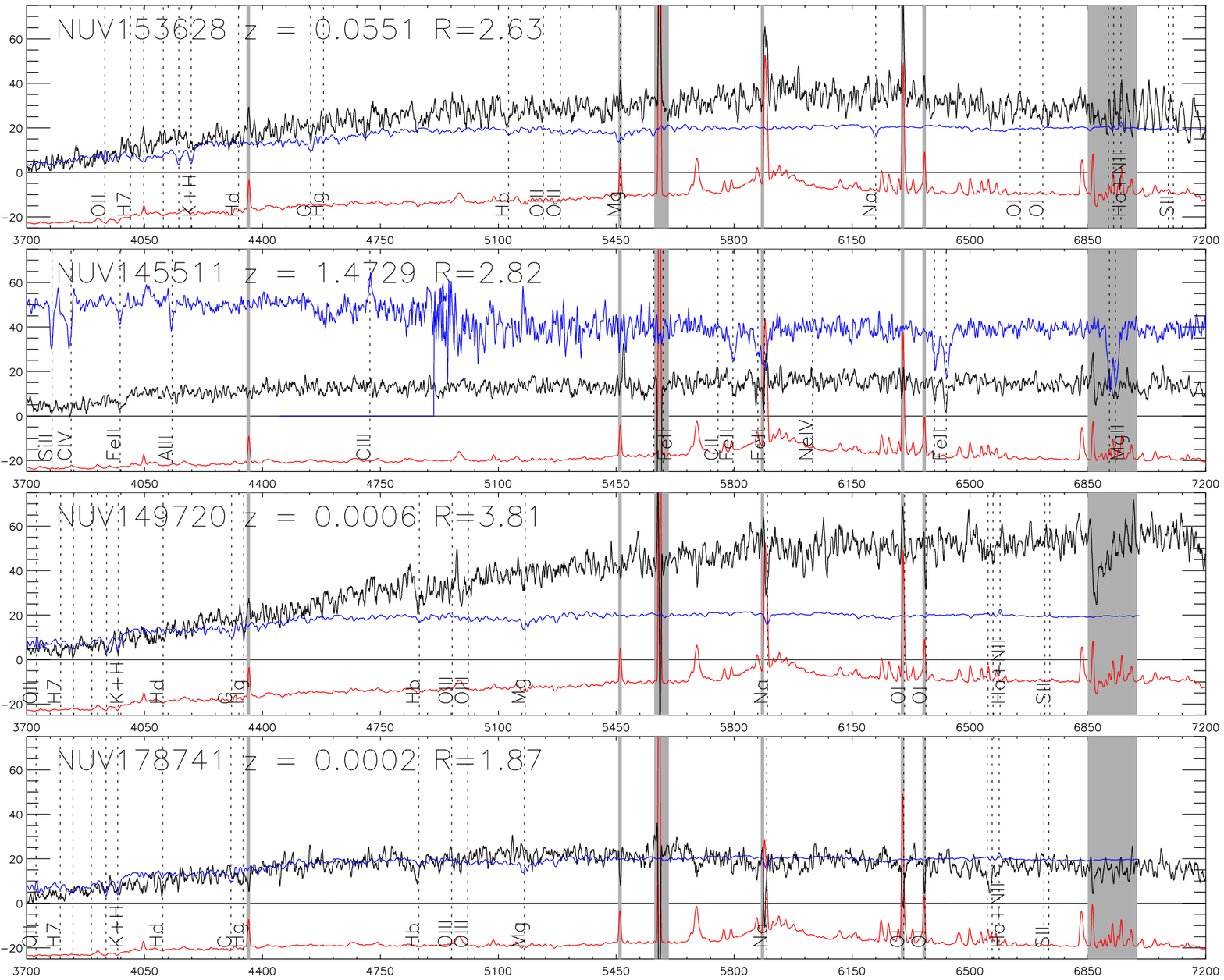}\vspace{-1.25in}
  \caption{Same as Figures~\ref{spec4} and \ref{spec5}, but this shows the low-$z$ interlopers
    and galactic stars \color }
  \label{spec6}
\end{figure*}

Among 21 spectra, 7 objects (2 are AGNs) are identified ($R>3.0$; 9 if $R>2.5$) at $z>1.5$, 2 objects are stars, 1 (2 with $R>2.5$)
is a $z<1.5$ interloper, and 11 are ambiguous (8 if $R>2.5$ is adopted). These MMT spectra are shown in
Figures~\ref{spec4}$-$\ref{spec6}, and their properties are listed in Table~\ref{table1}.\\
\indent The spectrum of a \Rc $\sim$ 22 $z\sim1.6$ LBG detected the \ion{Fe}{2} and \ion{Mg}{2} absorption lines, which
indicates that MMT is sensitive enough to detect luminous LBGs. In fact, since the surface density of bright LBGs is low,
slitmask instruments are not ideal for the bright end. However, the entire SDF can be observed with Hectospec, so all 
$\sim$150 $V_{\texttt{auto}}<23.0$ objects can be simultaneously observed.
%
%

\subsection{Additional Spectra with Subaru/MOIRCS}\label{2.4}
The {\it BzK} technique, which identifies galaxies with a wide range (old and young, dusty and unreddened)
of properties, could include objects
that would also be classified as \nuv-dropouts. As a check, cross-matching of spectroscopically identified
star-forming {\it BzK}'s with the \galex-SDF photometric catalog was performed. Spectra of BzKs were
obtained on 2007 May 3$-$4 with Subaru using the Multi-Object Infrared Camera and Spectrograph
\citep[MOIRCS;][]{ichikawa06}. 44 sources were targeted and 15 were identified by the presence of H$\alpha$
and [\ion{N}{2}] or [\ion{O}{2}], [\ion{O}{3}] and H$\beta$ emission. One of the 15 was not in the $B$-band
catalog. Among the 14 objects, 7 are also classified as \nuv-dropouts and were not previously identified
(i.e., LRIS or Hectospec targets). This included 5 galaxies at $z>1.5$ and 2 at $z=1-1.5$. Their properties
are included in Table~\ref{table1}. Among the 7 BzKs that did {\it not} meet the \nuv-dropout criteria, 2
are below $z=1.5$ and the other five are at high-$z$. For two of the high-$z$ BzKs, one was below the
$NUV-B=1.75$ cut because it is faint ($V>25.3$), thus not considered a \nuv-dropout, and the other missed
the $B-V=0.5$ selection by having $B-V = 0.53$.\footnote[12]{If the selection criteria were modified to
  include this object, no low-$z$ interlopers or stars would have contaminated the criteria. However, a
  $B-V\leq0.5$ is still adopted for simplicity.} The other three sources have low-$z$ neighboring sources
that are detected in the \nuv, which influences the \nuv~photometry to be brighter. The cause of confusion
is due to the poor resolution of \galex, which is discussed further in \S~\ref{4.3}. The details of these
observations and their results are deferred to \cite{hayashi08}.

\subsection{Summary of Observations}\label{2.5}
In order to probe $1.5<z<3$ with the Lyman break technique, deep ($>$100 ks) \galex/\nuv~imaging was obtained.
Spectroscopic observations from Keck and MMT independently confirm that most \nuv-dropouts (with their UV
continuum detected spectroscopically) are found to be at $1.5 < z < 2.7$.\\
\indent A summary of the number of LBGs, stars, and low-$z$ interlopers identified spectroscopically is
provided in Table~\ref{table2}. Among the spectra targeting \nuv-dropouts (i.e., excluding MOIRCS spectra),
53\% (30/57) were identified, and among those, 63\% are at $z>1.5$. Including seven objects with $R=2.5-3.0$, the
percentages are 65\% and 62\%, respectively. These statistics are improved with the final selection criteria
discussed in \S~\ref{3.2}.

\section{Photometric Selection of \nuv-dropouts}\label{3}
This section describes the \nuv~and optical photometric catalogs (\S~\ref{3.1}) and the methods for
merging the two catalogs. Then in \S~\ref{3.2}, $\sim$8000 \nuv-dropouts are empirically identified
with the spectroscopic sample to refine the selection criteria.

\subsection{Revised \nuv~Photometric Catalogs}\label{3.1}
Prior to any measurements, an offset ($\Delta\alpha=-0.39$\arcsec, $\Delta\delta=-0.18$\arcsec) in the
\nuv~image coordinates was applied to improve the astrometry for alignment with Suprime-Cam data. The
scatter in the astrometric corrections was found to be $\sigma_{\Delta\alpha}$=0.39\arcsec~and
$\sigma_{\Delta\delta}$=0.33\arcsec. This only results in a 0.01 mag correction for \nuv~measurements,
and is therefore neglected.\\
\indent The coordinates of $\sim$100000 SDF $B$-band sources
with $B_{\texttt{auto}}<27.5$ were used to measure \nuv~fluxes within a 3.39\arcsec~(2.26 pixels) radius
aperture with the \textsc{iraf/daophot} task, \texttt{phot}. For objects with \nuv~photometry below
the $3\sigma$ background limit, the $3\sigma$ value
is used. This limit is determined from the mode in an annulus with inner and outer radii of
22.5\arcsec~and 37.5\arcsec~(i.e., an area of 1200 pixels), respectively. For sources detected in the
\nuv, a point-source aperture correction of a factor of $\approx$1.83 is applied to obtain the ``total''
\nuv~flux. This correction was determined from the point spread function (PSF) of 21 isolated sources
distributed across the image. The \nuv~catalog is then merged with the $B$-band catalog from SExtractor
\citep[SE; ][]{bertin96} that contains $BV$\Rc$i\arcmin z\arcmin$ photometry.\\
\indent Throughout this paper, ``total'' magnitudes from the Suprime-Cam images are given by SE
\texttt{mag\_auto}, since the corrections between $B$-band Kron and the 5\arcsec~diameter magnitudes
were no greater than 0.03 mag for isolated (5\arcsec~radius), point-like
(SE \texttt{class\_star} $\geq$ 0.8) targets.\\
\indent The merged catalog was also corrected for galactic extinction based on the \cite{cardelli89}
extinction law. For the SDF, they are: $A$(\nuv) = 0.137, $A(B)$ = 0.067, $A(V)$ = 0.052, $A$(\Rc) = 0.043,
$A(i\arcmin)$ = 0.033, and $A(z\arcmin)$ = 0.025. Since the Galactic extinction for the SDF is low,
the amount of variation in A(\nuv) is no more than 0.02, so all \nuv~magnitudes are corrected by the
same value.

\subsection{Broad-band Color Selection}\label{3.2}
\begin{figure} 
  \epsscale{1.0}
  \plotone{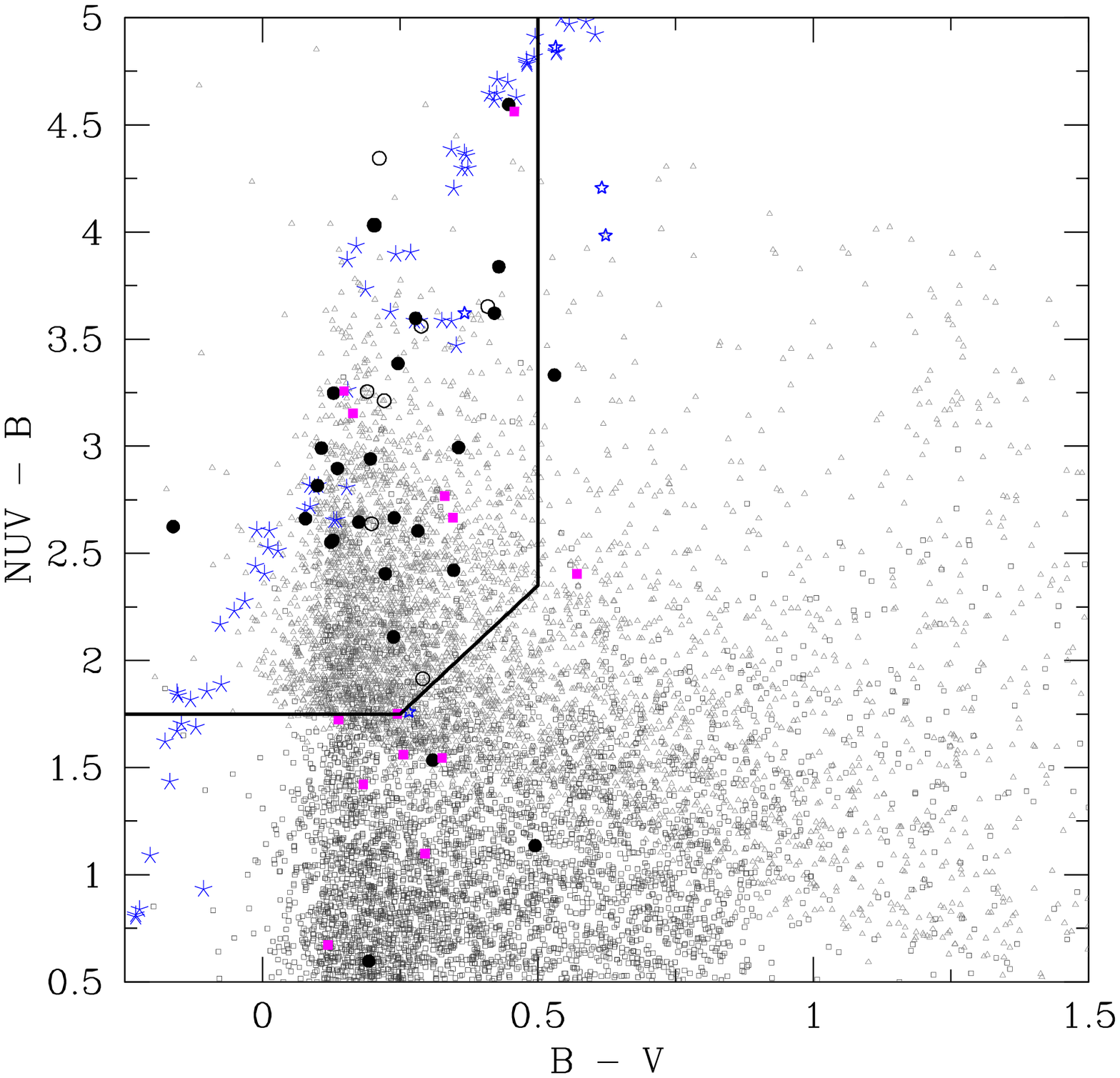}
  \caption{$NUV-B$ and $B-V$ colors for $22.0<V_{\texttt{auto}}<25.3$ sources. A total of $\sim$33,000
    sources are represented here, but only one-third are plotted, for clarity. Sources undetected (at
    the 3$\sigma$ level) in the \nuv~are shown as grey unfilled triangles while the detected sources
    are indicated as dark grey unfilled squares. Filled (unfilled) circles correspond to sources that
    have been confirmed as LBGs with (without) emission lines. Low-$z$ interlopers are shown as filled
    squares while stars are shown as unfilled stars. Skeletal stars represent Gunn-Stryker
    stars. \color}
  \label{select}
\end{figure}

Using the sample of spectroscopically confirmed $z>1.5$ LBGs, low-$z$ interlopers, and stars, the
color selection is optimized to minimize the number of interlopers while maximizing the number of
confirmed LBGs. In Figure~\ref{select}, known LBGs are identified in the $NUV-B$ versus $B-V$ diagram,
where the $NUV-B$ color is given by the ``total'' magnitude and the $B-V$ is the color within a
2\arcsec~aperture. The latter was chosen because of the higher S/N compared to larger apertures. The
final empirical selection criteria for the LBG sample are:
\begin{eqnarray}
NUV-B \geq 1.75,&\\
B-V \leq 0.50,&~{\rm and}\\
NUV-B \geq 2.4(B-V) + 1.15,
\end{eqnarray}
which yielded 7964 \nuv-dropouts with $21.90\leq V\leq25.30$. Among the Hectospec and LRIS spectra,
these selection criteria included all spectroscopic LBGs and excluded 4/5 stars and 4/6 (4/9 with $R>2.5$) interlopers.
Therefore, the fraction of \nuv-dropouts that are confirmed to be LBGs with the new selection
criteria is 86\% (the $R=2.5$ cut implies 79\%). Note that while the $B$-band catalog was used (since
the $B$ filter is closer in wavelength to the \nuv), the final magnitude selection was in $V$, to compare
with the rest-frame wavelength ($\approx$1700\AA) of $z\sim3$ LBGs in the $R$-band.\\
\indent To summarize, a \nuv-optical catalog was created, and it was combined with spectroscopic redshifts
to select 7964 \nuv-dropouts with $NUV-B\geq1.75$, $B-V\leq0.50$, $NUV-B\geq2.4(B-V)+1.15$, and
$21.90\leq V\leq25.30$. The spectroscopic sample indicates that 14\% of \nuv-dropouts are definite
$z\leq1.5$ interlopers. 

\section{CONTAMINATION AND COMPLETENESS ESTIMATES}\label{4}
Prior to constructing a normalized luminosity function, contaminating sources that are not LBGs must be
removed statistically. Section \ref{4.1} discusses how foreground stars are identified and removed,
which was found to be a 4$-$11\% correction. Section \ref{4.2} describes the method for estimating low-$z$
contamination, and this yielded a correction of $34\%\pm17\%$. These reductions are
applied to the number of \nuv-dropouts to obtain the surface density of $z\sim2$ LBGs. Monte Carlo
(MC) realizations of the data, to estimate the completeness and the effective volume of the survey,
are described in \S~\ref{4.3}. The latter reveals that the survey samples $z\approx1.8-2.8$.

\subsection{Removal of Foreground Stars}\label{4.1}
The \cite{GS} stellar track passes above the \nuv-dropout selection criteria box (as shown in
Figure~\ref{select}). This poses a problem, as objects that are undetected in the \nuv~can be
faint foreground stars. A simple cut to eliminate bright objects is not sufficient, because faint halo
stars exist in the SDF (as shown later). To reduce stellar contamination, additional photometric
information from the SExtractor $BVR_{\rm c}i$\arcmin$z$\arcmin~catalogs is used. The approach of
creating a ``clean'' sample of point-like sources, as performed by \cite{richmond05}, is followed. He
used the \texttt{class\_star} parameter and the difference ($\delta$) between the 2\arcsec~and
3\arcsec~aperture magnitudes for each optical image. A `1' is assigned when the \texttt{class\_star}
value is $0.90-1.00$ or $0.10<\delta<0.18$, and `0' otherwise for each filter. The highest score is
10 [(1+1)$\times$5], which 2623 $V_{\texttt{auto}}=21.9-26.0$ objects satisfied, and is referred to
as ``perfect'' point-like or ``rank 10'' object. These rank 10 objects will be used to define the
stellar locus, since contamination from galaxies is less of a problem for the most point-like sample.
Then objects with lower ranks that fall close to the stellar locus will also be considered as stars after
the locus has been defined.\\
\begin{figure*}[!htc] 
  \epsscale{1.1}
  \plottwo{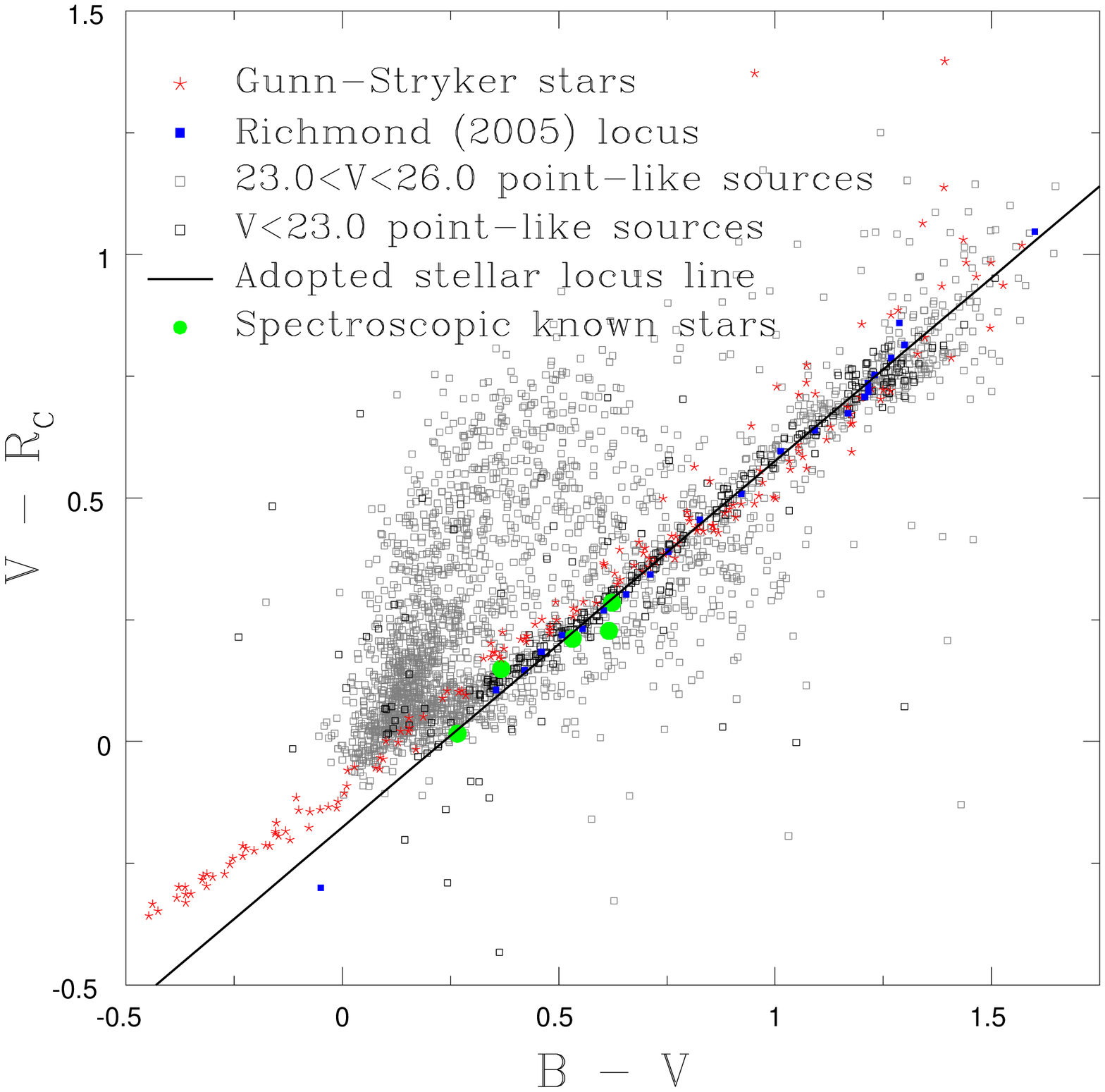}{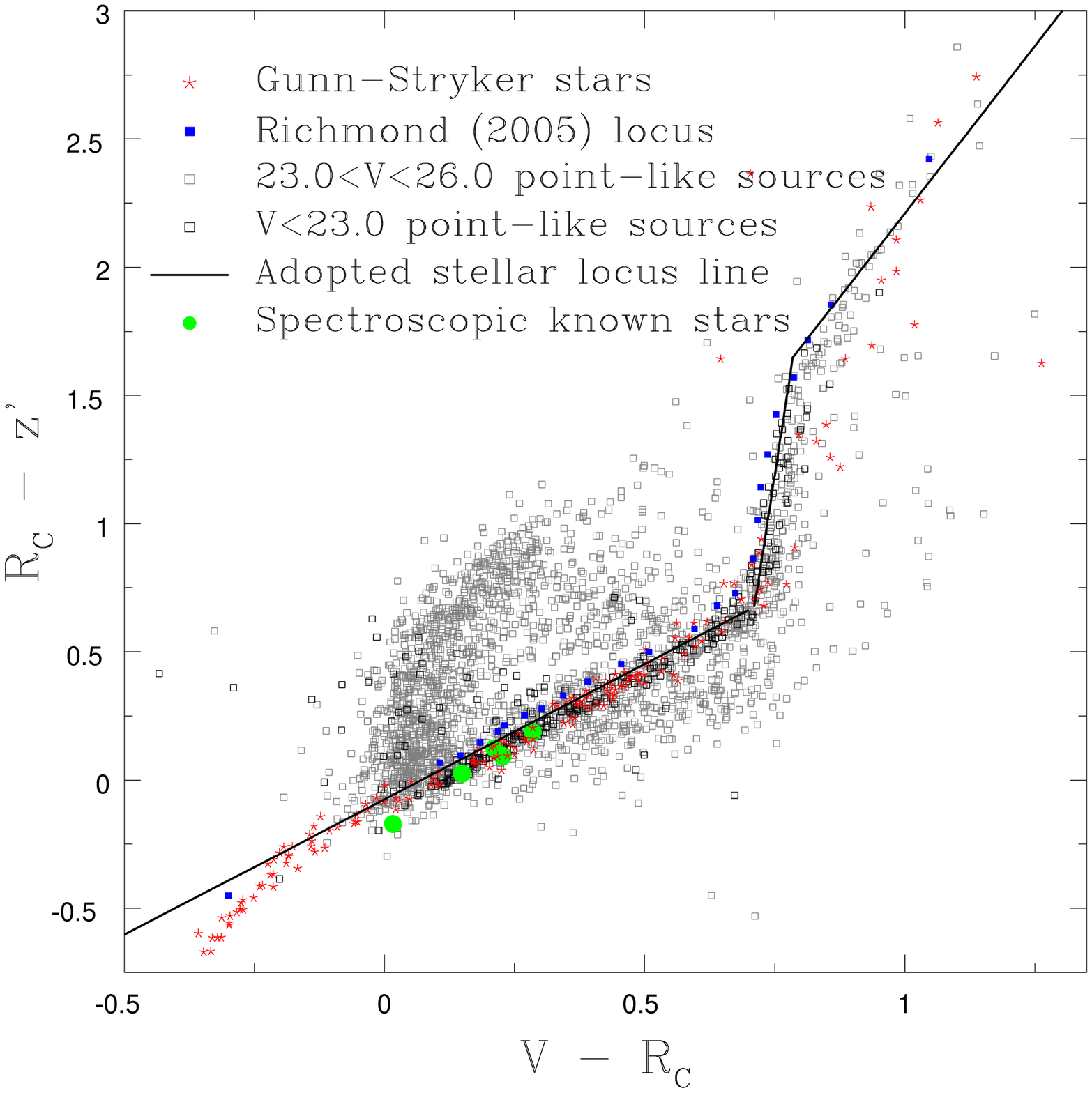}
  \caption{Two color-color diagrams for rank 10 point-like objects. Grey (small) and black (large)
    squares represent sources brighter than $V_{\texttt{auto}}=26.0$ and 23.0, respectively.
    The Gunn-Stryker stars are shown as stars, and the SDF stellar locus of \cite{richmond05}
    is shown as filled squares. The solid lines define the stellar locus for calculating $\Delta$ (see
    \S~\ref{4.1}). The five sources that have been spectroscopically identified to be stars are shown
    as filled green circles. \color}
  \label{fig3}
\end{figure*}
%
\indent Unfortunately, distant galaxies can also appear point-like, and must be distinguished from stars.
This is done by comparing their broad-band optical colors relative to the stellar locus. Figure~\ref{fig3}
shows the $B-V$, $V-R_{\rm C}$, and $R_{\rm C}-z\arcmin$ colors used in \cite{richmond05} for the ``clean''
sample. The stellar locus is defined by the solid black lines using brighter ($V\leq23.0$) sources.
Figure~\ref{fig3} shows differences in the colors between the
stellar locus defined for point-like SDF stars and those of Gunn-Stryker stars. \cite{richmond05}
states that this is due to metallicity, as the SDF and Gunn-Stryker stars are selected from the halo and
the disk of the Galaxy, respectively.\\
\indent For each object in the clean sample, the $V-R_{\rm C}$ color is used to predict the $B-V$ and
$R_{\rm C}-z\arcmin$ colors along the stellar locus (denoted by `S.L.' in the subscript of the colors
below). These values are then compared to the observed colors to determine the magnitude deviation
from the stellar locus, $\Delta = -[(B-V)_{\rm obs} - (B-V)_{\rm S.L.}] + [(R_{\rm C}-z\arcmin)_{\rm obs} -
  (R_{\rm C}-z\arcmin)_{\rm S.L.}]$. Therefore, an object with $\Delta\approx0$ mag is classified as a star.
This method is similar to what is done in \cite{richmond05}, where an object is considered a star if it
is located within the stellar locus ``tube'' in multi-color space. This approach provides stellar contamination
at faint magnitudes, which is difficult spectroscopically \citep{steidel03}. A histogram showing the distribution
of $\Delta$ in Fig.~\ref{fig4}a reveals two peaks: at $\Delta\approx0$ and 0.8 mag.
The comparison of $\Delta$ versus the $V$-band magnitude is shown in Fig.~\ref{fig4}b, and a source is
identified as a star if it falls within the selection criteria shown by the solid lines in this figure.
A total of 1431 stars $V\leq26.0$ are identified, while the remaining 1192 sources are classified as
galaxies. The surface density as a function of magnitude for the identified stars agrees with predictions
made by \cite{robin03} and other surface density measurements near the galactic pole. When the
\nuv-dropout selection criteria are applied\footnote[13]{The $B-V$ and $NUV-B$ color cuts limit the stellar
  sample to spectral types between A0 and G8.}, these numbers are reduced to 336 stars (i.e., a 4\%
contamination for the \nuv-dropout sample) and 230 galaxies with $21.9\leq V_{\texttt{auto}}\leq25.3$.\\
\indent Sources that are ranked $7-9$ are also considered and were classified as a star or a galaxy using the
above approach. Of those that met the \nuv-dropout criteria, 535 and 252 have the colors of stars and
galaxies, respectively. Thus, the photometric sample of \nuv-dropouts contains 7093 objects after
statistically removing 871 stars (11\% of the \nuv-dropout) that are ranked 7$-$10. The reasons for only
considering objects with a rank of 7 or greater are (1) the stellar contamination does not 
significantly increase by including rank 6 or rank 5 objects (i.e., another 128 rank 6 stars or
1.5\% and 143 rank 5 stars or 1.8\%), and (2) comparison of the surface density of rank 7$-$10 stars
with expectations from models showed evidence for possible contamination from galaxies at the faint end
($V>24.0$; A. Robin, priv. comm.), and the problem will worsen with rank 5 and 6 objects included. As
it will be apparent later in this paper, stellar contamination is small and not expected to significantly
alter any discussion of differences seen in the luminosity function. A hard upper limit by considering
objects of rank 1 and above as stars would imply an additional (rank 1 to 6) stellar contamination of
14.5\%.\\
%
\begin{figure*}[!htc] 
  \epsscale{1.1}
  \plottwo{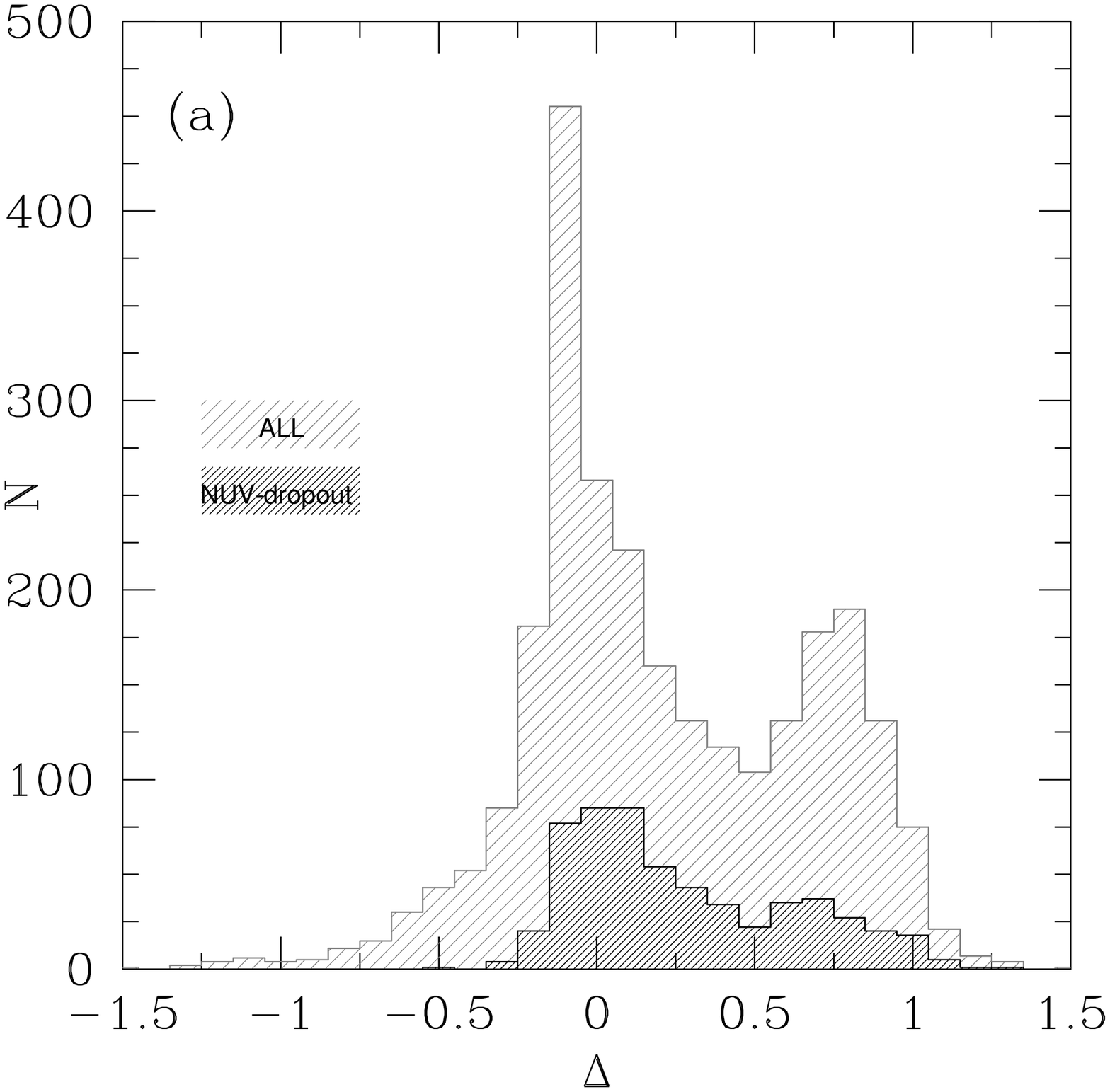}{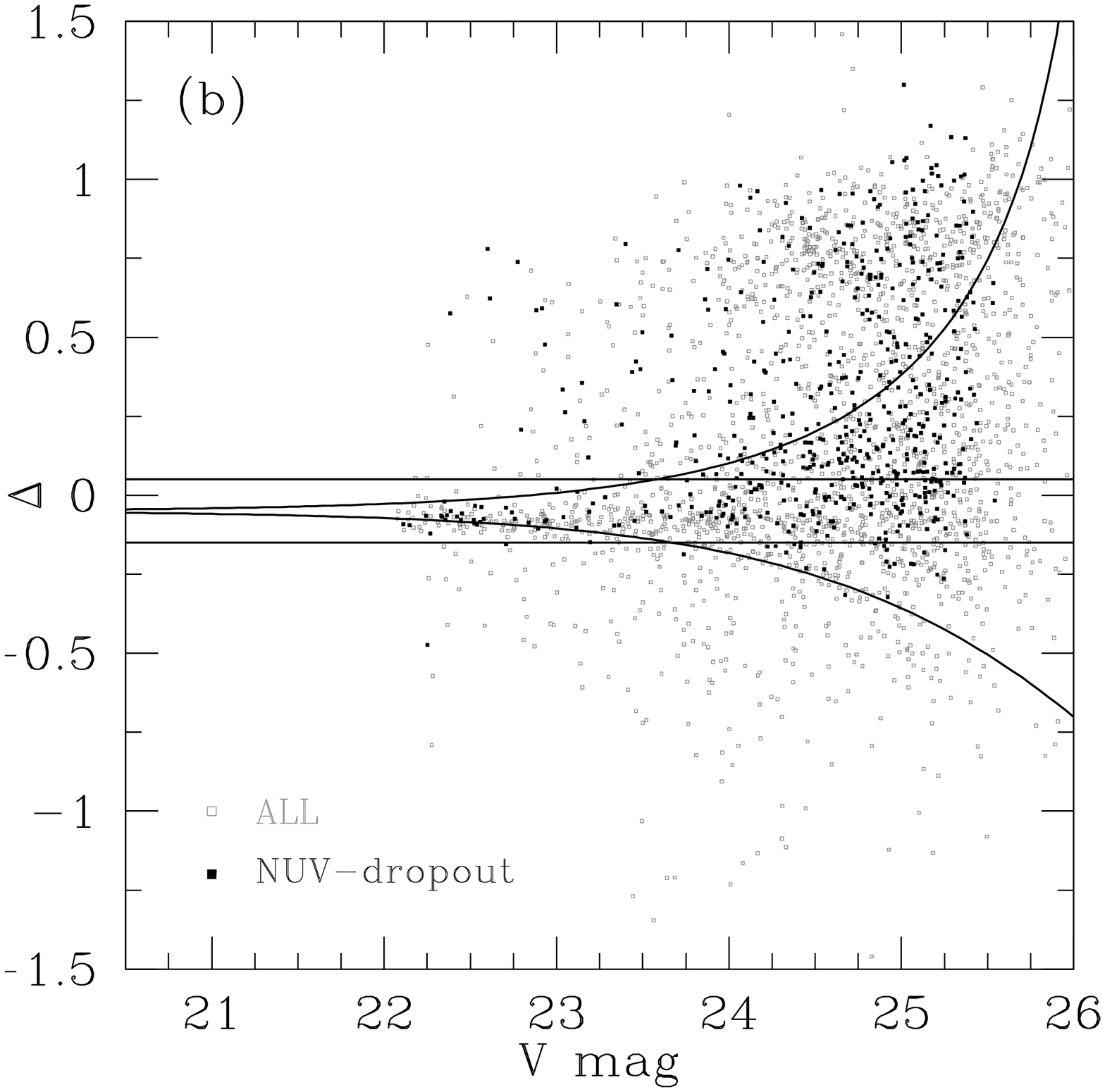}
  \caption{Photometric properties of rank 10 point-like objects. A histogram of $\Delta$ is shown in
    ({\it a}) while ({\it b}) plots $\Delta$ versus $V$-band Kron magnitude. The grey histogram and
    squares are for {\it all} point-like sources while those that satisfy the \nuv-dropout selection
    criteria are represented in black. The selection of foreground stars is given by the solid lines in
    ({\it b}). The horizontal solid lines represent a minimum $\Delta$ at the bright end while the two
    solid curves are the $\pm$3$\sigma$ criteria for $\Delta$, as given by
    $-2.5\log{[1\mp(f_{3\sigma B}^2+f_{3\sigma V}^2+f_{3\sigma R_{\rm C}}^2
	+f_{3\sigma z\arcmin}^2)^{0.5}/f_{V}]}$. Here $f_{X}$ is the flux density in the $X$ filter.}
  \label{fig4}
\end{figure*}
\indent Among the 5 sources spectroscopically determined to be stars, 3 of them (71239, 66611, and
149720) are classified as stars with the $\Delta$ method, and the other two stars (86900
and 178741) fall outside the $\Delta$ selection criteria. Among the known LBGs, 8 are rank 8$-$10
and 3 (166380, 78625, and 133660) are classified as {\it not} being stars.
Since the spectroscopic sample of rank 10 objects is small, additional spectra will be required to
further optimize the $\Delta$ technique. However, the spectroscopic sample (presented in this paper)
indicates that $3-7\%$ of \nuv-dropouts are stars, which is consistent with the $4-11\%$ derived with the
$\Delta$ method.

\subsection{Contamination from $z<1.5$ Interlopers}\label{4.2}
\begin{figure}[!htc] 
  \epsscale{1.0}
  \plotone{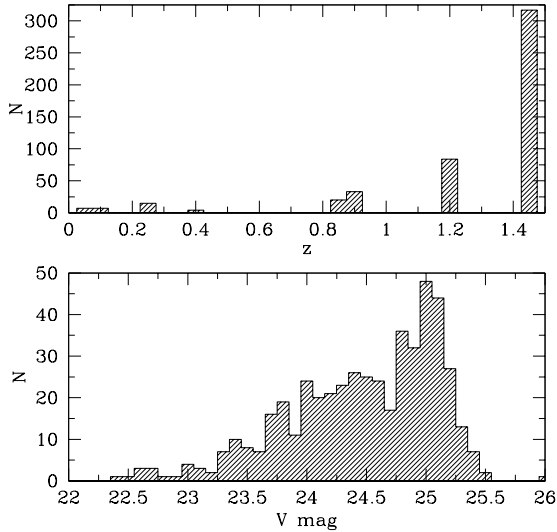}
  \caption{Redshift (top) and $V$-band magnitude (bottom) distributions of 487 NB emitters that meet the
    \nuv-dropout criteria. Note that the redshift bins are made larger to clearly show the histogram.}
  \label{NBemitters}
\end{figure}

One of the biggest concerns in any survey targeting a particular redshift range is contamination from
other redshifts. The spectroscopic sample of \nuv-dropouts shows that 5\% are definite $z<1.5$ galaxies.
This number increases to an upper value of 51\% if the ambiguous \nuv-dropouts (that meet the color
selection criteria) are all assumed to be low-$z$ interlopers. However, it is unlikely that all unidentified
\nuv-dropouts are low-$z$, since LBGs without \Lya~emission in their spectra\footnote[14]{Either because they
  do not possess \Lya~in emission or they are at too low of a redshift for \Lya\ to be observed.} are likely
missed.
A secondary independent approach for estimating low-$z$ contamination, which is adopted later in this
paper, is by using a sample of $z<1.5$
emission-line galaxies identified with narrow-band (NB) filters. Since a detailed description of this sample is
provided in \cite{ly07}, only a summary is given below:\\
A total of 5260 NB emitters are identified from their excess fluxes in the NB704, NB711, NB816, or NB921
filter either due to H$\alpha$, [\ion{O}{3}], or [\ion{O}{2}]~emission line in 12 redshift windows (some
overlapping) at $0.07\lesssim z\lesssim1.47$. These galaxies have emission line equivalent widths and
fluxes as small as 20\AA\ (observed) and a few $\times 10^{-18}$ erg s$^{-1}$ cm$^{-2}$, and are
as faint as $V=25.5-26.0$. Cross-matching was performed with the \nuv-dropout sample, which yielded 487 NB
emitters as \nuv-dropouts. The redshift and $V$-band magnitude distributions are shown in
Figure~\ref{NBemitters}. Note that most of the contaminating sources are at $1.0<z<1.5$, consistent with
the spectroscopic sample.

Since this sample represents a fraction of the $0.07\lesssim z \lesssim1.5$ redshift range, the above
results must be interpolated for redshifts in between the NB redshifts. It is assumed
that emission-line galaxies exist at all redshifts, and possess similar properties and number densities
to the NB emitters. One caveat of this approach is that blue galaxies that do {\it not} possess nebular
emission lines, may meet the \nuv-dropout selection.\footnote[15]{Red galaxies are excluded by
  the $B-V<0.5$ criterion.} The statistics of such objects are not well known, since spectroscopic surveys
are biased toward emission line galaxies, due to ease of identification. Therefore, these contamination
estimates are treated as lower limits. A further discussion of this approach is provided in \S~\ref{7}.\\
\indent Using the redshift distribution shown in Figure~\ref{NBemitters}, the number of objects per
comoving volume ($N/\Delta V$) is computed at each NB redshift window. For redshifts not included by the
NB filters, a linear interpolation is assumed. Integrating over the volume from $z=0.08$ to $z=1.5$
yields the total number of interlopers to be $2490\pm1260$, which corresponds to a contamination
fraction of \fcont = $0.34\pm0.17$. The error on \fcont~is from Poissonian statistics for each redshift
bin, and are added in quadrature during the interpolation step for other redshifts. This is also determined
as a function of magnitude (hereafter the ``mag.-dep.'' correction), since the redshift distribution
will differ between the bright and faint ends. The \fcont~($V$-band magnitude range) values are
$0.39\pm0.20$ (22.9$-$23.3), $0.40\pm0.21$ (23.3$-$23.7), $0.37\pm0.21$ (23.7$-$24.1), $0.31\pm0.16$
(24.1$-$24.5), $0.27\pm0.14$ (24.5$-$24.9), and $0.39\pm0.19$ (24.9$-$25.3).

\subsection{Modelling Completeness and Effective Volume}\label{4.3}
\begin{figure} 
  \epsscale{1.0}
  \plotone{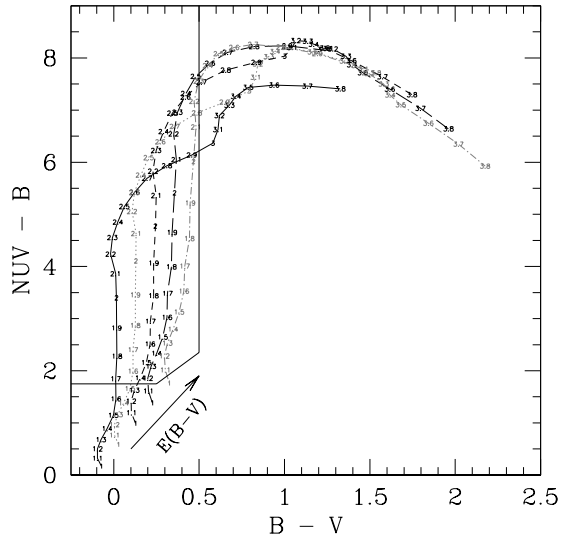}
  \caption{Modelled $NUV-B$ and $B-V$ colors for \nuv-dropouts. The solid, dotted, short-dashed,
    long-dashed, and dot short-dashed lines correspond to the spectral synthesis model described
    in \S~\ref{4.3} with $E(B-V)=0.0$, 0.1, 0.2, 0.3, and 0.4, respectively. The thick solid
    black lines represent the selection criteria in \S~\ref{3.2}. } 
  \label{lbgmodel}
\end{figure}

\begin{figure*} 
  \epsscale{0.35}
  \plotone{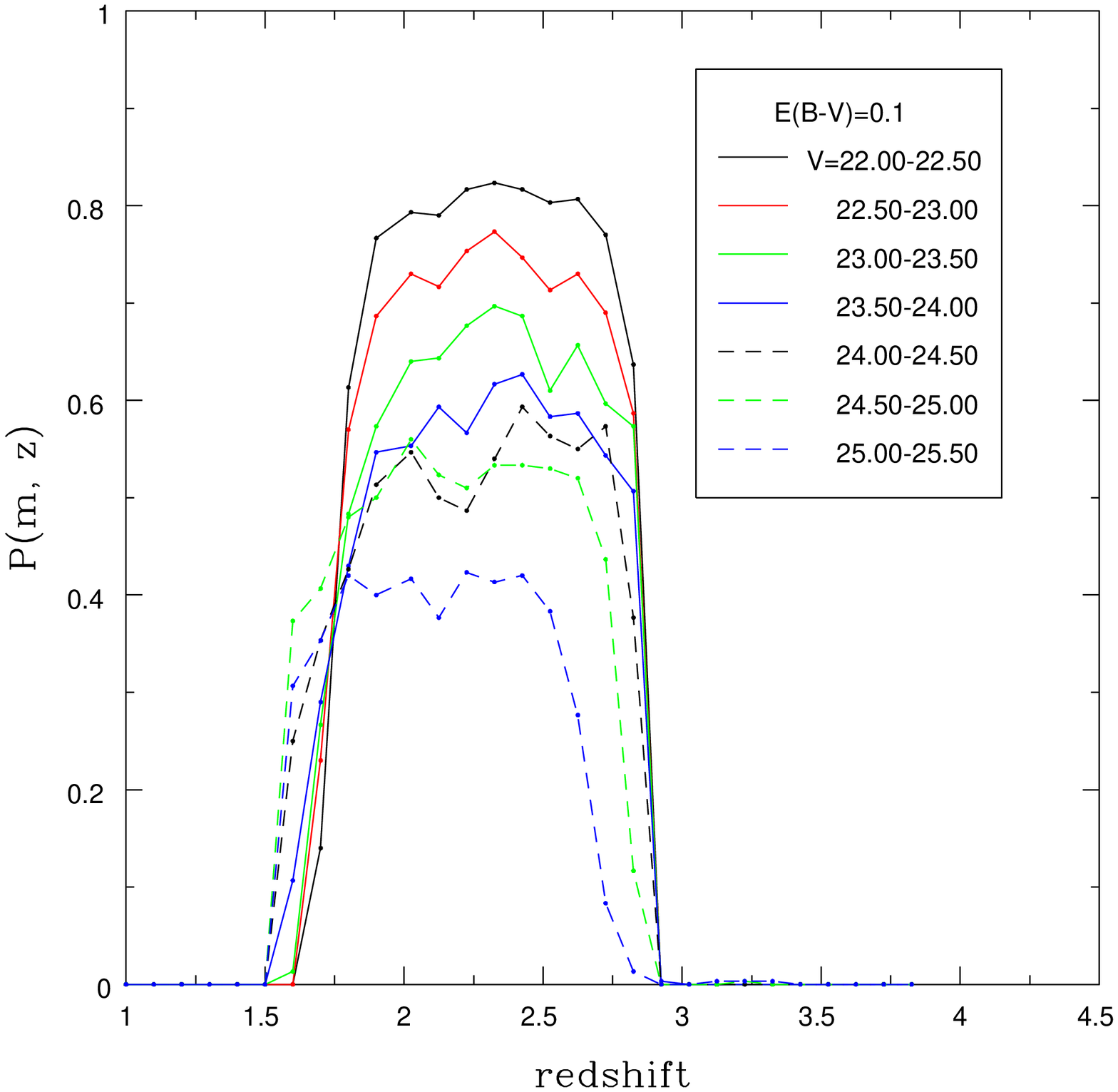}\plotone{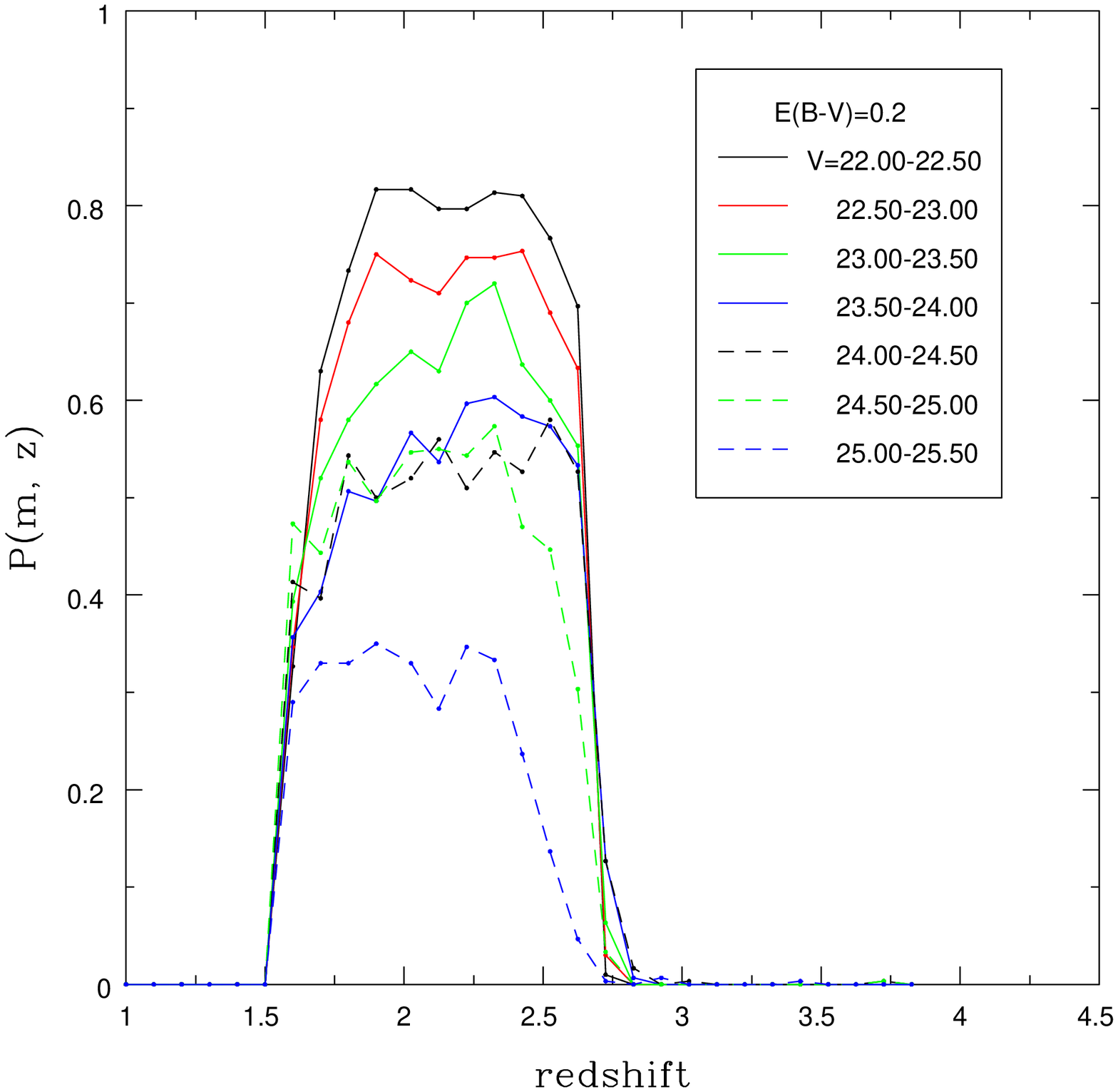}\plotone{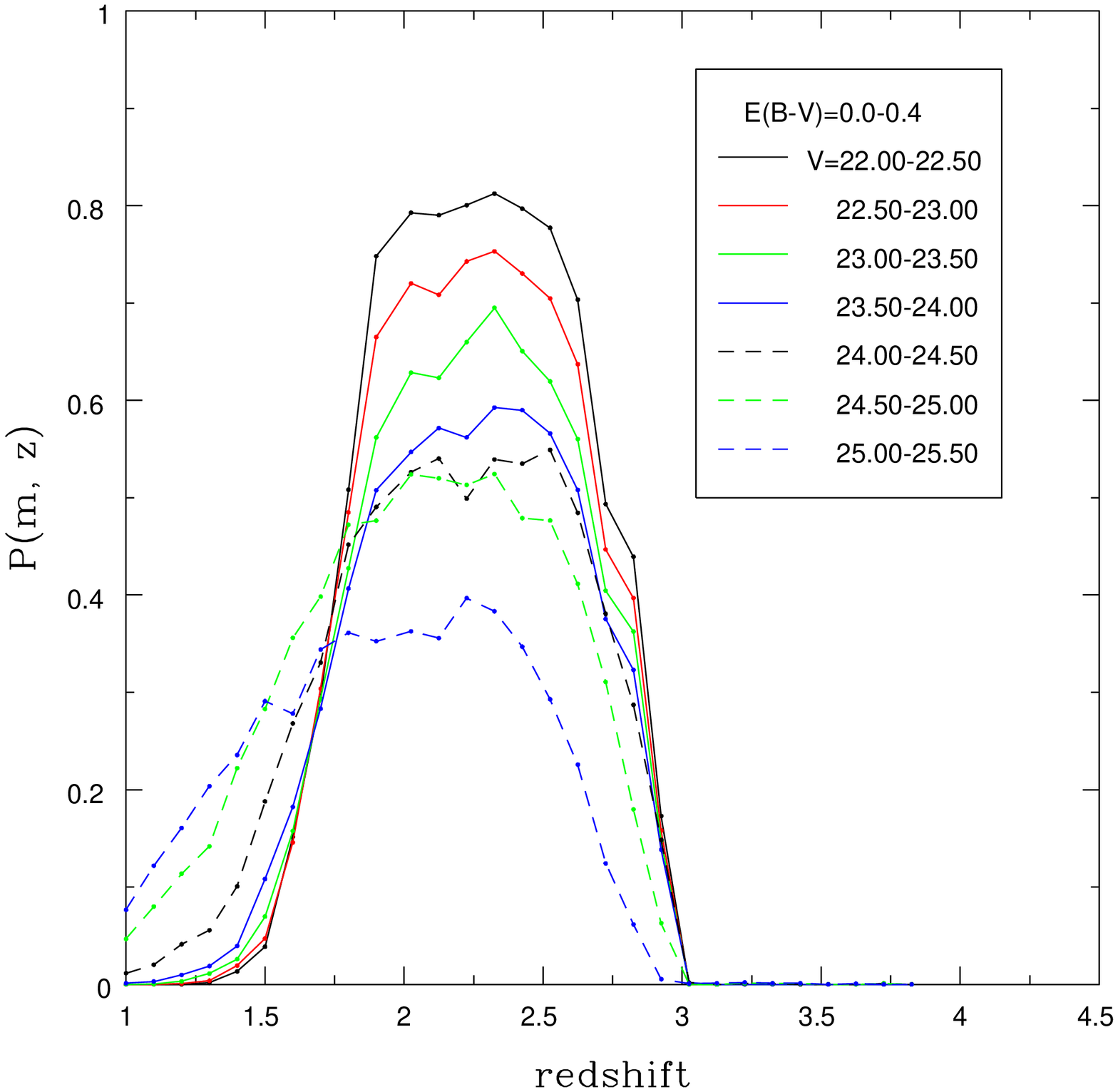}\\
  \caption{Monte Carlo completeness estimates as a function of redshift for different apparent
    magnitude. From left to right is the result for $\EBV=0.1$, 0.2, and $0.0-0.4$
    \citep[a weighted average assuming the $\EBV$ distribution of][]{steidel99}. \color}
  \label{MCsim}
\end{figure*}

In order to obtain an accurate LF for \nuv-dropouts, the completeness of the sample must be quantified.
This is accomplished with MC simulations to calculate $P(m,z)$, which is the probability that a galaxy
of apparent $V$-band magnitude $m$ and at redshift $z$ will be detected in the image, and will meet the
\nuv-dropout color selection criteria. The effective comoving volume per solid area is then given by
\begin{equation}
  \frac{V_{\rm eff}(m)}{\Omega} = \int dz P(m,z) \frac{dV(z)}{dz}\frac{1}{\Omega},
\end{equation}
where $dV/dz/\Omega$ is the differential comoving volume per $dz$ per solid area at redshift $z$.
Dividing the number of \nuv-dropouts for each apparent magnitude bin by $V_{\rm eff}$ will yield the
LF. This approach accounts for color selection biases, limitations (e.g., the depth and spatial resolution)
of the images \citep{steidel99}, and choice of apertures for ``total'' magnitude.\\
\indent In order to determine $P(m,z)$, a spectral synthesis model was first constructed from
\textsc{galaxev} \citep{bc03} by assuming a constant SFR with a Salpeter initial mass
function (IMF), solar metallicity, an age of 1Gyr, and a redshift between $z=1.0$ and $z=3.8$ with
$\Delta z=0.1$ increments. The model was reddened by assuming an extinction law following \cite{calzetti00}
with $\EBV=0.0-0.4$ (0.1 increments) and modified by accounting for IGM absorption following
\cite{madau95}. The latter was chosen over other IGM models \citep[e.g.,][]{bershady99} for consistency
with previous LBG studies. This model is nearly identical to that of \cite{steidel99}.\\
\indent Figure~\ref{lbgmodel} shows the redshift evolution of the $NUV-B$ and $B-V$ colors for this model.
These models were scaled to apparent magnitudes of $V=22.0-25.5$ in increments of 0.25. These 2175
($29\times15\times5$) artificial galaxies are randomly distributed across the \nuv, $B$, and $V$ images
with the appropriate spatial resolution (assumed to be point-like) and noise contribution with the IRAF
tasks \texttt{mkobject} (for optical images) and \texttt{addstar} (using the empirical \nuv~PSF).
Because of the poor spatial resolution of \galex, each iteration of 435 sources (for a given $E[B-V]$
value) was divided into three sub-iterations to avoid source confusion among the mock galaxies. The
artificial galaxies were then detected in the same manner as real sources. This process was repeated 100
times. Note that 21\% of artificial sources did not meet the \nuv-dropout criteria
(see e.g., Figure~\ref{MCsim}), as they were confused
with one or more nearby sources detected in the \nuv. This serves as an estimate for incompleteness due to
confusion, and is accounted for in the final LF. These results are
consistent with MOIRCS spectra that finds that $14-29$\% of BzKs with $z\geq1.5$ was
missed by \nuv-dropout selection criteria with nearby objects affecting the \nuv~flux. In addition, 
this simulation also revealed that among all mock LBGs with $z\leq1.5$, 30\% were photometrically scattered
into the selection criteria of \nuv-dropouts, which is consistent with the 34\% low-$z$ contamination
fraction predicted in \S~\ref{4.2}.\\
\indent Figure~\ref{MCsim} shows $P(m,z)$ as a function of magnitude for $\EBV=0.1$, 0.2, and $0.0-0.4$.
The latter is determined from a weighted average where the $\EBV$ distribution from \cite{steidel99} is
used for weighting each completeness distribution. This corresponds to an average $\EBV\sim0.15$. The
adopted comoving volume uses the weighted-average results. Table~\ref{table3} provides the effective
comoving volume per arcmin$^2$, the average redshift, the FWHM and standard deviation of the redshift
distribution for subsets of apparent magnitudes.
%

\subsection{Summary of Survey Completeness and Contamination}\label{4.4}
Using optical photometry, 871 foreground stars (i.e., a 11\% correction) were identified and excluded
to yield 7093 candidate LBGs. Then $z<1.5$ star-forming galaxies, identified with NB filters, were
cross-matched with the \nuv-dropout sample to determine the contamination fraction of galaxies at
$z<1.5$. Redshifts missed by the NB filters were accounted for by interpolating the number density
between NB redshifts, and this yielded $2490\pm1260$ interlopers, or a contamination fraction of
$0.34\pm0.17$.\\
\indent To determine the survey completeness, the $V_{\rm eff}$ was simulated. This consisted of generating
spectral synthesis models of star-forming galaxies, and then adding artificial sources with modelled
broad-band colors to the images. Objects were then detected and selected as \nuv-dropouts in the same
manner as the final photometric catalog. These MC simulations predict that the survey selects galaxies
at $z\sim2.28\pm0.33$ (FWHM of $z=1.8-2.8$), and has a maximum comoving volume of
$2.8\times10^3~h_{70}^{-3}$ Mpc$^3$ arcmin$^{-2}$.

\section{RESULTS}\label{5}
This section provides the key measurements for this survey: a $z\sim2$ rest-frame UV luminosity
function for LBGs (\S~\ref{5.1}), and by integrating this luminosity function, the luminosity and SFR
densities are determined (\S~\ref{5.2}).

\subsection{The 1700\AA\ UV Luminosity Function}\label{5.1}
\begin{figure} 
  \epsscale{1.0}
  \plotone{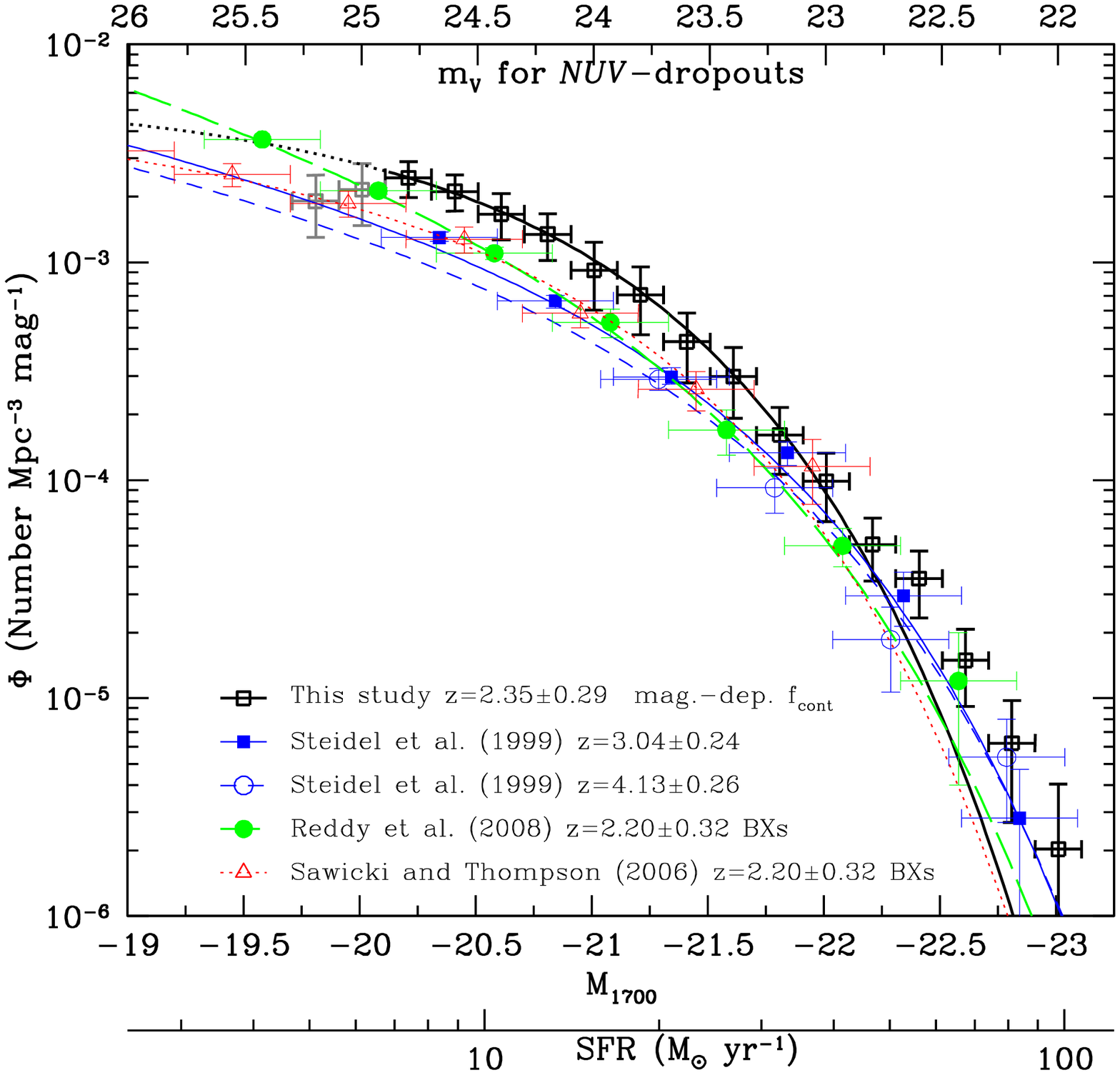}
  \caption{The {\it observed} $V$-band luminosity function for \nuv-dropouts. The LF of this work is
    shown by the thick black solid curve with unfilled squares. Grey points are those excluded from
    the MC fit. \cite{steidel99} measurements are shown as filled squares with solid thin curve
    ($z\sim3$) and opened circles with short-dashed thin curve ($z\sim4$). \cite{reddy08} BX results
    are shown as filled circles with long-dashed line, and \cite{ST06a} is represented by unfilled
    triangles and dotted line. Corrections to a common cosmology were made for \cite{steidel99}
    measurements, and SFR conversion follows \cite{kennicutt98}. \color }
  \label{Vlf}
\end{figure}

\begin{figure} 
  \epsscale{1.0}
  \plotone{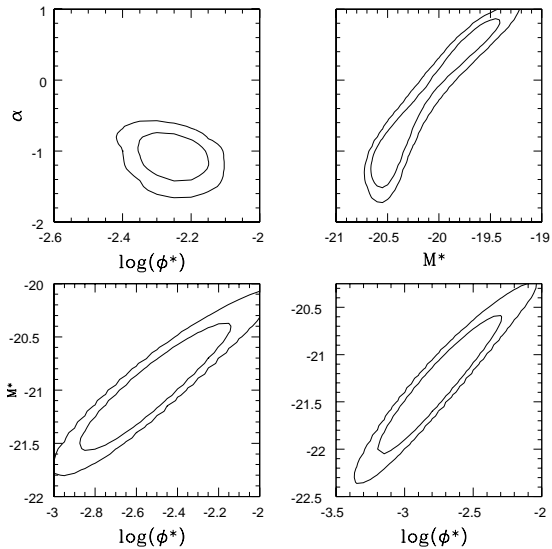}
  \caption{Confidence contours representing the best-fitting Schechter parameters for the LF. (Top)
    The mag.-dep. correction where the faint-end slope is a free parameter. The vertical axes show
    $\alpha$ while the horizontal axes show $\log{(\phi^{\star})}$ (left) and $M^{\star}$ (right).
    (Bottom) $M^{\star}$ vs. $\log{(\phi^{\star})}$ for $\alpha=-1.6$ (left) and $\alpha = -1.84$ (right).
    The inner and outer contours represent 68\% and 95\% confidence levels.}
  \label{contours}
\end{figure}

To construct a luminosity function, a conversion from apparent to absolute magnitude is needed. The
distance modulus is $m_{1700}-M_{1700} \approx 45.0$, where it is assumed that all the sources are at
$z\approx2.28$ and the K-correction term has been neglected, since it is no more than 0.08 mag. The
luminosity function is given by
\begin{equation}
\Phi(M_{1700}) = \frac{1}{\Delta m}\frac {N_{\rm raw}(1-f_{\rm contam})}{V_{\rm eff}(M_{1700})},
\end{equation}
where $N_{\rm raw}$ is the raw number of \nuv-dropouts within a magnitude bin ($\Delta m = 0.2$),
$V_{\rm eff}(M_{1700})$ is the effective comoving volume described in \S~\ref{4.3}, and \fcont~is the fraction
of \nuv-dropouts that are at $z<1.5$ (see \S~\ref{4.2}). The photometric LF is shown in Figure~\ref{Vlf}.
For the mag.-dep. \fcont~case, the adopted correction factor for $V\leq22.9$ is \fcont = 0.34
(the average over all magnitudes).\\
\indent Converting the \cite{schechter76} formula into absolute magnitude, the LF is fitted with the form:
\begin{equation}
  \Phi(M_{1700})dM_{1700} = \frac{2}{5}\ln{(10)}\phi^{\star} x^{\alpha+1}\exp{[-x]}dM_{1700},
\end{equation}
where $x \equiv 10^{-0.4(M_{1700}-M_{1700}^{\star})}$. In order to obtain the best fit, a MC simulation
was performed to consider the full range of scatter in the LF. Each datapoint was perturbed
randomly $5\times10^5$ times following a Gaussian distribution with $1\sigma$ given by the uncertainties
in $\Phi$. Each iteration is then fitted to obtain the Schechter parameters. This yielded for the
mag.-dep. \fcont~case: $M_{1700}^{\star}=-20.50\pm0.79$, $\log{\phi^{\star}}=-2.25\pm0.46$, and
$\alpha=-1.05\pm1.11$ as the best fit with 1$\sigma$ {\it correlated} errors. Since these
Schechter parameters are based on lower limits of low-$z$ contamination (see \S~\ref{4.2}), they
imply an upper limit on $\phi^{\star}$.
This luminosity function is plotted onto Figure~\ref{Vlf} as the solid black line, and the confidence
contours are shown in Figure~\ref{contours}.
With the faint-end slope fixed to $\alpha=-1.60$ \citep{steidel99} and $-1.84$ \citep{reddy08}, the
MC simulations yielded ($M_{1700}^{\star}$, $\log{\phi^{\star}}$) of ($-20.95\pm0.29$, $-2.50\pm0.17$)
and ($-21.30\pm0.35$, $-2.75\pm0.21$), respectively.


\subsection{The Luminosity and Star-Formation Rate Densities}\label{5.2}
The LF is integrated down to $M_{1700}=-20.11$---the magnitude where incompleteness is a problem---to
obtain a comoving {\it observed} specific luminosity density (LD) of
$\log{\mathcal{L}_{\rm lim}}=26.28\pm0.69$ erg s$^{-1}$ Hz$^{-1}$ Mpc$^{-3}$ at 1700\AA.
The conversion between the SFR and specific luminosity for 1500-2800\AA\ is
SFR$_{\rm UV}$(M$_{\sun}$ yr$^{-1}$) = $1.4\times10^{-28}L_{\nu}$(erg s$^{-1}$ Hz$^{-1}$), where a
Salpeter IMF with masses from $0.1-100M_{\sun}$ is assumed \citep{kennicutt98}. Therefore, the extinction-
(adopted $E[B-V]=0.15$ and Calzetti law) and completeness-corrected SFR density of $z\sim2$ LBGs is
$\log{\dot\rho_{star}}=-0.99\pm0.69$ $M_{\sun}$ yr$^{-1}$ Mpc$^{-3}$. Using the \cite{madau98} conversion
would decrease the SFR by $\sim$10\%. Integrating to $L=0.1L^{\star}_{z=3}$, where $L^{\star}_{z=3}$ is
$L^{\star}$ at $z\sim3$ \cite[$M^{\star}_{z=3}=-21.07$,][]{steidel99}, yields $\log{\mathcal{L}}=26.52\pm0.68$
erg s$^{-1}$ Hz$^{-1}$ Mpc$^{-3}$ or an extinction-corrected SFR density of
$\log{\dot\rho_{star}}=-0.75\pm0.68$~$M_{\sun}$ yr$^{-1}$ Mpc$^{-3}$.\footnote[16]{The above numbers are
  upper limits if the low-$z$ contamination fraction is higher than estimates described in \S~\ref{4.2}.}

\subsection{Summary of Results}\label{5.3}
A UV luminosity function was constructed and yielded a best Schechter fit of
$M_{1700}^{\star}=-20.50\pm0.79$, $\log{\phi^{\star}}=-2.25\pm0.46$, and $\alpha=-1.05\pm1.11$ for
$z\sim2$ LBGs. The UV specific luminosity density, above the survey limit, is
$\log{\mathcal{L}_{\rm lim}}=26.28\pm0.68$ erg s$^{-1}$ Hz$^{-1}$ Mpc$^{-3}$. Correcting for dust
extinction, this corresponds to a SFR density of $\log{\dot\rho_{star}}=-0.99\pm0.68$ $M_{\sun}$
yr$^{-1}$ Mpc$^{-3}$.

\section{COMPARISONS WITH OTHER STUDIES}\label{6}
Comparisons in the UV specific luminosity densities, LFs, and Schechter parameters can be made with
previous studies. First, a comparison is made between the $z\sim2$ LBG LF with $z\sim2$ BX and
$z\sim3$ LBG LFs. Then a discussion of the redshift evolution in the UV luminosity density and LF
(parameterized in the Schechter form) is given in \S~\ref{6.2}.\\
\indent The results are summarized in Figures~\ref{Vlf}, \ref{schechter}, and \ref{lumdens} and
Table~\ref{table4}. For completeness, three different UV specific luminosity densities are reported by
integrating the LF down to: (1) $0.1L^{\star}_{z=3}$; (2) $L_{\rm lim}$, the limiting depth of the survey;
and (3) $L=0$. The latter is the least confident, as it requires extrapolating the LF to the faint-end,
where in most studies, it is not well determined.

\subsection{UV-selected Studies at $z\sim2-3$}\label{6.1}
\begin{figure} 
  \epsscale{1.0}
  \plotone{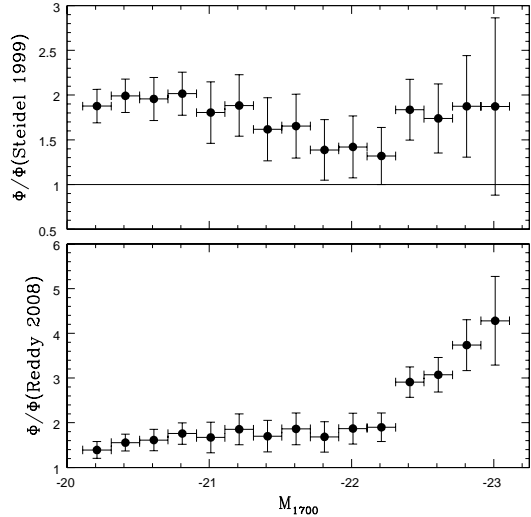}
  \caption{Comparisons of the LBG LF with other LFs. The ratios of the $z\sim2$ LBG LF to the Schechter
    fits of \cite{steidel99} LF and \cite{reddy08} are shown in the top and bottom panels, respectively.
    On average, the $z\sim2$ LBG LF is a factor of $1.7\pm0.1$ higher than these studies.}
  \label{LFcomp}
\end{figure}

In Figure~\ref{Vlf}, the $z\sim2$ LBG LF at the bright end is similar to those of LBGs from \cite{steidel99}
and BX galaxies from \cite{ST06a} and \cite{reddy08}; however, the faint end is systematically higher.
This is illustrated in Figure~\ref{LFcomp} where the ratios between the binned $z\sim2$ UV LF and the fitted
Schechter forms of \cite{steidel99} and \cite{reddy08} are shown. When excluding the four brightest and two
faintest bins, the \nuv-dropout LF is a factor of $1.7\pm0.1$ with respect to $z\sim3$ LBGs of
\cite{steidel99} and $z\sim2$ BX galaxies of \cite{reddy08} and \cite{ST06a}. 
The hard upper limit for stellar contamination (see \S~\ref{4.1}) would reduce this discrepancy to
a factor of $1.4\pm0.1$. There appears to be a trend that the ratio to \cite{reddy08} LF increases towards brighter
magnitudes. This is caused by the differences in the shape of the two LFs, particularly the faint-end slope.
The increase in the ratio is less noticeable when compared to \cite{steidel99}, which has a shallower faint-end
slope. Since the LFs of \cite{ST06a} and \cite{reddy08} are similar, the comparison of any results between the
\nuv-dropout and the BX selections will be made directly against \cite{reddy08}.\\
\indent All 11 points are $1-3\sigma$ from a ratio of 1. It has been assumed in this comparison that the
amount of dust extinction does not evolve from $z\sim3$ to $z\sim2$. Evidence supporting this assumption is:
in order for the {\it intrinsic} LBG LFs at $z\sim2$ and 3 to be consistent, the population of LBGs at
$z\sim2$ would have to be relatively {\it less} reddened by $\Delta E(B-V)=0.06$ (i.e., $E[B-V] = 0.09$
assuming a Calzetti extinction law). However, the stellar synthesis models, described previously, indicate
that $\EBV$ = 0.1 star-forming galaxies are expected to have observed $B-V\sim0.1$, and only 15\% of
\nuv-dropouts have $B-V \leq 0.1$. This result implies that dust evolution is unlikely to be the cause for
the discrepancy seen in the LFs.\\
%
%
\indent To compare the luminosity densities, the binned LF is summed. This is superior to integrating
the Schechter form of the LF as (1) no assumptions are made between individual LF values and for the
faint-end, and (2) the results do not suffer from the problem that Schechter parameters are affected
by small fluctuations at the bright- and faint-ends. The logarithm of the binned luminosity densities
for $-22.91<M_{1700}<-20.11$ are $26.27\pm0.16$ (this work), $26.02\pm0.04$ \citep{steidel99}, and
$26.08\pm0.07$ ergs s$^{-1}$ Hz$^{-1}$ Mpc$^{-3}$ \citep{reddy08}, which implies that the $z\sim2$ LBG
UV luminosity density is $0.25\pm0.16$ dex higher than the other two studies at the 85\% confidence
level.\\
\indent Since the low-$z$ contamination fraction is the largest contributor to the errors, more follow-up
spectroscopy will reduce uncertainties on the LF. This will either confirm or deny with greater statistical
significance that the luminosity density and LF of $z\sim2$ LBGs are higher than the $z\sim3$ LBGs and
$z\sim2$ BXs.

\subsection{Evolution in the UV Luminosity Function and Density}\label{6.2}
\begin{figure*} 
  \epsscale{0.9}
  \plotone{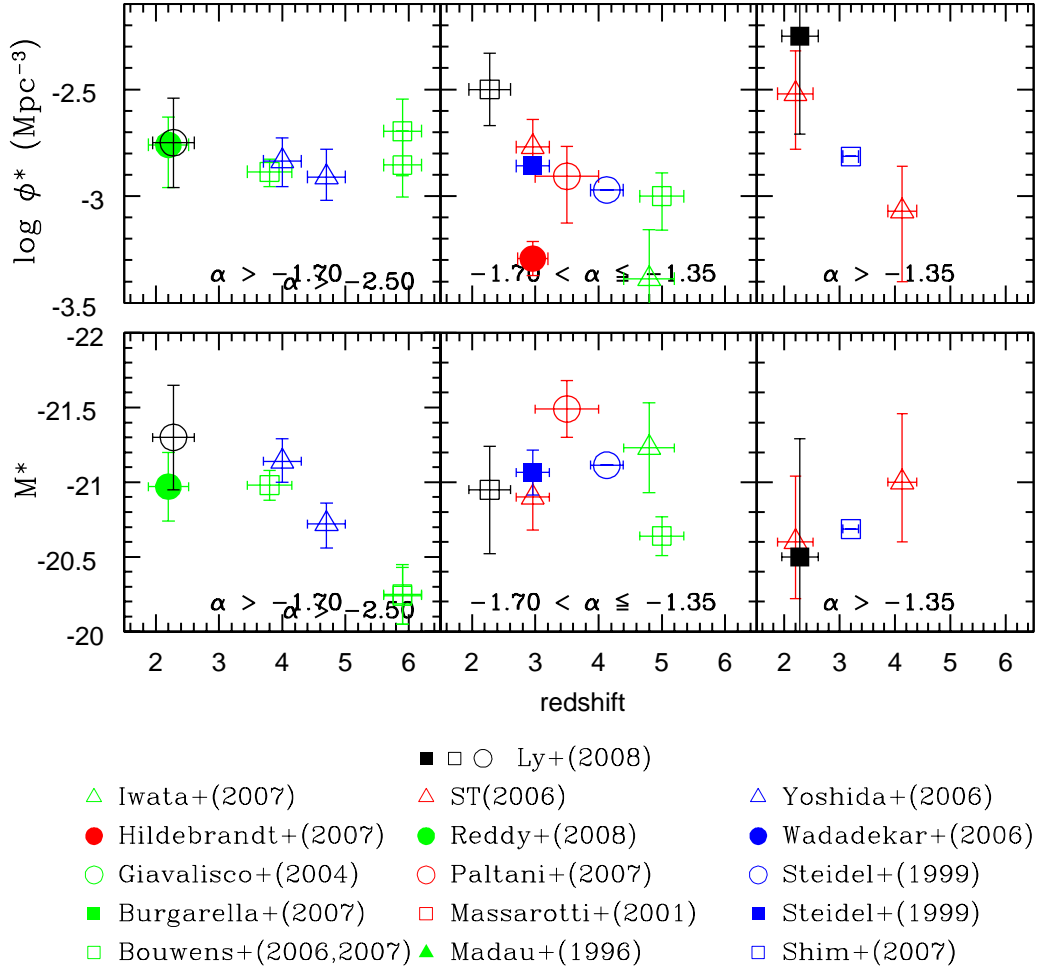}
  \caption{Compiled Schechter parameters of LBG and BX studies versus redshift. Top and bottom show the
    the normalization ($\phi^{\star}$), and the ``knee'' of the UV LF ($M^{\star}$), respectively.
    Measurements are grouped according to $\alpha$: $\leq-1.70$, between $-1.70$ and $-1.35$, and
    $>-1.35$. This \nuv-dropout work is shown as black filled square ($\alpha=-1.05$).
    The color and symbol conventions for studies in Figure~\ref{Vlf} are identical for this figure. In the
    legend, \cite{ST06a} is abbreviated as ``ST(2006)''. Some points are not shown here but have luminosity
    density measurements presented in Figure~\ref{lumdens}.}
  \label{schechter}
\end{figure*}

\begin{figure*} 
  \epsscale{1.1}
  \plottwo{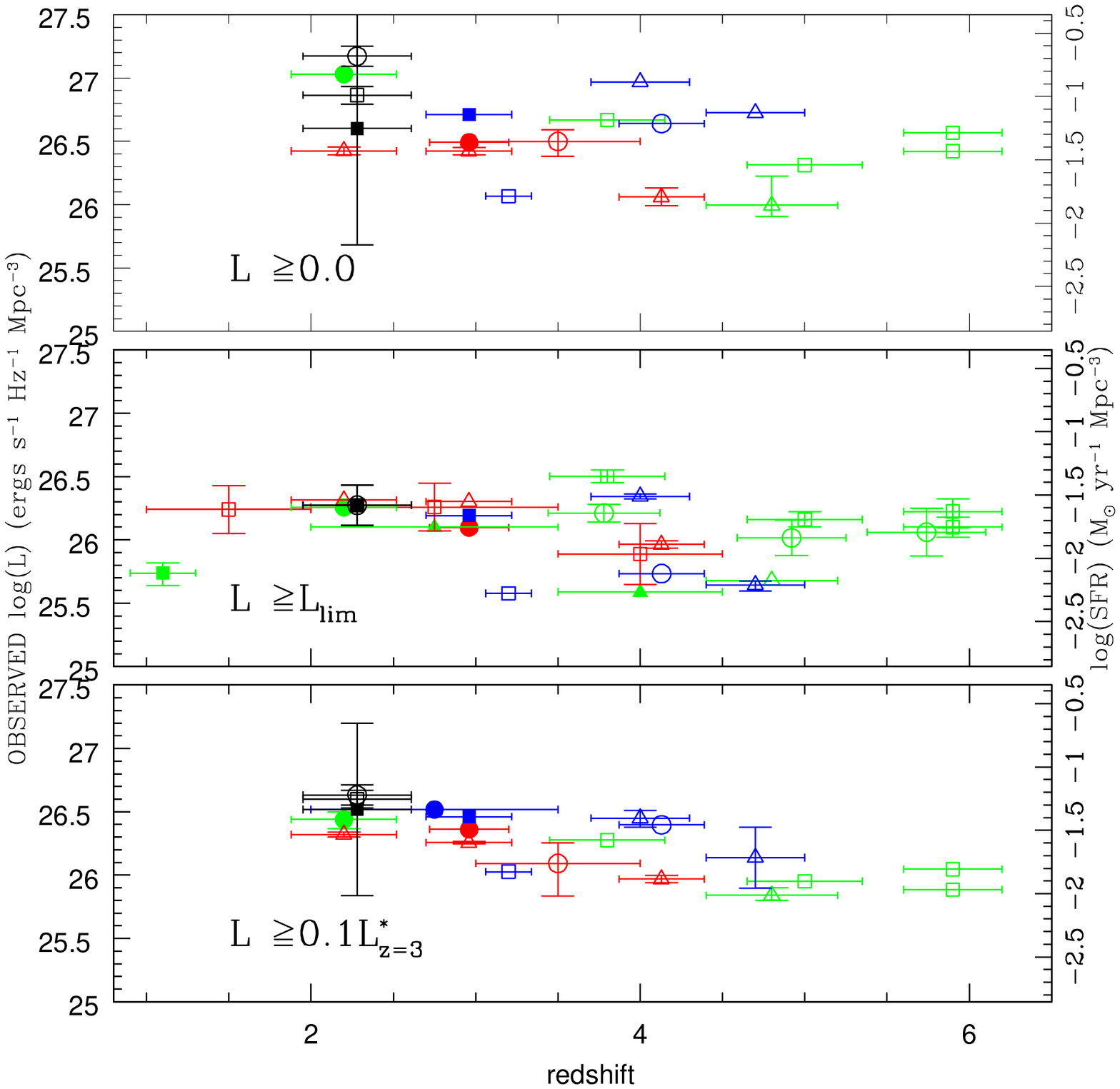}{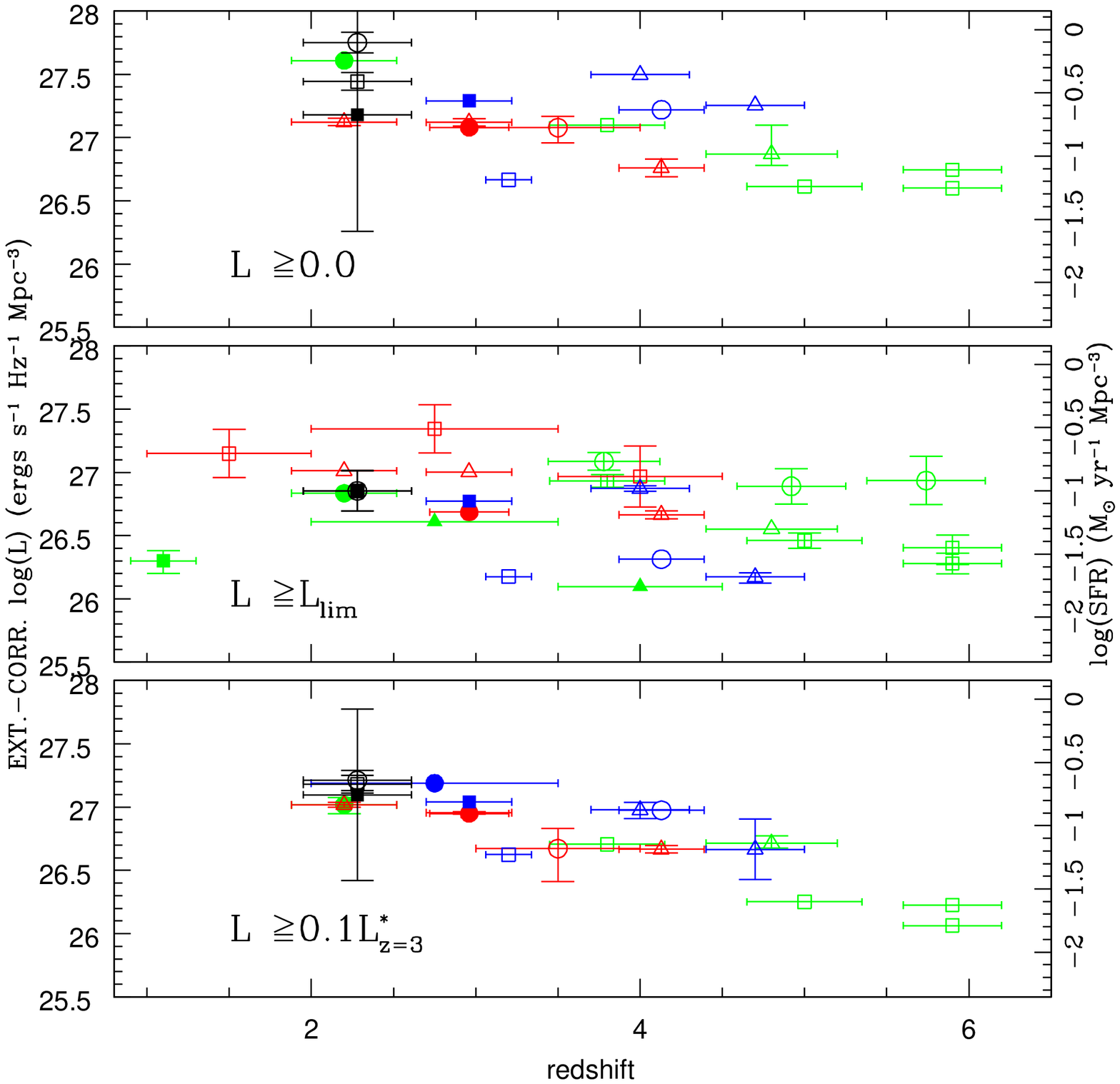}
  \caption{The observed ({\it left}) and extinction-corrected ({\it right}) UV specific luminosity
    densities as a function of redshift. The luminosity function is integrated to three different
    limits: $L=0$ (top panel), $L=L_{\rm lim}$ (the survey's limit; middle panel), and
    $L=0.1L^*_{z=3}$. The color and point-type schemes are the same as Figure~\ref{schechter}.
    The SFR densities are shown on the right axes following \cite{kennicutt98} conversion.
    For the $z\sim2$ LBG luminosity density integrated to $L=L_{\rm lim}$, only one value is shown,
    since all the fits with different $\alpha$ are almost identical.}
  \label{lumdens}
\end{figure*}

The Schechter LF parameters, listed in Table~\ref{table4}, are plotted as a function of redshift in
Figure~\ref{schechter}. There appears to be a systematic trend that $M^{\star}$ is less negative
(i.e., a fainter $L^{\star}$) by $\approx$1 mag at higher redshifts for surveys with $\alpha\leq-1.35$.
No systematic evolution is seen for $\phi^{\star}$, given the measurement uncertainties. Limited information
are available on the faint-end slope, so no analysis on its redshift evolution is provided. It is often
difficult to compare Schechter parameters, since they are correlated, and without confidence contours
for the fits of each study, the apparent evolution could be insignificant. A more robust measurement is the
product ($\phi^{\star}\times L^{\star}$), which is related to the luminosity density.\\
\indent The observed LDs, integrated to $0.1L^*_{z=3}$, show a slight increase of $\approx0.5$ dex from
$z\sim6$ to $z\sim3$. However, the two other luminosity densities appear to be flat, given the scatter
in the measurements of $\approx0.5-1.0$ dex. A comparison between $z\sim2$ and $z\sim5$ studies reveal
a factor of $3-6$ higher luminosity density at $z\sim2$. The extinction-corrected results for
$L_{\rm lim}=0$ and $L_{\rm lim}=0.1L^*_{z=3}$ show a factor of 10 increase from $z\sim6$
\cite{bouwens07}'s measurement to $z\sim2$. \cite{bouwens07} assumed a lower dust extinction correction.
If an average $E(B-V)=0.15$ with a Calzetti law is adopted, the rise in the extinction-corrected
luminosity density is $\approx3$.


\section{DISCUSSION}\label{7}
\begin{figure*} 
  \plottwo{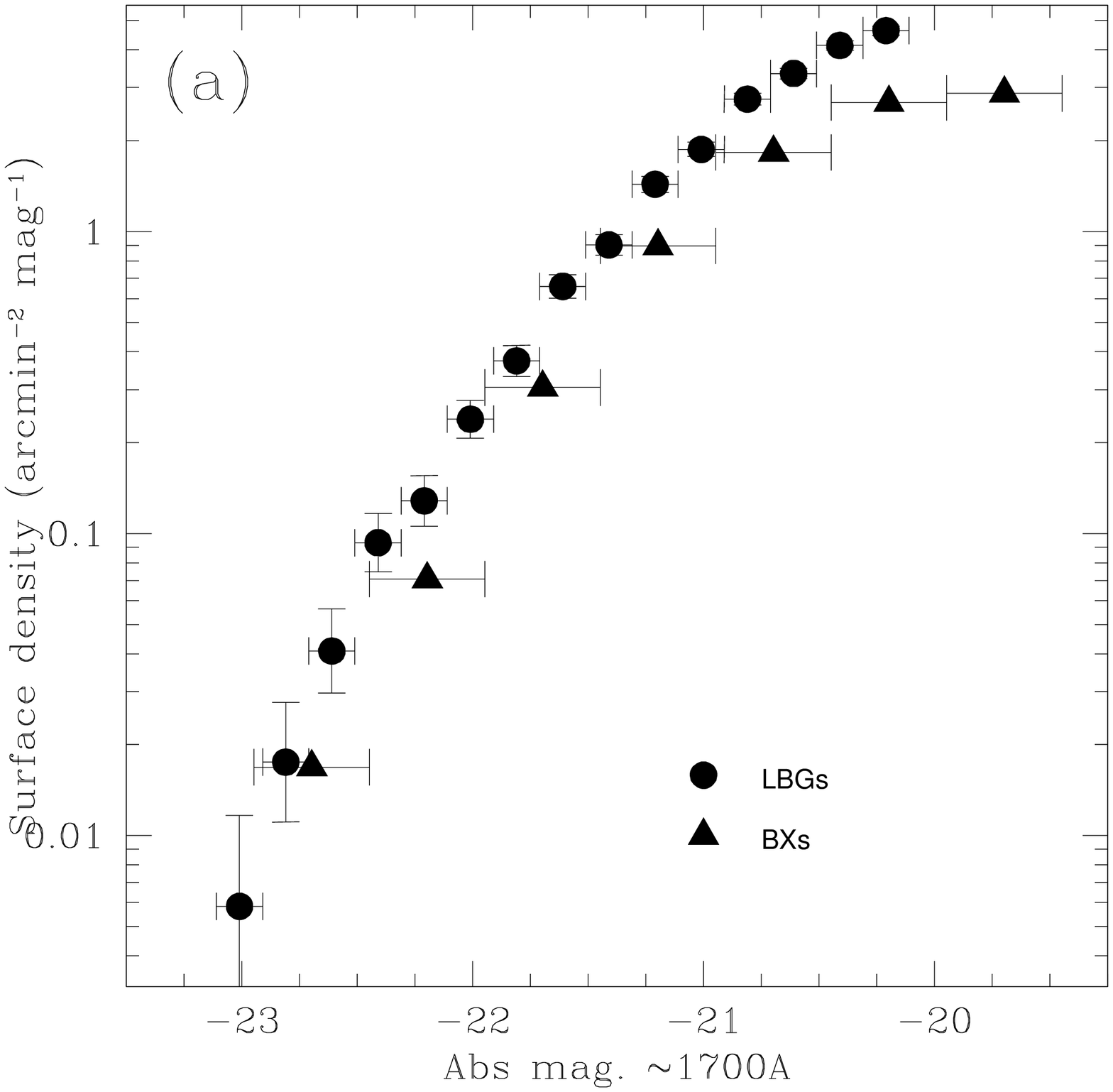}{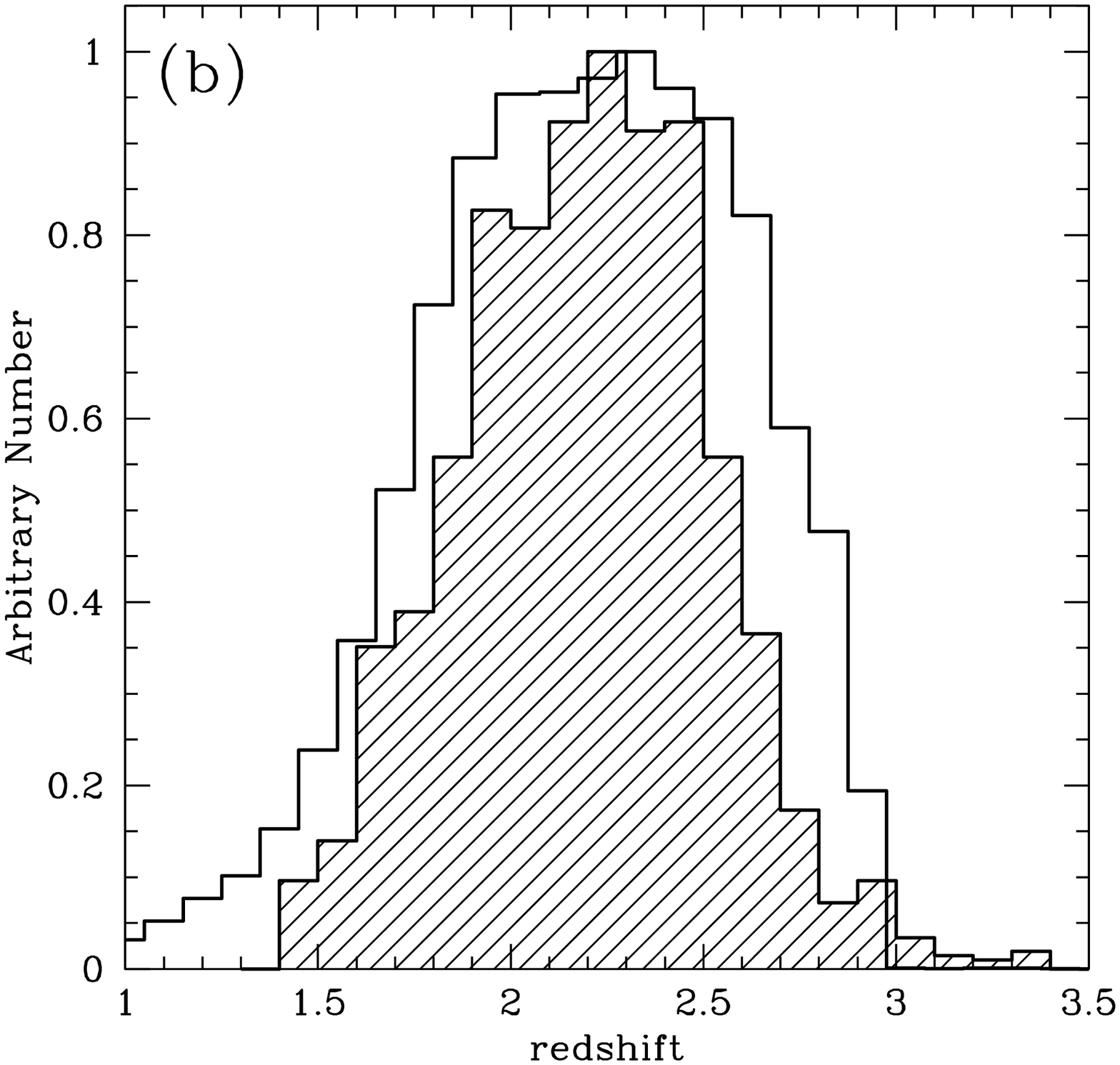}
  \caption{Surface densities and redshift distributions for $z\sim2$ BXs and LBGs. In ({\it a}), the
    surface densities of LBGs and BXs are shown as circles and triangles, respectively. Both studies
    have stellar and low-$z$ contamination corrections applied.
    This figure reveals that the LBG surface density is systematically higher than the BX's. The redshift
    distributions are shown in ({\it b}). The shaded (unshaded) histogram corresponds to BXs (LBGs).
    For the BX, the redshift distribution is obtained from \cite{reddy08} spectroscopic sample, while the
    LBG is determined from the MC simulations described in \S~{4.3} for all magnitudes. The similarities
    in redshifts surveyed by both studies and the higher surface density of LBGs indicate that the BX
    technique misses a fraction of LBGs.}
  \label{LBGBX}
\end{figure*}

In this section, the discrepancy between the UV LF of this study and two BX studies, shown in
\S~\ref{6.1}, is examined. Three possible explanations are considered:\\
\noindent{\bf 1. Underestimating low-$z$ contamination.} To estimate contamination, a large sample
of $z\lesssim1.5$ NB emitters was cross-matched with the \nuv-dropout sample. This method indicated
that $34\%\pm17\%$ of \nuv-dropouts are at $z<1.5$. However, it is possible that star-forming
galaxies at $z=1-1.5$ could be missed by the NB technique, but still be identified as \nuv-dropouts.
This would imply that the contamination rate was underestimated. To shift the \nuv-dropout LF to
agree with \cite{reddy08} and \cite{ST06a} would require that the contamination fraction be more than
60\%. However, the spectroscopic sample has yielded a large number of genuine LBGs and a similar
low-$z$ contamination (at least 21\% and at most 38\%). If the large (60\%) contamination rate is
adopted, it would imply that only 15 of 40 spectra (LRIS and Hectospec) are at $z>1.5$, which is
argued against at the 93\% confidence level (98\% with $R=2.5$ threshold), since 24 LBGs
(1.6 times as many) have been identified.
Furthermore, the LRIS and Hectospec observations independently yielded similar low contamination
fractions, and the MC simulation (that involved adding artificial LBGs to the images) independently
suggested 30\% contamination from $z\leq 1.5$.\\
\noindent{\bf 2. Underestimating the comoving effective volume.}
The second possibility is that $V_{\rm eff}$ was underestimated, as the spectral synthesis model
may not completely represent the galaxies in this sample, and misses $z\sim1-1.5$ galaxies. However,
a comparison between a top-hat $P(m,z)$ from $z=1.7-2.7$ versus $z=1.4-2.7$ ($z=1.0-2.7$) would
only decrease number densities by $\approx20$\% (37\%). Note that the latter value is consistent
with \fcont.\\
\noindent{\bf 3. Differences between LBG and BX galaxies selection.}
This study uses the Lyman break technique while other studies used the `BX' method to identify
$z\sim2$ galaxies. Because of differences in photometric selection, it is possible that the galaxy
population identified by one method does not match the other, but instead, only a fraction
of BX galaxies are also LBGs and vice versa. 
This argument is supported by the higher surface density of LBGs compared to BXs over 2.5 mag,
as shown in Figure~\ref{LBGBX}a. However, their redshift distributions, as shown in
Figure~\ref{LBGBX}b, are very similar.\\
\indent This scenario would imply that there is an increase in the LF and number density of LBGs
from $z\sim3$ to $z\sim2$, indicating that the comoving SFR density peaks at $z\sim2$, since
there is a decline towards $z\sim0$ from UV studies \cite[see][and references therein]{hopkins04}.
However, it might be possible that the selection (\nuv$-B-V$) of $z\sim2$ LBGs could include
more galaxies than the $U_nGR$ color selection used to find $z\sim3$ LBGs. Although no reason
exists to believe that $z\sim3$ LBG selection is more incomplete than at $z\sim2$ (nor is there
any evidence for such systematic incompleteness for $z>4$ LBGs), it is difficult to rule out
this possibility for certain. But if so, then the SFR density might not evolve.
In addition, the conclusion that $z\sim2$ is the peak in star-formation is based on UV selection techniques,
which are less sensitive at identifying dusty ($E[B-V]>0.4$) star-forming galaxies. However,
spectroscopic surveys have revealed that the sub-mm galaxy population peaks at $z\approx2.2$ \citep{chapman05},
which further supports the above statement that $z\sim2$ is the epoch of peak star-formation.

\section{CONCLUSIONS}\label{8}
By combining deep \galex/\nuv\ and optical Suprime-Cam imaging for the Subaru Deep Field, a
large sample of LBGs at $z\sim2$ has been identified as \nuv-dropouts. This extends the popular
Lyman break technique into the redshift desert, which was previously difficult due to the lack of
deep and wide-field UV imaging from space. The key results of this paper are:
\begin{enumerate}
  \setlength{\itemsep}{1pt}\setlength{\parskip}{0pt}\setlength{\parsep}{0pt}
\item Follow-up spectroscopy was obtained, and 63\% of identified galaxies are at $z=1.6-2.7$.
  This confirms that most \nuv-dropouts are LBGs. In addition, MMT/Hectospec will complement
  Keck/LRIS by efficiently completing a spectroscopic survey of the bright end of the LF.
\item Selecting objects with $NUV-B\geq1.75$, $B-V\leq0.5$, and $NUV-B\geq2.4(B-V)+1.15$ yielded
  7964 \nuv-dropouts with $V=21.9-25.3$. The spectroscopic sample implied that 50$-$86\% of
  \nuv-dropouts are LBGs.
\item Using broad-band optical colors and stellar classification, 871 foreground stars have been
  identified and removed from the photometric sample. This corresponds to a $4-11\%$ correction
  to the \nuv-dropout surface density, which is consistent with the $3-7\%$ from limited spectra
  of stars presented in this paper.
\item In addition, low-$z$ contamination was determined using a photometric sample of NB emitters at
  $z\lesssim1.47$. This novel technique indicated that the contamination fraction is (at least) on average
  $34\%\pm17\%$, which is consistent with the spectroscopic samples and predictions from MC simulations
  of the survey.
\item After removing the foreground stars and low-$z$ interlopers, MC simulations were performed to
  estimate the effective comoving volume of the survey. The UV luminosity function was constructed
  and fitted with a Schechter profile with $M_{1700}^{\star}=-20.50\pm0.79$,
  $\log{\phi^{\star}}=-2.25\pm0.46$, and $\alpha=-1.05\pm1.11$.
\item A compilation of LF and SFR measurements for UV-selected galaxies is made, and there appears to
  be an increase in the luminosity density: a factor of 3$-$6 ($3-10$) increase from $z\sim5$ ($z\sim6$)
  to $z\sim2$.
\item Comparisons between \nuv-dropouts with LBGs at $z\sim3$ \citep{steidel99} and BXs at $z\sim2$
  \citep{ST06a,reddy08} reveal that the LF is $1.7\pm0.1$ ($1.4\pm0.1$ if the hard upper limit of
  stellar contamination is adopted) times higher than these studies. The summed
  luminosity density for $z\sim2$ LBGs is 1.8 times higher at 85\% confidence (i.e., $0.25\pm0.16$ dex).
\item Three explanations were considered for the discrepancy with $z\sim2$ BX studies. The
  possibility of underestimating low-$z$ contamination is unlikely, since optical spectroscopy argues
  against the possibility of a high (60\%) contamination fraction at the 93\% confidence. Second,
  even extending the redshift range to increase the comoving volume is not sufficient to resolve the
  discrepancy. The final possibility, which cannot be ruled out, is that a direct comparison between
  BX-selected galaxies and LBG is not valid, since the selection criteria differ. It is likely
  that the BX method may be missing some LBGs. This argument is supported by the similar redshift
  distribution of BXs and LBGs, but the consistently higher surface density of LBGs over 2.5
  mag.
\item If the latter holds with future reduction of low-$z$ contamination uncertainties via spectroscopy,
  then the SFR density at $z\sim2$ is higher than $z\gtrsim3$ and $z\lesssim1.5$ measurements obtained
  via UV selection. Combined with sub-mm results \citep{chapman05}, it indicates that $z\sim2$ is the
  epoch where galaxy star-formation peaks.
\end{enumerate}\vspace{-0.75cm}

\acknowledgements
The Keck Observatory was made possible by the generous financial support of the W.M. Keck Foundation.
The authors wish to recognize and acknowledge the very significant cultural role and reverence that
the summit of Mauna Kea has always had within the indigenous Hawaiian community. We are most fortunate
to have the opportunity to conduct observations from this mountain. We gratefully acknowledge NASA's
support for construction, operation, and science analysis for the \galex~mission. This research was
supported, by NASA grant NNG-06GDD01G. We thank the Hectospec instrument and queue-mode scientists
and the MMT operators for their excellent assistance with the observations. Public Access MMT time
is available through an agreement with the National Science Foundation. C.L. thank A. Shapley,
M. Pettini, and S. Savaglio for providing their composite spectra, and S. C. Odewahn for providing
K. Adelberger's LRIS reduction code.

{\it Facilities:} \facility{Keck:I (LRIS)}, \facility{\galex}, \facility{MMT (Hectospec)},
\facility{Subaru (MOIRCS, Suprime-Cam)}

\begin{appendix}
\section{Individual Sources of Special Interest}\label{appendix}
\begin{figure}[!htc] 
  \epsscale{0.75}
  \plotone{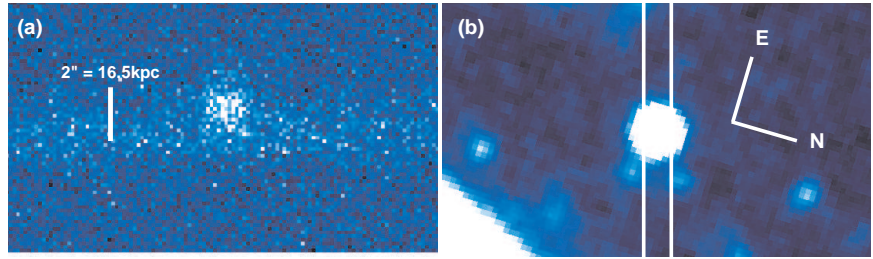}
  \caption{Optical images for 179350. (a) The 2-D spectrum with wavelength increasing to the right
    shows \Lya~emission offset by $\approx$1\arcsec~from the center of the continuum. The vertical
    white line corresponds to 2\arcsec. (b) The Suprime-Cam $B$-band image centered on the targeted
    source shows that there are no sources in the direction of the extended emission. The two white
    vertical lines correspond to the slit, so (b) is rotated to have the same orientation as (a),
    and the vertical scales are the same. \color}
  \label{extlya}
\end{figure}

In most cases, the confirmed LBGs showed no unique spatial or spectral properties. However,
3 cases are worth mentioning in more detail.\\
\textbf{1. SDFJ132431.8+274214.3 (179350)}. Upon careful examination of the 2-D spectra, it appears
that the \Lya~emission from this source is offset by $\approx$1.1\arcsec~(9 kpc at 107\arcdeg~east
of north) from the continuum emission, which is shown in Figure~\ref{extlya}a. The extended emission
appears in the individual exposures of $15-30$ minutes. The deep ($3\sigma=28.45$) $B$-band image
(Figure~\ref{extlya}b) reveals that there are no sources in this direction and at this distance,
assuming that the continuum emission in the spectrum corresponds to the bright source in the
$B$-band image. The two sources located below the bright object in Figure~\ref{extlya}b are too faint
for their continuum emission to be detected with LRIS. Also, absorption features seen in the 1-D
spectra (see Figure~\ref{spec1}a) are at nearly the same redshift as \Lya. This indicates that
the \Lya~emission is associated with the targeted source, rather than a secondary nearby companion.

Extended \Lya~emission galaxies are rare
\cite[e.g.,][have the largest sample of 41 objects]{saito06}, and the extreme cases are extended on
larger ($\sim100$ kpc) scales, such as LAB1 and LAB2 of \cite{steidel00}. In addition, extended
\Lya~emission has been seen in some cases that show evidence for energetic galactic winds
\citep{hesse03}. Either this source is a fortuitous discovery from a dozen spectra, or perhaps a
fraction of \nuv-dropouts have extended \Lya~emission. The physical significance of this source is not
discussed here, given limited information.

\begin{figure}[!btc] 
  \vspace{9cm}
  \includegraphics{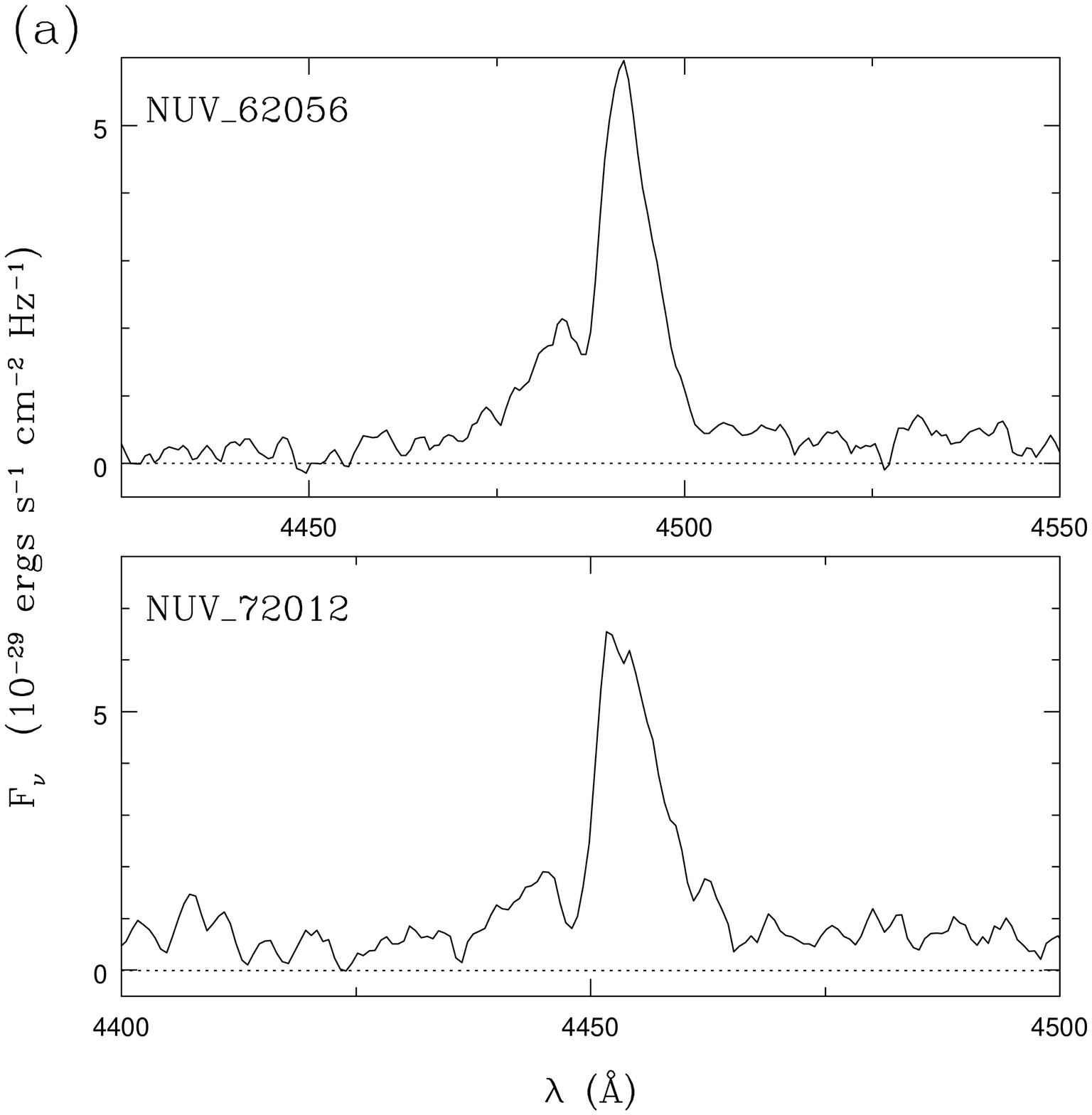}
  \includegraphics{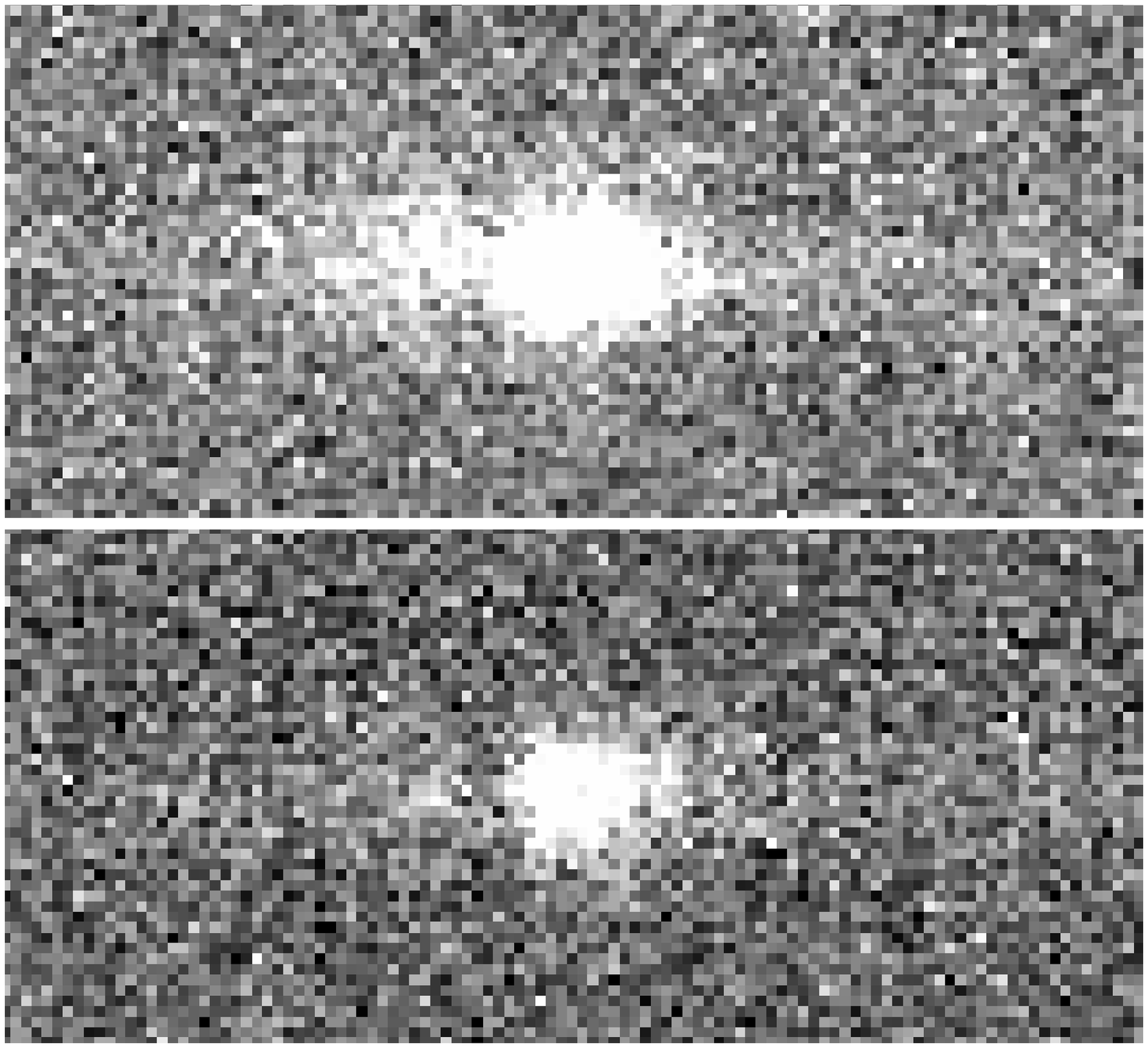}
  {{}}
  \caption{(a) One- and (b) two-dimensional spectra for 62056 (top) and 72012 (bottom) centered on the
    \Lya~emission. These objects appear to show weak emission blue-ward of \Lya. See \S~\ref{appendix}
    for a discussion. \color}
  \label{bluelya}
\end{figure}

\begin{figure}[!htc] 
  \epsscale{0.5}
  \plotone{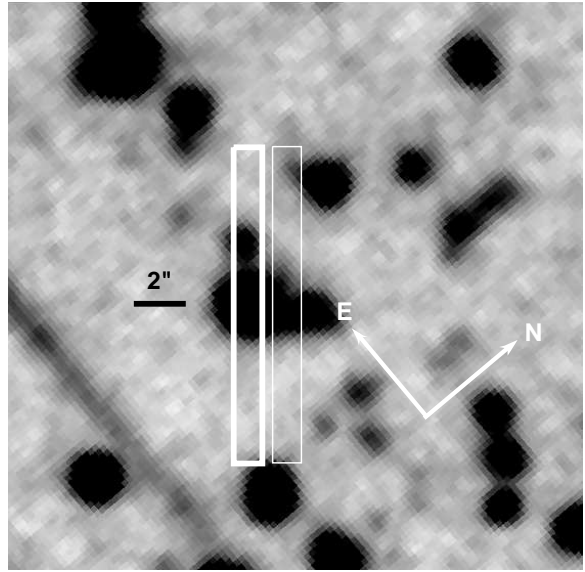}
  \caption{The $B$-band image cropped to 20\arcsec~on a side and centered on 62056. The white box with
    thick lines is the LRIS slit intended to target the bright object. However, a 1.5\arcsec~offset of
    the slit in the north-west direction (as shown by the thin white box) may explain the blue excess
    seen in the 1-D and 2-D spectra (Figure~\ref{bluelya}) by including both objects.}
  \label{62056}
\end{figure}

\textbf{2. SDFJ132452.9+272128.5 (62056)}. The 1- and 2-D spectra for this source reveal an asymmetric
emission line, as shown in Figure~\ref{bluelya}a, but with a weak ``bump'' about 10\AA\ blue-ward from
the peak of \Lya~emission. The $B$-band image (see Figure~\ref{62056}) shows two nearby sources where
one is displaced $\approx$2\arcsec~nearly in the direction of the slit orientation while the other
source is displaced in the direction perpendicular to the slit orientation. It may be possible that
the blue excess is originating from the latter source due to a slight misalignment of the slit to fall
between the two sources (i.e., they are physically near each other). To confirm this hypothesis, 
spectroscopy with a 90\arcdeg~rotation of the slit would show two sources with \Lya~emission
$\approx$800 km s$^{-1}$ apart.


\begin{figure}[!htc] 
  \epsscale{0.75}
  \plotone{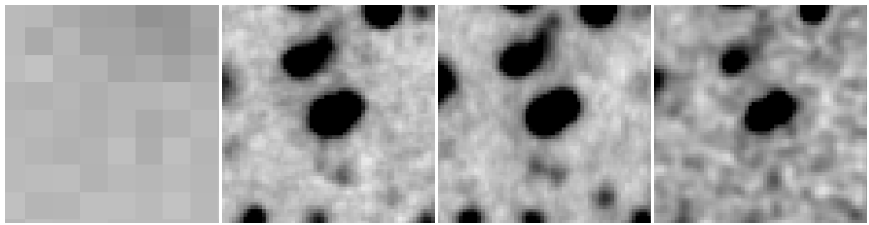}
  \caption{Postage stamp images (10\arcsec~on a side) for 72012. From left to right is \nuv, $B$, \Rc, and
    NB704. North is up and east is to the left. The source on the right shows a weak excess in NB704
    relative to the broad-band images.}
  \label{72012}
\end{figure}

\textbf{3. SDFJ132450.3+272316.24 (72012)}. This object is not listed in Table~\ref{table1}, as it
was serendipitously discovered. The slit was originally targeting a narrow-band (NB) emitter. The
LRIS-R spectrum showed an emission line at 7040\AA, but the blue-side showed a strong emission line
that appears asymmetric at $\approx$4450\AA. One possibility is that the 4450\AA\ feature is \Lya,
so that the 7040\AA\ emission line is the redshifted \ion{C}{3}]~$\lambda$1909, but at $z=2.6634$,
\ion{C}{3}] is expected at $\approx$6994\AA. This $\approx$40\AA\ difference is not caused
by poor wavelength calibration, as night sky and arc-lamps lines are located where they are
expected in both the blue and red spectra. In Figure~\ref{72012}, the $B$-band image reveals two sources,
one of which is moderately brighter in the NB704 image, as expected for a NB704 emitter. These two
sources were too close for SExtractor to deblend, but the coordinate above has been corrected. Because
the NB704 emitter is a foreground source, the measured \nuv~flux for the other source is affected, and
results in a weak detected source in the \nuv. Thus, this source is missed by the selection criteria
of the ver. 1 catalog and those described in \S~\ref{3.2}. It is excluded from the spectroscopic sample
discussed in \S~\ref{2}.

 This source is of further interest because it also shows a blue excess bump (shown in 
Figure~\ref{bluelya}) much like 62056, but weaker. This blue bump does not correspond to a different
emission line with the same redshift as the 7040\AA\ emission line. Since the bump is 10\AA\ from the 
strong \Lya~emission, it is likely associated with the source producing \Lya. Both 62056 and 72012 were
obtained on the second mask. These blue bumps are not due to a misalignment of single exposures when
stacking the images together, as other equally bright sources in the mask with emission lines do not show
a secondary blue peak. Other studies have also seen dual peak \Lya~emission profiles
\citep[e.g.,][]{tapken04,tapken07,cooke08,verhamme08}. In addition, high resolution spectra of 9 LBGs
have also revealed 3 cases with double-peaked \Lya~profile \citep{shapley06}, which indicates that such
objects may not be rare.

\end{appendix}

\clearpage \newpage 
\LongTables
\newcommand{\A}{\tablenotemark{a}}
\newcommand{\B}{\tablenotemark{b}}
\newcommand{\C}{\tablenotemark{c}}
\newcommand{\D}{\tablenotemark{d}}
\newcommand{\pa}{\phm{\D}}
\newcommand{\pb}{\phm{]}}
\newcommand{\po}{\phm{1}}
\newcommand{\pt}{$>$}
\newcommand{\pc}{\phm{\pt}}
\begin{landscape}
\begin{deluxetable}{clccrcccccccrcrr}
  \tabletypesize{\scriptsize}
  \tablewidth{0pc}
  \tablecaption{Properties of Spectroscopically Targeted \nuv-dropouts and BzKs}
  \tablehead{
  \colhead{$B$-band ID\A}&\colhead{Name (SDF)}&\multicolumn{8}{c}{UV and Optical measurements}& &\multicolumn{5}{c}{Spectroscopic measurements}\\
   \cline{3-10} \cline{12-16}
   & & \colhead{$NUV-B$}&\colhead{$B-V$}&\colhead{$NUV$}&\colhead{$B$}&\colhead{$V$}&\colhead{$R_{\rm C}$}&\colhead{$i$\arcmin}&\colhead{$z$\arcmin}& &
   \colhead{redshift}&\colhead{$R$}&\colhead{Temp.\B}&\colhead{F(\Lya)}&\colhead{EW$_{\rm o}$(\Lya)}\\
  \colhead{(1)}&\colhead{(2)}&\colhead{(3)}&\colhead{(4)}&\colhead{(5)}&\colhead{(6)}&\colhead{(7)}&\colhead{(8)}&\colhead{(9)}&\colhead{(10)}& &
  \colhead{(11)}&\colhead{(12)}&\colhead{(13)}&\colhead{(14)}&\colhead{(15)}}
  \startdata
  \multicolumn{15}{c}{With emission lines}\\
    179350L& 132431.8+274214.28& \pt2.662& 0.078& \pt27.121& 24.459& 24.366& 24.379& 24.430& 24.500& & 2.0387\pa&  9.72& 4$^{5}$     & 157\pa & 5.38\pa\\
    170087L& 132428.6+274037.95& \pt2.562& 0.128& \pt27.126& 24.564& 24.498& 23.991& 23.880& 23.859& & 2.2992\pa& 12.29& 4$^{5}$     & 80.4\pa&58.20\pa\\
  \po62056L& 132452.9+272128.50& \pt2.991& 0.107& \pt27.165& 24.174& 24.161& 24.103& 24.182& 24.263& & 2.6903\pa& 34.21& 4$^{3,5}$   & 66.4\C &37.12\C \\
  \po60962L& 132436.7+272118.67& \pt2.896& 0.136& \pt27.164& 24.268& 24.110& 23.993& 23.917& 23.527& & 1.9098\pa&  3.06& 4$^{3,5}$   & 20.3\pa& 7.14\pa\\
  \po96658L& 132521.5+272730.24& \pt2.605& 0.282& \pt27.158& 24.553& 24.310& 24.193& 24.220& 24.485& & 2.5639\pa&  3.99& 5$^{3,4}$   &  9.5\pa& 5.21\pa\\
  \po87890L& 132520.3+272559.22& \pt3.597& 0.278& \pt27.161& 23.564& 23.334& 23.298& 23.335& 23.362& & 2.5747\pa&  9.84& 2$^{3,4,5}$ & 28.9\pa& 5.56\pa\\
  \po92076L& 132507.6+272303.44& \pt2.666& 0.239& \pt27.143& 24.477& 24.256& 24.130& 24.115& 24.184& & 2.1720\pa&  3.93& 3$^{2,5}$   & 13.4\pa& 6.52\pa\\
  \po89984L& 132506.8+272620.75& \pt3.386& 0.246& \pt27.169& 23.783& 23.567& 23.516& 23.436& 23.309& & 2.0894\pa&  3.30& 2$^{1}$     &  5.6\pa& 4.11\pa\\
  \po94093L& 132457.7+272703.10& \pt3.248& 0.129& \pt27.165& 23.917& 23.770& 23.693& 23.674& 23.705& & 2.0025\pa&  6.87& 5$^{2,3,4}$ & 56.9\pa&20.71\pa\\
  \po82392L& 132454.4+272503.97& \pt2.941& 0.196& \pt27.166& 24.225& 24.037& 24.025& 24.084& 24.054& & 2.6527\pa& 28.57& 4$^{2,3,5}$ & 112\pa &45.56\pa\\
    139014M& 132417.5+273512.63& \pc2.110& 0.238& \pc26.131& 24.021& 23.863& 23.569& 23.283& 23.109& & 1.750\pa &\ldots&\ldots       & \ldots &\ldots\\
    140830M& 132422.4+273530.21& \pt2.816& 0.100& \pt27.158& 24.342& 24.264& 24.024& 23.821& 23.390& & 1.504\pa &\ldots&\ldots       & \ldots &\ldots\\
    142813M& 132414.8+273552.41& \pt2.552& 0.124& \pt27.160& 24.608& 24.420& 24.401& 24.194& 23.647& & 2.018\pa &\ldots&\ldots       & \ldots &\ldots\\
    143960M& 132425.5+273603.42& \pt2.421& 0.347& \pt27.161& 24.740& 24.436& 24.393& 23.965& 23.682& & 1.872\pa &\ldots&\ldots       & \ldots &\ldots\\
    166380M& 132410.4+273958.51& \pt2.994& 0.356& \pt27.278& 24.284& 23.951& 23.870& 23.745& 23.532& & 2.013\pa &\ldots&\ldots       & \ldots &\ldots\\
    166078M& 132418.2+273954.46& \pt3.332& 0.530& \pt27.138& 23.806& 23.290& 23.081& 22.884& 22.628& & 2.044\pa &\ldots&\ldots       & \ldots &\ldots\\
    158464M& 132419.6+273842.92& \pc0.597& 0.193& \pt25.235& 24.638& 24.485& 24.162& 23.887& 23.513& & 1.506\pa &\ldots&\ldots       & \ldots &\ldots\\
    170958M& 132415.8+274043.52& \pt1.136& 0.495& \pt27.121& 25.985& 25.489& 25.079& 24.912& 24.690& & 1.710\pa &\ldots&\ldots       & \ldots &\ldots\\
    171558M& 132409.1+274052.82& \pc0.094& 0.248& \pt25.145& 25.051& 24.842& 24.675& 24.304& 23.936& & 1.796\pa &\ldots&\ldots       & \ldots &\ldots\\
    188586M& 132417.8+274405.52& \pc1.535& 0.309& \pc25.720& 24.185& 24.102& 23.725& 23.328& 22.886& & 1.719\pa &\ldots&\ldots       & \ldots &\ldots\\
  \po78625H& 132343.4+272426.33& \pc2.625& -0.162& \pc25.171& 22.546& 22.704& 22.225& 22.211& 22.153& & 1.6755\pa& 2.30& AGN         & \ldots &\ldots\\ 
    175584H& 132504.3+274147.60& \pc2.405& 0.223& \pc25.284& 22.879& 22.643& 22.341& 22.084& 21.723& & 2.3902\pa&  2.69& 4$^{3}$     & \ldots &\ldots\\
    169311H& 132440.0+274040.27& \pt3.838& 0.429& \pt27.142& 23.304& 22.949& 22.847& 22.890& 22.875& & 2.6693\pa&  6.72&3$^{1,2,4,5}$ & \ldots &\ldots\\
    144397H& 132422.5+273612.47& \pc3.621& 0.421& \pc26.231& 22.970& 22.541& 22.386& 22.336& 22.326& & 2.6421\pa&  2.60& 1           & \ldots &\ldots\\
    133660H& 132507.0+273413.84& \pc2.646& 0.175& \pc25.785& 23.139& 22.970& 23.009& 22.673& 22.609& & 1.9345\pa&  2.52& AGN         & \ldots &\ldots\\\hline
  \multicolumn{15}{c}{Absorption line systems}\\
    186254L& 132442.0+274334.89& \pt4.032& 0.203& \pt27.145& 23.113& 22.910& 22.833& 22.727& 22.560& & 1.7550\pa&  3.15& 6$^{2}$     & \ldots &\ldots\\
  \po62351L& 132447.2+272135.84& \pt3.213& 0.221& \pt27.179& 23.966& 23.750& 23.712& 23.610& 23.462& & 1.7921\pa&  6.61& 6$^{1,2}$   & \ldots &\ldots\\
    144516H& 132350.8+273614.52& \pc1.915& 0.291& \pc25.014& 23.099& 22.804& 22.638& 22.243& 21.955& & 1.7488\pa&  5.79& 5$^{6}$     & \ldots &\ldots\\
    182284H& 132348.4+274301.74& \pc3.560& 0.288& \pc26.004& 22.444& 22.198& 21.962& 21.774& 21.384& & 1.5926\pa&  3.22& 7$^{6}$     & \ldots &\ldots\\\hline
  \multicolumn{15}{c}{$Z<1.5$ interlopers and stars}\\
    179764L& 132442.6+274220.19& \pc1.421& 0.183& \pc25.673& 24.252& 24.054& 23.970& 23.602& 23.391& & 1.0139\pa&  9.32&[7]$^{4,5,6}$& \ldots &\ldots\\
  \po63771L& 132452.9+272147.91& \pt2.767& 0.331& \pt27.163& 24.396& 24.058& 23.832& 23.383& 22.922& & 1.0965\pa& 10.37&[7]$^{4,5}$  & \ldots &\ldots\\
  \po68765L& 132444.4+272237.13& \pt1.561& 0.256& \pt27.183& 25.622& 25.336& 24.711& 24.401& 24.404& & 0.6898\pa&  6.29&[5]$^{4,6,7}$& \ldots &\ldots\\
  \po48542L& 132434.6+271901.63& \pc1.725& 0.138& \pc26.109& 24.384& 24.191& 23.904& 23.680& 23.361& & 1.4220\pa&  3.77&[7]$^{4,5}$  & \ldots &\ldots\\
    104403L& 132508.4+272853.98& \pc1.544& 0.326& \pc26.110& 24.566& 24.294& 24.102& 23.752& 23.580& & 0.9921\pa& 10.22&[7]$^{4,5,6}$& \ldots &\ldots\\
    136893M& 132424.1+273447.28& \pt2.666& 0.346& \pt27.157& 24.491& 24.145& 23.700& 23.300& 22.769& & 1.479\pa &\ldots&\ldots       & \ldots &\ldots\\
    137114M& 132416.4+273455.52& \pc1.751& 0.245& \pc25.626& 23.875& 23.669& 23.348& 23.064& 22.504& & 1.174\pa &\ldots&\ldots       & \ldots &\ldots\\
    163292M& 132423.2+273923.57& \pc1.098& 0.295& \pc26.278& 25.180& 24.875& 24.544& 24.263& 23.777& & 1.498\pa &\ldots&\ldots       & \ldots &\ldots\\
    191435M& 132422.3+274421.71& \pc0.673& 0.119& \pc23.899& 23.226& 23.100& 22.959& 22.788& 22.427& & 1.250\pa &\ldots&\ldots       & \ldots &\ldots\\
    145511H& 132429.9+273635.92& \pc3.258& 0.148& \pc25.944& 22.686& 22.539& 22.337& 22.176& 21.765& & 1.4729\pa&  2.82& 7           & \ldots &\ldots\\
  \po71239L& 132453.1+272307.35& \pt3.984& 0.623& \pt27.183& 23.199& 22.574& 22.301& 22.175& 22.088& &-0.0008\pa&  6.32&[2]$^{1,3}$  & \ldots &\ldots\\
  \po66611L& 132446.5+272218.81& \pt4.863& 0.532& \pt27.178& 22.315& 21.783& 21.581& 21.485& 21.440& &-0.0018\pa&  9.96&[2]$^{1,3}$  & \ldots &\ldots\\
  \po86900L& 132511.5+272303.44& \pt4.206& 0.616& \pt27.161& 22.955& 22.341& 22.131& 22.060& 22.033& &-0.0015\pa&  4.65&[1]$^{2,3}$  & \ldots &\ldots\\
    149720H& 132407.7+273704.83& \pc3.621& 0.367& \pc26.215& 22.594& 22.227& 22.094& 22.061& 22.048& & 0.0006\pa&  3.81&[9]          & \ldots &\ldots\\
    178741H& 132515.4+274212.36& \pc1.760& 0.266& \pc24.286& 22.526& 22.262& 22.268& 22.338& 22.432& & 0.0002\pa&  1.87&[8]          & \ldots &\ldots\\\hline
  \multicolumn{15}{c}{Ambiguous \nuv-dropouts}\\
    185177L& 132442.7+274319.52& \pt3.181& 0.128& \pt27.145& 23.964& 23.837& 23.703& 23.630& 23.286& & 2.6739\D &  2.28& 3           & \ldots &\ldots\\
    165834L& 132431.0+273954.97& \pt4.227& 0.181& \pt27.147& 22.920& 22.762& 22.728& 22.649& 22.607& & 2.1348\D &  2.41&1,6          & \ldots &\ldots\\
  \po56764L& 132449.4+272029.14& \pt2.591& 0.159& \pt27.171& 24.580& 24.320& 24.317& 24.346& 24.192& & 0.2473\D &  2.49&[1]          & \ldots &\ldots\\
  \po80830L& 132503.3+272445.16& \pt2.522& 0.226& \pt27.148& 24.626& 24.414& 24.380& 24.360& 24.351& & 2.0855\D &  2.22&[6]          & \ldots &\ldots\\
  \po96927L& 132523.0+272734.22& \pt3.255& 0.190& \pt27.162& 23.907& 23.707& 23.621& 23.537& 23.384& & 2.6436\D &  2.98& 1$^{6}$     & \ldots &\ldots\\ 
           &                   &         &      &          &       &       &       &       &       & & 1.0713\pa&  2.68&[3]          & \ldots &\ldots\\
  \po92942L& 132515.7+272653.97& \pt2.638& 0.198& \pt27.160& 24.522& 24.253& 24.280& 24.260& 24.236& & 2.1863\D &  2.64& 6           & \ldots &\ldots\\ 
           &                   &         &      &          &       &       &       &       &       & & 0.9496\pa&  2.44&[3]          & \ldots &\ldots\\
    169090L& 132420.8+274025.74& \pt3.154& 0.164& \pt27.137& 23.983& 23.798& 23.660& 23.461& 23.223& & 0.0932\D &  2.51&[2]$^{1,3}$  & \ldots &\ldots\\ 
  \po86765L& 132453.3+272545.01& \pt2.404& 0.571& \pt27.176& 24.772& 24.215& 24.078& 24.025& 23.759& & 0.4717\D &  2.55&[1]$^{2,3}$  & \ldots &\ldots\\ 
    137763H& 132406.1+273502.82& \pc2.547& 0.102& \pc25.458& 22.911& 22.821& 22.798& 22.801& 22.522& & 0.1367\D &  1.89&[8]          & \ldots &\ldots\\
  \po92150H& 132505.3+272646.15& \pc3.903& 0.498& \pc26.516& 22.613& 22.117& 21.934& 21.864& 21.855& & \ldots   &\ldots&\ldots       & \ldots &\ldots\\ 
    176626H& 132352.4+274152.41& \pc4.172& 0.482& \pc26.852& 22.680& 22.201& 21.972& 21.896& 21.837& & 1.6906\D &  2.64& 6           & \ldots &\ldots\\ 
    166856H& 132442.1+274005.24& \pt4.160& 0.305& \pt27.286& 23.126& 22.820& 22.625& 22.469& 22.137& & 0.0071\D &  2.39&[1]          & \ldots &\ldots\\
    146434H& 132524.8+273631.59& \pc1.864& 0.091& \pc24.721& 22.857& 22.755& 22.676& 22.630& 22.470& & 2.1028\D &  2.29& 6           & \ldots &\ldots\\ 
    183911H& 132439.6+274311.57& \pc2.367& 0.157& \pc25.361& 22.994& 22.821& 22.562& 22.396& 22.043& & 2.0213\D &  2.97& 5           & \ldots &\ldots\\
  \po78733H& 132346.4+272426.09& \pc4.126& 0.367& \pc27.164& 23.038& 22.683& 22.586& 22.567& 22.575& & 1.8112\D &  2.15& 7           & \ldots &\ldots\\
  \po66488H& 132520.4+272220.46& \pt4.595& 0.447& \pt27.165& 22.570& 22.126& 21.960& 21.897& 21.833& & 2.4233\D &  2.89& 5           & \ldots &\ldots\\
    190498H& 132516.9+274417.26& \pc3.652& 0.409& \pc26.211& 22.559& 22.148& 21.999& 21.969& 21.930& & 2.3434\D &  2.84& 6           & \ldots &\ldots\\ 
           &                   &         &      &          &       &       &       &       &       & & 0.8932\pa&  2.79&[3]          & \ldots &\ldots\\
    190947H& 132346.0+274419.92& \pt4.344& 0.212& \pt27.140& 22.796& 22.597& 22.423& 22.259& 22.081& & 1.7273\D &  2.89& 3           & \ldots &\ldots\\ 
           &                   &         &      &          &       &       &       &       &       & & 0.9612\pa&  2.38&[3]          & \ldots &\ldots\\
    153628H& 132514.5+273743.52& \pt4.563& 0.457& \pt27.268& 22.705& 22.243& 22.066& 21.993& 21.962& & 0.0551\D &  2.63&[2]          & \ldots &\ldots\\\hline  
   \multicolumn{15}{c}{Undetected \nuv-dropouts}\\
    174747L& 132436.7+274129.12& \pc1.951& 0.247& \pc26.332& 24.381& 24.184& 24.080& 24.028& 23.969& & \ldots   &\ldots&\ldots       & \ldots &\ldots\\
    182447L& 132429.1+274249.80& \pt2.603& 0.385& \pt27.105& 24.502& 24.126& 23.642& 23.192& 22.584& & \ldots   &\ldots&\ldots       & \ldots &\ldots\\
    180088L& 132421.6+274223.22& \pt2.720& 0.129& \pt27.120& 24.400& 24.262& 24.198& 24.040& 23.766& & \ldots   &\ldots&\ldots       & \ldots &\ldots\\
    172253L& 132414.3+274100.26& \pt2.911& 0.224& \pt27.120& 24.209& 24.017& 23.970& 23.960& 23.853& & \ldots   &\ldots&\ldots       & \ldots &\ldots\\
    184387L& 132414.6+274308.58& \pt2.362& 0.173& \pt27.137& 24.775& 24.606& 24.275& 24.109& 23.586& & \ldots   &\ldots&\ldots       & \ldots &\ldots\\
  \po63360L& 132433.2+272142.21& \pt2.353& 0.443& \pt27.167& 24.814& 24.275& 23.840& 23.600& 23.346& & \ldots   &\ldots&\ldots       & \ldots &\ldots\\
    113109L& 132514.6+273028.10& \pt2.478& 0.209& \pt27.158& 24.680& 24.417& 24.408& 24.252& 24.298& & \ldots   &\ldots&\ldots       & \ldots &\ldots\\
  \po94367L& 132459.1+272709.00& \pt3.386& 0.246& \pt27.169& 23.783& 23.567& 23.516& 23.436& 23.309& & \ldots   &\ldots&\ldots       & \ldots &\ldots\\
   \enddata
   \label{table1}
   \tablecomments{Identified sources are based on an $R>3.0$ criterion (exceptions are AGNs, stars, and those with emission lines).
     Col. (1) is the $B$-band catalog ID, Col. (2) is the J2000 coordinates, and magnitudes and colors are given in
     Cols. (3) to (10). The cross-correlated redshifts and $R$-values from \texttt{xcsao} are provided in Cols. (11) and (12). The
     template yielding the highest $R$-value is given in Col. (13), where 1-4 correspond to the four spectra of \cite{steidel03}
     from strongest \Lya~absorption to emission, and 5 and 6 refer to the \cite{shapley03} composite and cB58 spectra, respectively.
     `7' corresponds to the rest-frame NUV spectra from \cite{savaglio04}, and 'AGN' refers to a SDSS QSO template. For interlopers,
     the six SDSS composite spectra presented in \cite{yip04} correspond to [1] to [6] from strongest absorption-line to strongest
     emission-line systems. [7], [8], and [9] correspond to the \textsc{rvsao} templates ``femtemp'', ``EA'', and ``eatemp'',
     respectively. For objects with \Lya~emission, the line flux (in units of 10$^{-18}$~ergs~s$^{-1}$~cm$^{-2}$) and rest-frame
     EWs (in units of \AA) are given in Cols. (14) and (15), respectively. These were measured using the \texttt{splot} routine.}
   \tablenotetext{a}{The character following the ID number corresponds to the spectrograph used: `H'=Hectospec, `L'=LRIS, `M'=MOIRCS.}
   \tablenotetext{b}{Values in superscript correspond to other templates, which yielded similar cross-correlated velocities with $R\geq2.5$.}
   \tablenotetext{c}{As discussed in \S~\ref{appendix}, this source shows an unusual \Lya~profile, so the reported flux and rest-frame
   EW excluded the blue excess by deblending in \textsc{IRAF} \texttt{splot}.}
   \tablenotetext{d}{The values reported here correspond to the best cross-correlated results, but may be wrong due to the
   low S/N of the spectra.} 
\end{deluxetable}
\clearpage
\end{landscape}

\clearpage \newpage
\begin{deluxetable}{lcccccc}
  \tablewidth{0pc}
  \tablecaption{Summary of Spectroscopic Observations}
  \tablehead{
    \colhead{Instrument} &\colhead{Total} &\colhead{LBGs [AGNs]} &\colhead{$z\leq1.5$}
    &\colhead{stars} &\colhead{Ambiguous} &\colhead{Undetected}\\
    \colhead{(1)}&\colhead{(2)}&\colhead{(3)}&\colhead{(4)}&\colhead{(5)}&\colhead{(6)}&\colhead{(7)}}
  \startdata
  LRIS      &  36 (28) \{4\} & 12 \{2\}          &  5 (1) \{2\} & 3 (0) &  8 (7) \{-4\}  & 8 (8)\\
  Hectospec &  21 (20) \{3\} &  7 [2] \{2\}      &  1 (1) \{1\} & 2 (1) & 11 (11) \{-3\} & 0 (0)\\
  MOIRCS    &  44            & 10 (5)            &  4 (2)       & 0     & \ldots         &\ldots\\\hline
  Total     & 101            & 29 (24) [2] \{4\} & 10 (4) \{3\} & 5 (1) & 19 (18) \{-7\} & 8 (8)\\
  \vspace{-3mm}
  \enddata
  \tablecomments{Sources with $z>1.5$ are classified as ``LBG''. Values in square
    brackets are those that appear to be AGNs, and those in parentheses meet the final selection criteria
    in \S~\ref{3.2}. Values in curly brackets represent LBGs that are reclassified as ``identified'' if a
    lower ($R=2.5$) threshold is adopted rather than a $R=3.0$ cut. None of the LBGs and AGNs was missed
    by the final selection criteria.}
  \label{table2}
\end{deluxetable}

\begin{deluxetable}{ccccccccccccc}
  \tabletypesize{\scriptsize}
  \tablewidth{0pc}
  \tablecaption{Effective Volume Estimates}
  \tablehead{
    \colhead{$V$ mag}&\colhead{$z_{avg}$}&\colhead{$\sigma_z$}&\colhead{$FWHM(z)$}&\colhead{$V_{\rm eff}/d\Omega$}&
    \colhead{$z_{avg}$}&\colhead{$\sigma_z$}&\colhead{$FWHM(z)$}&\colhead{$V_{\rm eff}/d\Omega$}&
    \colhead{$z_{avg}$}&\colhead{$\sigma_z$}&\colhead{$FWHM(z)$}&\colhead{$V_{\rm eff}/d\Omega$}\\
    \colhead{(1)}&\colhead{(2)}&\colhead{(3)}&\colhead{(4)}&\colhead{(5)}&\colhead{(6)}&
    \colhead{(7)}&\colhead{(8)}&\colhead{(9)}&\colhead{(10)}&\colhead{(11)}&\colhead{(12)}&\colhead{(13)}}
  \startdata
            & \multicolumn{4}{c}{$E(B-V)=0.0$}     & \multicolumn{4}{c}{$E(B-V)=0.1$}     & \multicolumn{4}{c}{$E(B-V)=0.2$}\\
  22.0-22.5 & 2.422 & 0.312 & 1.860$-$2.962 & 2.93 & 2.313 & 0.318 & 1.757$-$2.860 & 2.89 & 2.150 & 0.304 & 1.627$-$2.667 & 2.69\\
  22.5-23.0 & 2.420 & 0.312 & 1.865$-$2.961 & 2.67 & 2.304 & 0.323 & 1.746$-$2.859 & 2.67 & 2.148 & 0.307 & 1.613$-$2.668 & 2.48\\
  23.0-23.5 & 2.418 & 0.317 & 1.862$-$2.962 & 2.36 & 2.304 & 0.328 & 1.738$-$2.864 & 2.40 & 2.146 & 0.312 & 1.592$-$2.664 & 2.23\\
  23.5-24.0 & 2.401 & 0.333 & 1.831$-$2.969 & 2.12 & 2.290 & 0.339 & 1.717$-$2.863 & 2.20 & 2.166 & 0.320 & 1.585$-$2.682 & 1.97\\
  24.0-24.5 & 2.379 & 0.361 & 1.772$-$2.974 & 2.14 & 2.261 & 0.355 & 1.645$-$2.846 & 2.10 & 2.157 & 0.331 & 1.570$-$2.684 & 1.93\\
  24.5-25.0 & 2.312 & 0.369 & 1.680$-$2.910 & 2.14 & 2.199 & 0.349 & 1.575$-$2.774 & 2.02 & 2.104 & 0.314 & 1.561$-$2.631 & 1.81\\
  25.0-25.5 & 2.220 & 0.352 & 1.603$-$2.788 & 1.80 & 2.133 & 0.325 & 1.569$-$2.659 & 1.44 & 2.043 & 0.292 & 1.560$-$2.487 & 1.01\\\hline
            & \multicolumn{4}{c}{$E(B-V)=0.3$}     & \multicolumn{4}{c}{$E(B-V)=0.4$}     & \multicolumn{4}{c}{$E(B-V)=0.0-0.4$}\\
  22.0-22.5 & 2.074 & 0.290 & 1.563$-$2.569 & 2.41 & 1.956 & 0.236 & 1.553$-$2.359 & 1.78 & 2.278 & 0.333 & 1.752$-$2.837 & 2.79\\
  22.5-23.0 & 2.079 & 0.287 & 1.573$-$2.567 & 2.16 & 1.952 & 0.233 & 1.559$-$2.352 & 1.57 & 2.274 & 0.335 & 1.740$-$2.834 & 2.56\\
  23.0-23.5 & 2.088 & 0.291 & 1.573$-$2.574 & 1.88 & 1.957 & 0.230 & 1.562$-$2.350 & 1.40 & 2.273 & 0.338 & 1.741$-$2.832 & 2.28\\
  23.5-24.0 & 2.077 & 0.295 & 1.567$-$2.572 & 1.79 & 1.952 & 0.232 & 1.557$-$2.349 & 1.34 & 2.268 & 0.345 & 1.711$-$2.839 & 2.07\\
  24.0-24.5 & 2.060 & 0.293 & 1.559$-$2.559 & 1.77 & 1.934 & 0.235 & 1.553$-$2.313 & 1.28 & 2.248 & 0.360 & 1.610$-$2.834 & 2.02\\
  24.5-25.0 & 2.002 & 0.279 & 1.557$-$2.422 & 1.29 & 1.887 & 0.237 & 2.131$-$2.243 & 0.56 & 2.194 & 0.354 & 1.574$-$2.762 & 1.91\\
  25.0-25.5 & 1.898 & 0.266 & 1.550$-$2.004 & 0.35 & 1.967 & 0.334 & 1.962$-$2.062 & 0.08 & 2.139 & 0.336 & 1.571$-$2.652 & 1.30\\
  \vspace{-3mm}
  \enddata
  \label{table3}
  \tablecomments{Results for MC simulations described in \S~\ref{4.3} for different assumed $E(B-V)$ values. Col. (1) lists the
    apparent $V$-band magnitude. Cols. (2), (6), and (10) show the average redshift ($z_{avg}$), and Cols. (3), (7),
    and (11) list the redshift $1\sigma$ uncertainties. Cols. (4), (8), and (12) give the FWHM of the redshift distribution, and
     the effective comoving volume per area in units of $10^3$ $h_{70}^{-3}$ Mpc$^3$ arcmin$^{-2}$ are in Cols. (5), (9),
     and (13).}
\end{deluxetable}

\clearpage \newpage
\newcommand{\ul}[2]{^{+#1}_{-#2}}
\newcommand{\E}{\tablenotemark{e}}
\newcommand{\F}{\tablenotemark{f}}
\newcommand{\pn}{\phantom{\tablenotemark{a}}}
\begin{landscape}
\begin{deluxetable}{lcrrcccccccccc}
  \tablewidth{0pc}
  \tabletypesize{\scriptsize}
  \tablecaption{Compilation of UV Luminosity Functions and Luminosity Densities}
  \tablehead{
    \colhead{Reference}&\colhead{$z$}&\colhead{$N$}&\colhead{Area}&\colhead{$\lambda_{\rm rest}$}&\colhead{$C_E$}&\colhead{$C_L$}&\colhead{$C_{\Phi}$}&
    \colhead{$\log{\phi_{\star}}$}&\colhead{$M_{\star}$}&\colhead{$\alpha$}&\multicolumn{3}{c}{$\log{\mathcal{L_{\rm obs}}}$}\\
    & & & & & & & & & & & \colhead{$L\geq0.1L_{z=3}$}&\colhead{$L\geq L_{\rm lim}$}&\colhead{$L\geq0$}\\ 
    \cline{12-14}
    \colhead{(1)}&\colhead{(2)}&\colhead{(3)}&\colhead{(4)}&\colhead{(5)}&\colhead{(6)}&\colhead{(7)}&\colhead{(8)}&\colhead{(9)}&\colhead{(10)}&\colhead{(11)}&\colhead{(12)}&\colhead{(13)}&\colhead{(14)}}
  \startdata
  \cite{bouwens06}    &5.90$\pm$0.30        &  506& 344.2&1350&  1.51&1.000&1.000&-2.69$\ul{0.15}{0.21}$   &-20.25$\pm$0.20        &-1.73$\ul{0.21}{0.20}$   &26.047                 &26.224$\ul{0.10}{0.13}$&26.567                 \\
  \cite{bouwens07}    &3.80$\pm$0.35        & 4671& 347  &1600& 2.692&1.000&1.000&-2.89$\ul{0.06}{0.07}$   &-20.98$\pm$0.10        &-1.73$\pm$0.05        &26.276                 &26.503$\pm$0.05        &26.668                 \\
                      &5.00$\pm$0.35        & 1416& 367  &1600& 1.995&1.000&1.000&-3.00$\ul{0.11}{0.16}$   &-20.64$\pm$0.13        &-1.66$\pm$0.09        &25.952                 &26.161$\pm$0.06        &26.313                 \\
                      &5.90$\pm$0.30        &  627& 396  &1350&  1.51&1.000&1.000&-2.85$\ul{0.16}{0.15}$   &-20.24$\pm$0.19        &-1.74$\pm$0.16        &25.885                 &26.100$\pm$0.08        &26.421                 \\
  \cite{burgarella07} &1.10$\pm$0.20        &  420& 947  &1800& 3.651&1.000&1.000&\ldots                &\ldots                 &\ldots                &\ldots                &25.738$\ul{0.08}{0.10}$&\ldots                \\
  \cite{foucaud03}    &3.20$\ul{0.02}{0.30}$& 1294&1700  &1900&\ldots&1.000&1.000&\ldots                &\ldots                 &\ldots                &\ldots                &\ldots                &\ldots                \\
  \cite{giavalisco04} &3.78$\pm$0.34        & 1115& 316  &1500&  7.48&1.000&1.000&\ldots                &\ldots                 &-1.60\F               &\ldots                &26.212$\pm$0.07        &\ldots                \\
                      &4.92$\pm$0.33        &  275& 316  &1500&  7.48&1.000&1.000&\ldots                &\ldots                 &-1.60\F               &\ldots                &26.017$\pm$0.14        &\ldots                \\
                      &5.74$\pm$0.36        &  122& 316  &1500&  7.48&1.000&1.000&\ldots                &\ldots                 &-1.60\F               &\ldots                &26.061$\pm$0.19        &\ldots                \\
  \cite{hildebrandt07}&2.96$\pm$0.24        &14283& 9.99 &1650&  3.88&1.000&1.000&-3.29$\pm$0.08\E      &-22.43$\pm$0.11\E     &-1.60\E\F            &26.362                 &26.097                 &26.492                 \\
  {\bf\cite{iwata07}} &4.80$\pm$0.40        &  853&1290  &1500&  7.48&1.000&1.000&-3.39$\ul{0.23}{0.53}$   &-21.23$\pm$0.30        &-1.45$\ul{0.38}{0.32}$   &25.841$\ul{0.06}{0.04}$&25.679                 &25.995$\ul{0.23}{0.09}$\\
  \cite{madau96}      &2.75$\pm$0.75        &   69& 4.65 &1620&  3.23&1.108&0.712&\ldots                &\ldots                 &\ldots                &\ldots                &$<$26.101              &\ldots                \\
                      &4.00$\pm$0.50        &   14&4.65  &1630&  3.23&1.170&0.662&\ldots                &\ldots                 &\ldots                &\ldots                &$<$25.588              &\ldots                \\
  \cite{mass01}       &1.50$\pm$0.50        &  315&5.31  &1500&  8.10&0.980&0.844&\ldots                &\ldots                 &\ldots                &\ldots                &26.240$\pm$0.19        &\ldots                \\
                      &2.75$\pm$0.75        &  232&5.31  &1500& 12.17&1.108&0.712&\ldots                &\ldots                 &\ldots                &\ldots                &26.259$\pm$0.19        &\ldots                \\
                      &4.00$\pm$0.50        &   54&5.31  &1500& 12.00&1.170&0.662&\ldots                &\ldots                 &\ldots                &\ldots                &25.889$\pm$0.24        &\ldots                \\
  \cite{paltani07}    &3.50$\pm$0.50        &  113&1720  &1700&  3.80&1.000&1.000&-2.91$\ul{0.14}{0.22}$   &-21.49$\pm$0.19        &-1.40\F               &26.093$\ul{0.16}{0.26}$&\ldots                &26.499$\ul{0.09}{0.12}$\\
  \cite{reddy08}      &2.20$\pm$0.32        &10007&1925.8&1700& 3.796&1.000&1.000&-2.76$\ul{0.13}{0.20}$   &-20.97$\pm$0.23        &-1.84$\pm$0.11        &26.439$\ul{0.06}{0.07}$&26.256                 &27.030                 \\
  \cite{ST06a,ST06b}  &2.20$\pm$0.32        & 2417& 169  &1700&  5-15&1.000&1.000&-2.52$\ul{0.20}{0.26}$   &-20.60$\ul{0.38}{0.44}$   &-1.20$\ul{0.24}{0.22}$   &26.320$\pm$0.02        &26.314                 &26.424$\pm$0.03        \\
                      &2.96$\pm$0.26        & 1481& 169  &1700&  5-15&1.000&1.000&-2.77$\ul{0.13}{0.09}$   &-20.90$\ul{0.22}{0.14}$   &-1.43$\ul{0.17}{0.09}$   &26.257$\pm$0.01        &26.303                 &26.422$\pm$0.03        \\
                      &4.13$\pm$0.26        &  427& 169  &1700&  5-15&1.000&1.000&-3.07$\ul{0.21}{0.33}$   &-21.00$\ul{0.40}{0.46}$   &-1.26$\ul{0.40}{0.36}$   &25.969$\pm$0.03        &25.965$\pm$0.03        &26.061$\pm$0.07        \\
  \cite{shima05}      &5.90$\pm$0.30        &   12& 767  &1425& 4.358&1.000&1.000&\ldots                &\ldots                 &\ldots                &\ldots                &24.447$\ul{0.11}{0.15}$&\ldots                \\
  \cite{shim07}       &3.20$\pm$0.14        & 1088&9468  &1550& 3.964&1.000&1.000&-2.81\E              &-20.69\E              &-0.83\E\F            &26.027                 &25.578                 &26.067                 \\
  \cite{steidel99}    &2.96$\pm$0.26        & 1270&1046  &1700& 3.796&1.126&0.693&-2.86                 &-21.07$\pm$0.15        &-1.60$\pm$0.13        &26.462                 &26.193                 &26.711                 \\
                      &4.13$\pm$0.26        &  207& 828  &1700& 3.796&1.175&0.658&-2.97                 &-21.11                 &-1.60\F               &26.397                 &25.734                 &26.640                 \\
  \cite{wadadekar06}  &2.75$\pm$0.75        &  125& 5.67 &1850& 4.70 &0.862&1.249&\ldots                &\ldots                 &\ldots                &26.518                 &\ldots                &\ldots                \\
  \cite{yoshida06}    &4.00$\pm$0.30        & 3808& 875  &1500& 3.381&1.000&1.000&-2.84$\ul{0.11}{0.12}$   &-21.14$\ul{0.14}{0.15}$   &-1.82$\pm$0.09        &26.450$\ul{0.06}{0.07}$&26.343$\pm$0.02        &26.968                 \\
                      &4.70$\pm$0.30        &  539& 875  &1500& 3.381&1.000&1.000&-2.91$\ul{0.13}{0.11}$   &-20.72$\ul{0.16}{0.14}$   &-1.82\F               &26.138$\pm$0.24        &25.645$\ul{0.03}{0.05}$&26.726                 \\
  This work           &2.28$\pm$0.33        & 7093&857.5 &1700& 3.796&1.000&1.000&-2.25$\pm$0.46         &-20.50$\pm$0.79        &-1.05$\pm$1.11        &26.519$\pm$0.68        &26.276$\pm$0.68        &26.601$\pm$0.92        \\
                      &                     &     &      &    &      &     &     &-2.50$\pm$0.17         &-20.95$\ul{0.43}{0.29}$   &-1.60\F               &26.601$\pm$0.07        &26.293$\pm$0.07        &26.864$\pm$0.07        \\
                      &                     &     &      &    &      &     &     &-2.75$\pm$0.21         &-21.30$\pm$0.35        &-1.84\F               &26.633$\pm$0.08        &26.305$\pm$0.08        &27.172$\pm$0.08        \\
  \vspace{-3mm}
  \enddata
  \tablecomments{Cols. (1) through (5) list the reference, the redshift, the sample size,
    the area (arcmin$^{2}$), and the rest-wavelength of measurements (\AA). Dust extinction correction is provided
    in Col. (6), and corrections to a common cosmology for luminosity and number density are shown in Cols. (7) and (8)
.    Schechter LF parameters are listed in Cols. (9)-(11) in units of Mpc$^{-3}$ for number density, and finally the
    {\it observed} luminosity densities (erg s$^{-1}$ Hz$^{-1}$ Mpc$^{-3}$) are provided in Cols. (12)-(14)
for three different limits of integration.}
  \label{table4}
  \tablenotetext{e}{These values were not reported in the paper, but was obtained by fitting their LF with data
    provided in the paper or by the authors. For \cite{shim07}, the results are not well constrained given the limited
    number of degrees of freedom.}
  \tablenotetext{f}{This value was kept fixed, while the other Schechter parameters were fitted.}
\end{deluxetable}
\clearpage
\end{landscape}

\clearpage


\begin{thebibliography}{}
\bibitem[Adelberger et al.(2004)]{adelberger04}
  Adelberger, K.~L., Steidel, C.~C., Shapley, A.~E., Hunt, M.~P., Erb, D.~K., Reddy, N.~A., \& 
  Pettini, M.\ 2004, \apj, 607, 226 
\bibitem[Bershady et al.(1999)]{bershady99}
  Bershady, M.~A., Charlton, J.~C., \& Geoffroy, J.~M.\ 1999, \apj, 518, 103
\bibitem[Bertin \& Arnouts(1996)]{bertin96}
  Bertin, E., \& Arnouts, S.\ 1996, \aaps, 117, 393 
\bibitem[Bouwens et al.(2006)]{bouwens06}
  Bouwens, R.~J., Illingworth, G.~D., Blakeslee, J.~P., \& Franx, M.\ 2006, \apj, 653, 53
\bibitem[Bouwens et al.(2007)]{bouwens07}
  Bouwens, R.~J., Illingworth, G.~D., Franx, M., \& Ford, H.\ 2007, \apj, 670, 928
\bibitem[Burgarella et al.(2007)]{burgarella07}
  Burgarella, D., et al.\ 2007, \mnras, 380, 986
\bibitem[Bruzual \& Charlot(2003)]{bc03}
  Bruzual, G., \& Charlot, S.\ 2003, \mnras, 344, 1000
\bibitem[Calzetti et al.(2000)]{calzetti00}
  Calzetti, D., Armus, L., Bohlin, R.~C., Kinney, A.~L., Koornneef, J., \& Storchi-Bergmann, T.\ 
  2000, \apj, 533, 682
\bibitem[Cardelli et al.(1989)]{cardelli89}
  Cardelli, J.~A., Clayton, G.~C., \& Mathis, J.~S.\ 1989, \apj, 345, 245 
\bibitem[Chapman et al.(2005)]{chapman05}
  Chapman, S.~C., Blain, A.~W., Smail, I., \& Ivison, R.~J.\ 2005, \apj, 622, 772 
\bibitem[Cooke et al.(2008)]{cooke08}
  Cooke, J., Barton, E.~J., Bullock, J.~S., Stewart, K.~R., \& Wolfe, A.~M.\ 2008, \apjl, 681, L57 
\bibitem[Daddi et al.(2004)]{daddi04}
  Daddi, E., Cimatti, A., Renzini, A., Fontana, A., Mignoli, M., Pozzetti, L., Tozzi, P., \& 
  Zamorani, G.\ 2004, \apj, 617, 746
\bibitem[Erb et al.(2003)]{erb03}
  Erb, D.~K., Shapley, A.~E., Steidel, C.~C., Pettini, M., Adelberger, K.~L., Hunt, M.~P.,
  Moorwood, A.~F.~M., \& Cuby, J.-G.\ 2003, \apj, 591, 101
\bibitem[Fabricant et al.(2005)]{fabricant05}
  Fabricant, D., et al.\ 2005, \pasp, 117, 1411
\bibitem[Foucaud et al.(2003)]{foucaud03}
  Foucaud, S., et  al.\ 2003, \aap, 409, 835
\bibitem[Giavalisco(2002)]{giavalisco02}
  Giavalisco, M.\ 2002, \araa, 40, 579
\bibitem[Giavalisco et al.(2004)]{giavalisco04}
  Giavalisco, M., et al.\ 2004, \apjl, 600, L103
\bibitem[Gunn \& Stryker(1983)]{GS}
  Gunn, J.~E., \& Stryker, L.~L.\ 1983, \apjs, 52, 121
\bibitem[Hayashi et al.(2007)]{hayashi07}
  Hayashi, M., Shimasaku, K., Motohara, K., Yoshida, M., Okamura, S., \& Kashikawa, N.\ 2007,
  \apj, 660, 72
\bibitem[Hayashi et al.(2008)]{hayashi08}
  Hayashi, M., et al.\ 2009, \apj, 691, 140 
\bibitem[Hopkins(2004)]{hopkins04}
  Hopkins, A.~M.\ 2004, \apj, 615, 209
\bibitem[Mas-Hesse et al.(2003)]{hesse03}
  Mas-Hesse, J.~M., Kunth, D., Tenorio-Tagle, G., Leitherer, C., Terlevich, R.~J., 
  \& Terlevich, E.\ 2003, \apj, 598, 858
\bibitem[Hildebrandt et al.(2007)]{hildebrandt07}
  Hildebrandt, H., et al.\ 2007, \aap, 462, 865
\bibitem[Ichikawa et al.(2006)]{ichikawa06} 
  Ichikawa, T., et al.\ 2006, \procspie, 6269, 
\bibitem[Iwata et al.(2007)]{iwata07}
  Iwata, I., Ohta, K., Tamura, N., Akiyama, M., Aoki, K., Ando, M., Kiuchi, G., 
  \& Sawicki, M.\ 2007, \mnras, 376, 1557 
\bibitem[Iye et al.(2004)]{iye04}
  Iye, M., et al.\ 2004, \pasj, 56, 381 
\bibitem[Kashikawa et al.(2004)]{kashik04}
  Kashikawa, N., et  al.\ 2004, \pasj, 56, 1011
\bibitem[Kennicutt(1998)]{kennicutt98}
  Kennicutt, R.~C.\ 1998, \araa, 36, 189 
\bibitem[Kurtz \& Mink(1998)]{kurtz98}
  Kurtz, M.~J., \& Mink, D.~J.\ 1998, \pasp, 110, 934 
\bibitem[Ly et al.(2007)]{ly07}
  Ly, C., et al.\ 2007, \apj, 657, 738
\bibitem[Madau(1995)]{madau95}
  Madau, P.\ 1995, \apj, 441, 18
\bibitem[Madau et al.(1996)]{madau96}
  Madau, P., Ferguson, H.~C., Dickinson, M.~E., Giavalisco, M., Steidel, C.~C., \& Fruchter,
  A.\ 1996, \mnras, 283, 1388 
\bibitem[Madau et al.(1998)]{madau98}
  Madau, P., Pozzetti, L., \& Dickinson, M.\ 1998, \apj, 498, 106
\bibitem[Malkan et al.(1996)]{malkan96}
  Malkan, M.~A., Teplitz, H., \& McLean, I.~S.\ 1996, \apjl, 468, L9
\bibitem[Martin et al.(2005)]{martin05}
  Martin, D.~C., et al.\ 2005, \apjl, 619, L1 
\bibitem[Massarotti et al.(2001)]{mass01}
  Massarotti, M., Iovino, A., \& Buzzoni, A.\ 2001, \apjl, 559, L105 
\bibitem[Miyazaki et al.(2002)]{miyazaki02}
  Miyazaki, S., et al.\ 2002, \pasj, 54, 833 
\bibitem[Moorwood et al.(2000)]{moorwood00}
  Moorwood, A.~F.~M., van der Werf, P.~P., Cuby, J.~G., \& Oliva, E.\ 2000, \aap, 362, 9 
\bibitem[Morrissey et al.(2007)]{morrissey07}
  Morrissey, P., et al.\ 2007, \apjs, 173, 682 
\bibitem[Oke(1974)]{oke74}
  Oke, J.~B.\ 1974, \apjs, 27, 21
\bibitem[Oke \& Gunn(1983)]{oke83}
  Oke, J.~B., \& Gunn, J.~E.\ 1983, \apj, 266, 713
\bibitem[Oke et al.(1995)]{oke95} 
  Oke, J.~B., et al.\ 1995, \pasp, 107, 375 
\bibitem[Paltani et al.(2007)]{paltani07}
  Paltani, S., et al.\ 2007, \aap, 463, 873
\bibitem[Pettini et al.(2000)]{pettini00}
  Pettini, M., Steidel, C.~C., Adelberger, K.~L., Dickinson, M., \& Giavalisco, M.\ 2000, \apj, 
  528, 96 
\bibitem[Reddy et al.(2008)]{reddy08}
  Reddy, N.~A., Steidel, C.~C., Pettini, M., Adelberger, K.~L., Shapley, A.~E., Erb, D.~K., 
  \& Dickinson, M.\ 2008, \apjs, 175, 48
\bibitem[Richmond(2005)]{richmond05}
  Richmond, M.\ 2005, \pasj, 57, 969
\bibitem[Robin et al.(2003)]{robin03}
  Robin, A.~C., Reyl{\'e}, C., Derri{\`e}re, S., \& Picaud, S.\ 2003,
  \aap, 409, 523 
\bibitem[Saito et al.(2006)]{saito06}
  Saito, T., Shimasaku, K., Okamura, S., Ouchi, M., Akiyama, M., \& Yoshida, M.\ 2006, \apj, 648, 54
\bibitem[Savaglio et al.(2004)]{savaglio04}
  Savaglio, S., et al.\ 2004, \apj, 602, 51
\bibitem[Sawicki \& Thompson(2006a)]{ST06a}
  Sawicki, M., \& Thompson, D.\ 2006a, \apj, 642, 653 
\bibitem[Sawicki \& Thompson(2006b)]{ST06b}
  Sawicki, M., \& Thompson, D.\ 2006b, \apj, 648, 299
\bibitem[Schechter(1976)]{schechter76}
  Schechter, P.\ 1976, \apj, 203, 297
\bibitem[Shapley et al.(2003)]{shapley03}
  Shapley, A.~E., Steidel, C.~C., Pettini, M., \& Adelberger, K.~L.\ 2003, \apj, 588, 65 
\bibitem[Shapley et al.(2006)]{shapley06}
  Shapley, A.~E., Steidel, C.~C., Pettini, M., Adelberger, K.~L., \& Erb, D.~K.\ 2006,
  \apj, 651, 688
\bibitem[Shim et al.(2007)]{shim07}
  Shim, H., Im, M., Choi, P., Yan, L., \& Storrie-Lombardi, L.\ 2007, \apj, 669, 749 
\bibitem[Shimasaku et al.(2005)]{shima05}
  Shimasaku, K., Ouchi, M., Furusawa, H., Yoshida, M., Kashikawa, N., \& Okamura, S.\ 2005, \pasj, 
  57, 447
\bibitem[Steidel et al.(1999)]{steidel99}
  Steidel, C.~C., Adelberger, K.~L., Giavalisco, M., Dickinson, M., \& Pettini, M.\ 1999, \apj,
  519, 1 
\bibitem[Steidel et al.(2000)]{steidel00}
  Steidel, C.~C., Adelberger, K.~L., Shapley, A.~E., Pettini, M., Dickinson, M., \& Giavalisco,
  M.\ 2000, \apj, 532, 170
\bibitem[Steidel et al.(2003)]{steidel03}
  Steidel, C.~C., Adelberger, K.~L., Shapley, A.~E., Pettini, M., Dickinson, M., \& 
  Giavalisco, M.\ 2003, \apj, 592, 728
\bibitem[Tapken et al.(2004)]{tapken04}
  Tapken, C., Appenzeller, I., Mehlert, D., Noll, S., \& Richling, S.\ 2004, \aap, 416, L1 
\bibitem[Tapken et al.(2007)]{tapken07}
  Tapken, C., Appenzeller, I., Noll, S., Richling, S., Heidt, J., Meink{\"o}hn, E., \& Mehlert,
  D.\ 2007, \aap, 467, 63
\bibitem[Verhamme et al.(2008)]{verhamme08}
  Verhamme, A., Schaerer, D., Atek, H., \& Tapken, C.\ 2008, \aap, 491, 89 
\bibitem[van Dokkum et al.(2004)]{dokkum04}
  van Dokkum, P.~G., et al.\ 2004, \apj, 611, 703
\bibitem[van der Werf et al.(2000)]{werf00}
  van der Werf, P.~P., Moorwood, A.~F.~M., \& Bremer, M.~N.\ 2000, \aap, 362, 509 
\bibitem[Wadadekar et al.(2006)]{wadadekar06}
  Wadadekar, Y., Casertano, S., \& de Mello, D.\ 2006, \aj, 132, 1023
\bibitem[Yip et al.(2004)]{yip04}
  Yip, C.~W., et al.\ 2004, \aj, 128, 585
\bibitem[Yoshida et al.(2006)]{yoshida06}
  Yoshida, M., et al.\ 2006, \apj, 653, 988
\end{thebibliography}
\end{document}